\documentclass[12pt]{article}
\pdfoutput=1
\usepackage[margin=1in]{geometry}
\usepackage{amsmath, amssymb}
\usepackage{cite}
\usepackage{graphicx}
\usepackage[font=small,labelfont=bf]{caption}
\usepackage[normalem]{ulem}
\usepackage{makecell}

\usepackage{slashed}
\usepackage{cancel}
\usepackage{bm}
\usepackage{verbatim}
\usepackage{tabularx}
\usepackage{soul}
\allowdisplaybreaks

\usepackage{enumerate}

\usepackage[utf8]{inputenc}
\usepackage{graphicx}
\usepackage{caption}
\usepackage{hhline}
\usepackage{amsmath}
\usepackage{amsfonts}
\usepackage{amssymb}
\usepackage[dvipsnames]{xcolor}
\RequirePackage{luatex85}
\usepackage{braket}
\usepackage{array} 
\usepackage{pict2e}
\usepackage{multirow}
\usepackage[compat=1.1.0]{tikz-feynman}
\usepackage[super]{nth}
\usepackage{float}
\pagecolor{white}
\definecolor{babyblueeyes}{rgb}{0.63, 0.79, 0.95}
\definecolor{carminepink}{rgb}{0.92, 0.3, 0.26}

\usepackage{stackengine}
\stackMath

\definecolor{amethyst}{rgb}{0.6, 0.4, 0.8}
\definecolor{antiquefuchsia}{rgb}{0.57, 0.36, 0.51}
\definecolor{blue-violet}{rgb}{0.54, 0.17, 0.89}

\definecolor{DarkRed}{rgb}{0.6,0,0}
\definecolor{DarkGreen}{rgb}{0,0.6,0}
\definecolor{DarkOrange}{rgb}{1,0.35,0}
\definecolor{DarkBlue}{rgb}{0,0,0.6}

\usepackage{setspace}
\usepackage{authblk}
\usepackage{physics}
\usepackage{subcaption}
\usepackage{multirow}

\usepackage{bbm}

\usepackage{dsfont}

\usepackage{etoolbox}

\let\OLDthebibliography\thebibliography
\renewcommand\thebibliography[1]{
	\OLDthebibliography{#1}
	\setlength{\parskip}{3.0pt plus 2.5pt minus 1.0pt}
	\setlength{\itemsep}{3.0pt plus 2.5pt minus 1.0pt}
}

\usepackage[colorlinks=true,linktocpage=true,linkcolor=DarkOrange,citecolor=blue,urlcolor=blue,pagebackref=true]{hyperref}

\usepackage{cleveref}
\crefname{table}{Table}{Tables}
\crefname{equation}{Eq.}{Eqs.}
\crefname{appendix}{App.}{Apps.}
\crefname{section}{Sec.}{Secs.}
\crefname{figure}{Fig.}{Figs.}

\definecolor{goodgreen}{rgb}{0,.6,0.4}

\newcommand{\mq}{m_q} 
\newcommand{\qD}{q_{D}} 
\newcommand{\aD}{\alpha_{D}} 
\newcommand{\LD}{\Lambda_{D}} 
\newcommand{\piD}{\pi_{D}} 
\newcommand{\rhoD}{\rho_{D}} 
\newcommand{\Zp}{Z^\prime} 
\newcommand{\Nc}{N_c} 
\newcommand{\Nf}{N_f} 

\makeatletter

\newcounter{subsubsubsection}[subsubsection]

\renewcommand{\thesubsubsubsection}{%
  \thesubsubsection.\arabic{subsubsubsection}%
}

\newcommand{\subsubsubsection}[1]{%
  \refstepcounter{subsubsubsection}%
  \par
  \addvspace{3.25ex\@plus 1ex \@minus .2ex}%
  \noindent
  {\normalfont\normalsize\bfseries
    \thesubsubsubsection\quad #1\par}%
  \nobreak
  \vspace{1.5ex\@plus .2ex}%
  \addcontentsline{toc}{subsubsubsection}{%
    \protect\numberline{\thesubsubsubsection}#1%
  }%
}

\newcommand*\l@subsubsubsection{\@dottedtocline{4}{7.0em}{4.1em}}

\makeatother

\title{{\vskip 1em} Rich Phenomenology from Simple Ingredients: \\ A Review of Confining Dark Sectors \\ \vskip 1em}

\author{Pouya Asadi$^1$, Austin Batz$^{2,3}$, Graham D. Kribs$^{2,4}$\\ \vskip 0em
{\small \color{purple} 
\texttt{pasadi@ucsc.edu,
abatz@uoregon.edu, 
kribs@uoregon.edu},}
\\\vskip 1.5em
{\small \textit{${}^1$Department of Physics and Santa Cruz Institute for Particle Physics,} \\ 
\textit{University of California Santa Cruz, Santa Cruz, CA 95064, USA}\\\vskip 0.3cm
\textit{${}^{2}$Institute for Fundamental Science and Department of Physics,}\\
\textit{University of Oregon, Eugene, OR 97403, USA} \\ \vskip 0.3cm
\textit{${}^3$Theoretical Physics Group, Lawrence Berkeley National Laboratory,} \\
\textit{Berkeley, CA 94720, USA
} \\\vskip 0.3cm
\textit{${}^4$Theoretical Physics Department, CERN, 1211 Geneva 23, Switzerland}
}}

\usepackage[absolute,overlay]{textpos}

\date{}

\begin{document}

\graphicspath{{figs/}}

\begin{textblock*}{4cm}(14cm,2.5cm)

\raggedleft

{CERN-TH-2026-156}

\end{textblock*}

\maketitle
\begin{abstract}
We review theories with confining dark sectors and their implications for dark matter, cosmology, phenomenology, and unsolved Standard Model puzzles. Models with new strongly-coupled non-Abelian gauge interactions can lead to a variety of dark matter candidates (dark mesons, baryons, glueballs, etc.), as well as mechanisms to generate its abundance and symmetries that explain its stability. There are also many potential discovery channels, including direct detection, indirect detection, astrophysical observables, and colliders, as well as correlations between different experiments. We compile a broad conceptual overview of the literature on this topic, aimed at both theorists looking for which questions remain unanswered and experimentalists looking for novel search opportunities. While the theoretical landscape is vast, there are both unifying features and calculational techniques that apply to various regimes. We particularly highlight applications to explaining the similarity of visible and dark matter energy densities, i.e.~the \emph{abundance similarity puzzle}. We advocate further exploration of this class of theories in the effort to uncover physics beyond the Standard Model.

\end{abstract}

\newpage

\tableofcontents

\begin{spacing}{1.15}

\newpage 

\section{Introduction}

The overwhelming gravitational evidence for dark matter (DM) has driven an extensive effort to uncover its particle nature and possible non-gravitational interactions, see Ref.~\cite{Cirelli:2024ssz} for a recent review. 
Constructing viable DM models and exploring their implications have therefore become central in modern particle phenomenology. Two of the earliest and still most-studied DM candidates are the Weakly Interacting Massive Particle (WIMP), interacting with Standard Model (SM) via the weak force \cite{Lee:1977ua}, and the quantum chromodynamics (QCD) axion \cite{Preskill:1982cy}. Each of these is particularly compelling due to both their minimality and their relations to solutions to other puzzles, namely the electroweak hierarchy problem in the case of WIMPs and the strong $CP$ problem for axions. However, as WIMP parameter space has become increasingly constrained \cite{Arcadi:2024ukq} and experiments have begun probing the canonical QCD axion models \cite{Baryakhtar:2025jwh}, there is motivation to study alternative theories with qualitatively different phenomenology.

A natural class of models to consider are those with a confining dark sector: a set of particles beyond the SM which interact via some non-Abelian gauge group that becomes strongly-coupled at an infrared (IR) scale. An early example was technicolor, wherein the electroweak scale is dynamically generated by dimensional transmutation in a strongly-coupled sector.
Analogously to the WIMP emerging as a DM candidate from supersymmetry (SUSY), technicolor was the first model to provide a DM candidate in the form of a dark baryon that is naturally stabilized by a conserved dark baryon number symmetry \cite{Nussinov:1985xr,Chivukula:1989qb,Barr:1990ca,Bagnasco:1993st,Gudnason:2006ug,Gudnason:2006yj,Ryttov:2008xe,Foadi:2008qv}.
While technicolor itself suffers from stringent electroweak precision constraints and the lack of observed techniparticles, it provided an early template for how dark sector confinement can dynamically generate mass scales and supply stable DM candidates protected by accidental symmetries.
The majority of visible matter energy density is made of hadrons that have a mass set by the QCD confinement scale and that are stabilized by an accidental baryon number symmetry, so a dark sector with its own strong dynamics and stabilizing symmetries is arguably the \emph{DM framework most closely aligned} with the structure of the SM\@.

The generalization from technicolor models to confining dark sectors is analogous to other cases where canonical models become particular manifestations of a broader organizing principle. WIMPs are members of a larger class of simple models where the DM candidate communicates with the SM via some ``portal'' interaction \cite{Pospelov:2007mp}, and similarly the QCD axion is one of many axion-like particles (ALPs).
Furthermore, experimental developments have prompted a broadening of the scope of searches to include new mass ranges and interaction structures, alongside the development of alternative theoretical frameworks.
In this expanded view, it is particularly well-motivated to pursue models that yield qualitatively novel phenomenology, thereby inspiring new search strategies, as well as those addressing additional theoretical puzzles beyond providing an origin of dark matter itself.
These developments have contributed to confining DM models gaining more popularity over the last two decades.\footnote{Such DM relics are also referred to as \textit{composite DM} states; 
while weakly-bounded composite states can be viable DM candidates as well \cite{Kaplan:2009de}, in this review, we focus on DM candidates that are bound states of a confining force and use the terms composite strongly-coupled DM or confining DM interchangeably. }
Dark hadron DM models include a plethora of possibilities from dark baryons to dark mesons and dark glueballs. 
These generalizations have revealed novel solutions to various puzzles as well as a variety of potential experimental signatures.

In the near absence of empirical guidance on the particle nature of DM, the range of viable candidates is virtually unbounded. This includes even so-called “nightmare” scenarios, in which DM interacts with the SM only gravitationally \cite{Morrison:2020yeg}. While such frameworks are logically consistent, it is important to recall a suggestive empirical hint for the existence of non-gravitational DM-SM interactions: the striking similarity between the present-day DM and SM energy densities. We refer to this as the ``\textit{abundance similarity puzzle},'' or simply the \textit{abundance puzzle} throughout the rest of this review. 
While the popular DM models discussed above originally emerged as off-shoots of broader beyond the SM frameworks motivated by other shortcomings of the SM, the study of DM particle properties has increasingly come into its own. 
As this direction matures, it is only natural that the abundance puzzle takes a more central role.

Considering that minimal changes in model parameters could easily give rise to orders-of-magnitude changes in both the visible and the DM abundance, the similarity of the energy density of DM and matter, 
that was established no later than the time of matter-radiation equality, is one of the most surprising cosmic observations.
The fact that these \textit{a priori} unrelated energy densities, which could span many orders of magnitude, happen to be within an $\mathcal{O}(1)$ factor of each other is the \emph{only suggestive hint} implying that there should be some non-gravitational interaction between the two sectors that correlates their abundances.

Let us briefly review past work where confining dark sectors provide a framework for addressing puzzles in the SM beyond DM itself. 
As mentioned earlier, confining strongly-coupled DM models originated from technicolor solutions to the hierarchy problem, with many other solutions to the hierarchy problem also involving a dark confining force \cite{Frigerio:2012uc,Marzocca:2014msa,Barnard:2014tla}.
They have also been studied as the DM candidate in supersymmetric solutions to the hierarchy problem (potentially as part of the SUSY-breaking messenger mechanism) and give rise to a host of correlated signals across many different experiments \cite{Banks:2005hc,Hamaguchi:2007rb,Hamaguchi:2008rv,Mardon:2009gw,Hamaguchi:2009db}.
New composite sectors also provide high-quality QCD axion solutions to the strong $CP$ problem
\cite{Kim:1984pt,Randall:1992ut,Dobrescu:1996jp,Flacke:2006ad,Redi:2016esr,DiLuzio:2017tjx,Lillard:2017cwx,Lillard:2018fdt,Cox:2019rro,Gherghetta:2020ofz,Contino:2021ayn,Gherghetta:2025fip,Gherghetta:2025kff,Agrawal:2025mke,Azatov:2025mep} and cogenesis of axion DM and SM abundances \cite{Asadi:2025cvm}.
Closely related are warped extra-dimensional models, which are holographically dual to strongly-coupled four-dimensional confining theories and have long provided a compelling framework for addressing the SM flavor hierarchy problem
\cite{Arkani-Hamed:1999ylh,Huber:2000ie,Csaki:2008qq}.
Although DM is not always the central focus of these studies, such extra-dimensional or composite-sector frameworks naturally contain additional neutral states, some of which can be cosmologically stable and serve as DM candidates, see e.g.~Refs.~\cite{Redi:2012ha,Contino:2021ayn}. 

A characteristic challenge in the realm of confining dark sectors is non-perturbative dynamics in the IR\@. The formation, spectrum, and interactions of dark hadrons are all subject to uncertainty and vary greatly across parameter space. Existing tools such as lattice simulations (and others described below) are useful, but these dark sectors serve as motivation for further developing methods to understand confinement. 
Model-specific studies are especially important for collider searches for confining dark sectors because their signals may not be captured by a purely effective field theory (EFT) description, e.g.~SM EFT (SMEFT). 
For instance, if the confining scale is within the reach of collider probes, 
production could proceed into the deconfined phase (dark quarks and dark gluons) while the signals arise from the decays of dark hadrons in the confined phase.  Even if the dark sector fields are heavy, the strongly-coupled nature of confined dark sectors implies the usual perturbative intuition for matching and power counting can become problematic, for example in the presence of large anomalous dimensions that render a canonical dimension operator-truncated SMEFT unreliable.

We now summarize the main motivations for studying confining dark sectors from a high-level perspective.

\begin{figure}
    \centering
    \resizebox{0.51\textwidth}{!}{
    \includegraphics[width=0.5\linewidth]{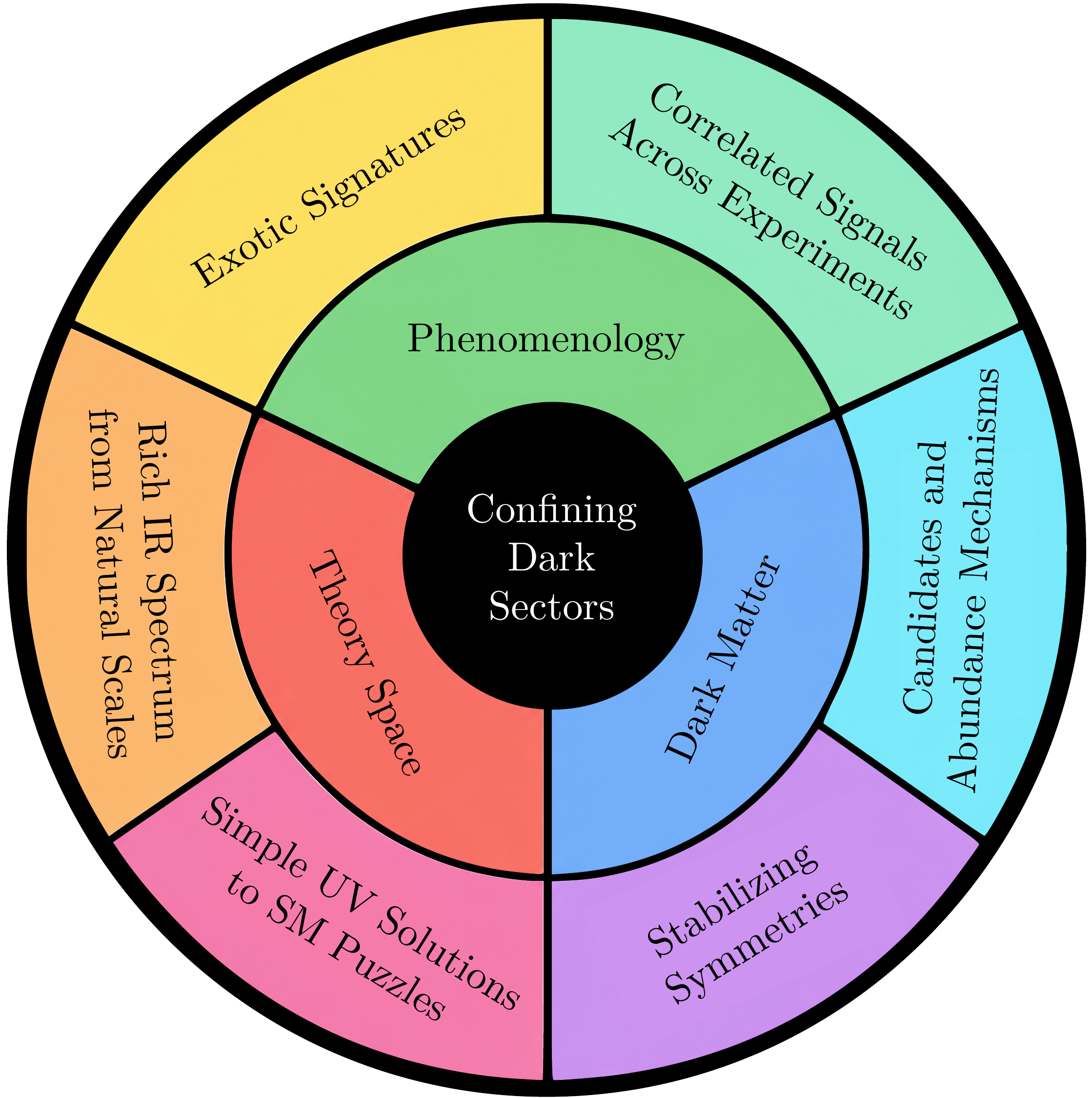}
    }
    \caption{
    \label{fig:propaganda}
    Overview of motivations and common features of confining dark sectors. They can be used to solve a variety of theoretical puzzles, and the rich dark hadron spectrum leads to opportunities in theory modeling, exciting experimental signatures, and various naturally stable DM candidates. While the space of theories and applications is vast, commonalities in analysis methods and correlations among experimental signals motivate treating confining dark sectors as a unified subject. }
\end{figure}

\begin{itemize}
\item 
\textbf{Dark Matter Candidate(s) ---} 
It is imperative to develop and study the phenomenology of confining dark sector models
in order to fully cover  the vast range of dark matter detection possibilities.  WIMPs and axions have well-developed models and experimental search techniques, while confining dark sectors can produce radically different dark matter candidates.

\item \textbf{The Abundance Puzzle ---} 
Confining dark sectors provide a variety of ways to achieve the dark matter relic abundance.  There are many examples of 
models that can utilize well-known symmetric mechanisms including freeze-out, 
freeze-in, cannibalism, etc., as well as several possibilities for generating an asymmetric abundance.
The ``golden prize'' of
simultaneously explaining the baryon asymmetry \cite{Sakharov:1967dj} with the DM energy density is also possible in a rather small number of confined dark sector models.  This  further motivates the continued investigation of models of confined dark sectors that can address the abundance puzzle.

\item \textbf{Rich and Varied Phenomenology ---} A vast space of experimental signatures can arise from confining dark sectors, including some that are not targeted by traditional searches. Correlations between signals in different experiments are also common and can potentially act as a smoking gun signature. From a signal-driven perspective, particular models can serve as benchmarks to probe the broader space, analogously to the Simplified Model paradigm used in the SUSY context. 

\item \textbf{Solutions to Other SM Puzzles ---} 
The electroweak hierarchy problem may be addressed in strongly-coupled composite Higgs theories, the axion quality problem may be addressed with a composite axion, the hierarchy of Yukawa couplings could arise from partial compositeness, etc.

\item \textbf{Natural Scales and New Symmetries ---} Confinement within a dark sector provides a natural new scale that may be paired with technically natural scales such as dark fermion masses.  Dark flavor symmetries can provide an explanation for DM stability.

\item \textbf{Avenue for Studying Confinement ---} 
Finally, studying confining dark sectors enables one to study confinement from a phenomenological perspective. Given the many challenges of non-perturbative dynamics, which are core to understanding QCD, studies of these SM and BSM sectors naturally reinforce each other.

\end{itemize}
A pictorial presentation of these motivations is shown in \cref{fig:propaganda}.

Throughout this review, unless stated otherwise, we focus on dark sectors in which the non-Abelian gauge group is $\mathrm{SU}(\Nc)$ and the matter content consists of $\Nf$ fermions in the fundamental representation. We denote the dark confinement scale, dark pions, dark quarks, and dark quark mass by $\LD$, $\piD$, $\qD$, and $m_q$, respectively.

The following sections are structured as follows. 
We begin in \cref{sec:models} with a broad overview of how different dark hadrons can manifest as DM candidates. In \cref{sec:simplifiedUV}, we describe how these dark sectors can serve as UV completions of various well-known DM abundance mechanisms. In \cref{sec:cosmo}, we provide examples on how confining dark sectors can impact cosmological history in various ways. We discuss the DM detection searches 
in \cref{sec:detection} 
and collider phenomenology in \cref{sec:pheno}.
\cref{sec:calc_tools} is dedicated to calculational techniques or tools used to make predictions in the strongly-coupled regime. We conclude in \cref{sec:outlook} and discuss some future directions.

\section{Dark Matter Candidates}
\label{sec:models}

\begin{figure}[t]
    \centering

    \begin{subfigure}[t]{0.32\textwidth}
        \centering
        \includegraphics[width=0.6\linewidth]{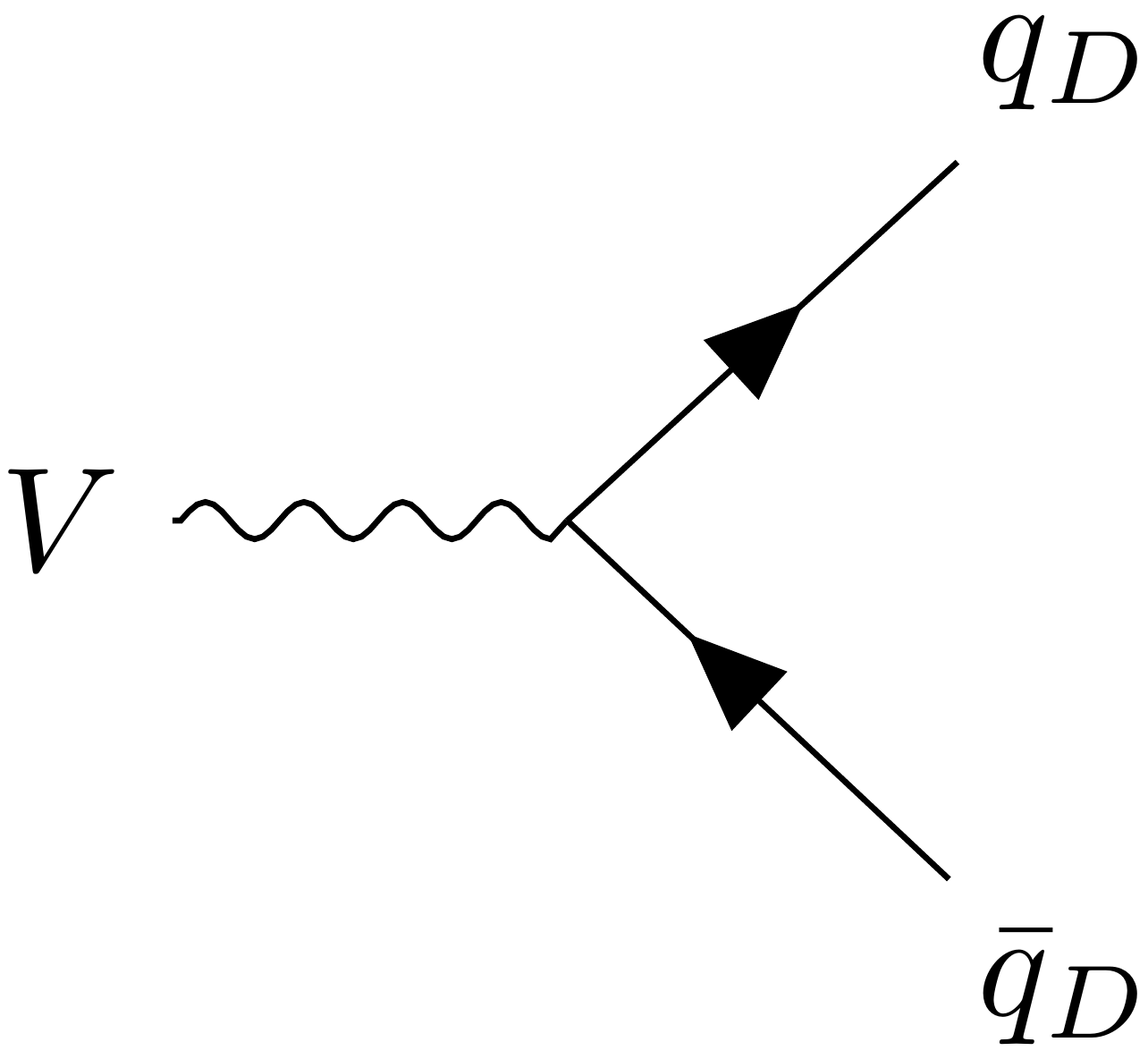}
        \caption{{\bf SM charge:} $\qD$ couples directly to SM gauge bosons $V$. Charged dark hadrons must be sufficiently heavy to evade constraints. If $\qD$ is heavy compared to $\LD$, it generates effective interactions between gauge sectors when integrated out.
        }
        \label{fig:qqV}
    \end{subfigure}
    \hspace{5em}
    \begin{subfigure}[t]{0.32\textwidth}
        \centering
        \includegraphics[width=0.6\linewidth]{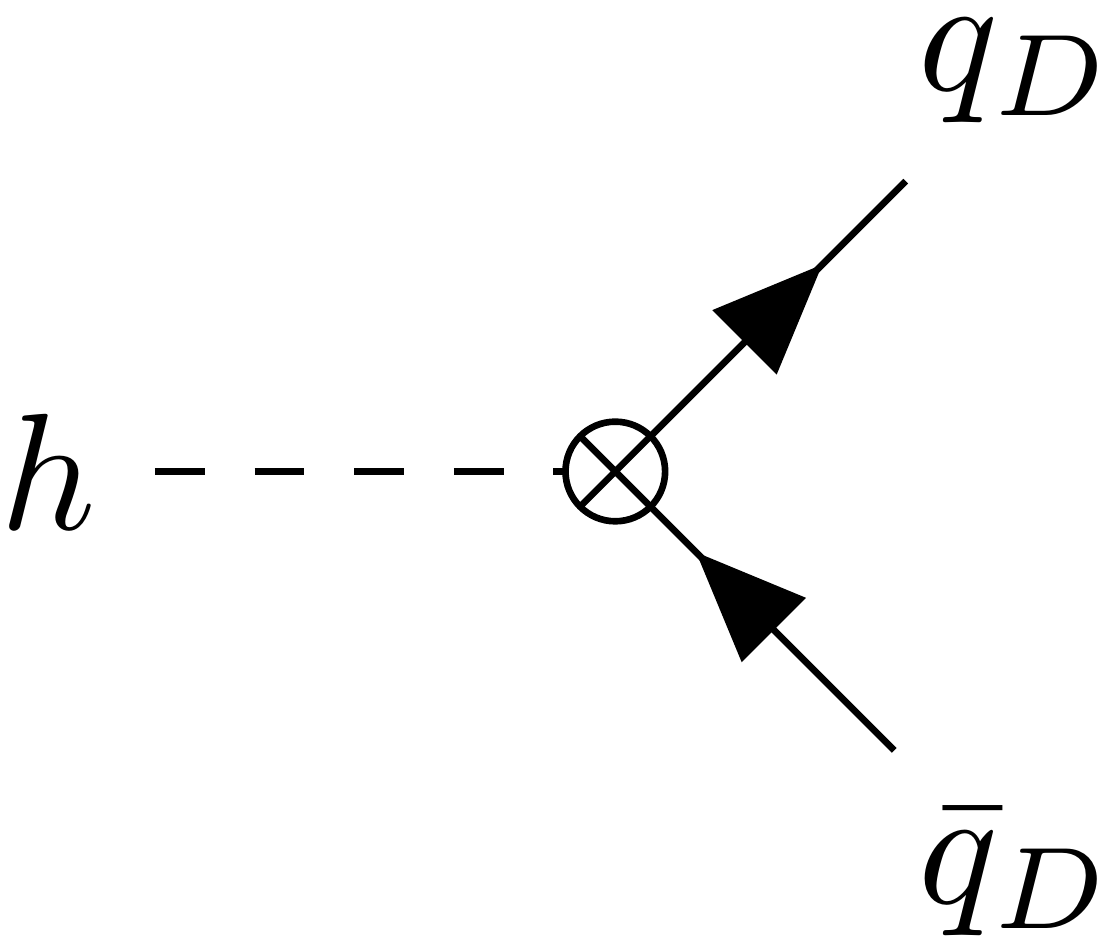}
        \caption{{\bf Higgs:} $\qD$ couples either directly to the SM Higgs or indirectly via some higher-dimensional operators or the Higgs mixing with some other scalar. Heavy dark quarks can induce higher-dimensional operators coupling the Higgs to dark gluons as well.}
        \label{fig:qqh}
    \end{subfigure}

    \vspace{1em}

    \begin{subfigure}[t]{0.32\textwidth}
        \centering
        \includegraphics[width=0.6\linewidth]{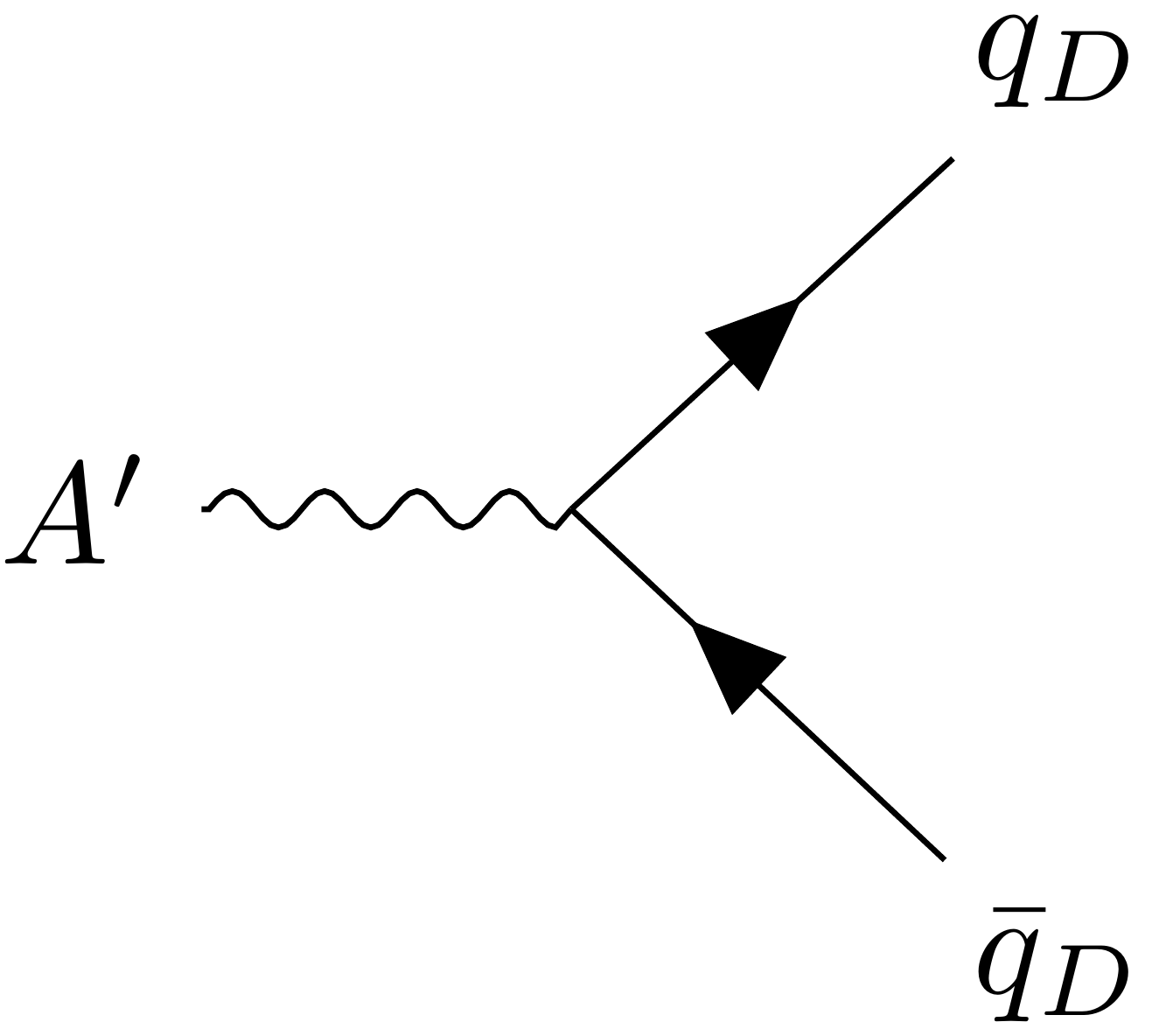}
        \caption{{\bf New vector:} $\qD$ couples to a vector boson besides the dark gluon, typically a $Z^\prime$ or dark photon $A^\prime$ from a spontaneously broken U(1). The new vector may, for example, couple directly to SM matter or mix with U(1)$_Y$.}
        \label{fig:qqVp}
    \end{subfigure}    
    \hfill
    \begin{subfigure}[t]{0.32\textwidth}
        \centering
        \includegraphics[width=0.6\linewidth]{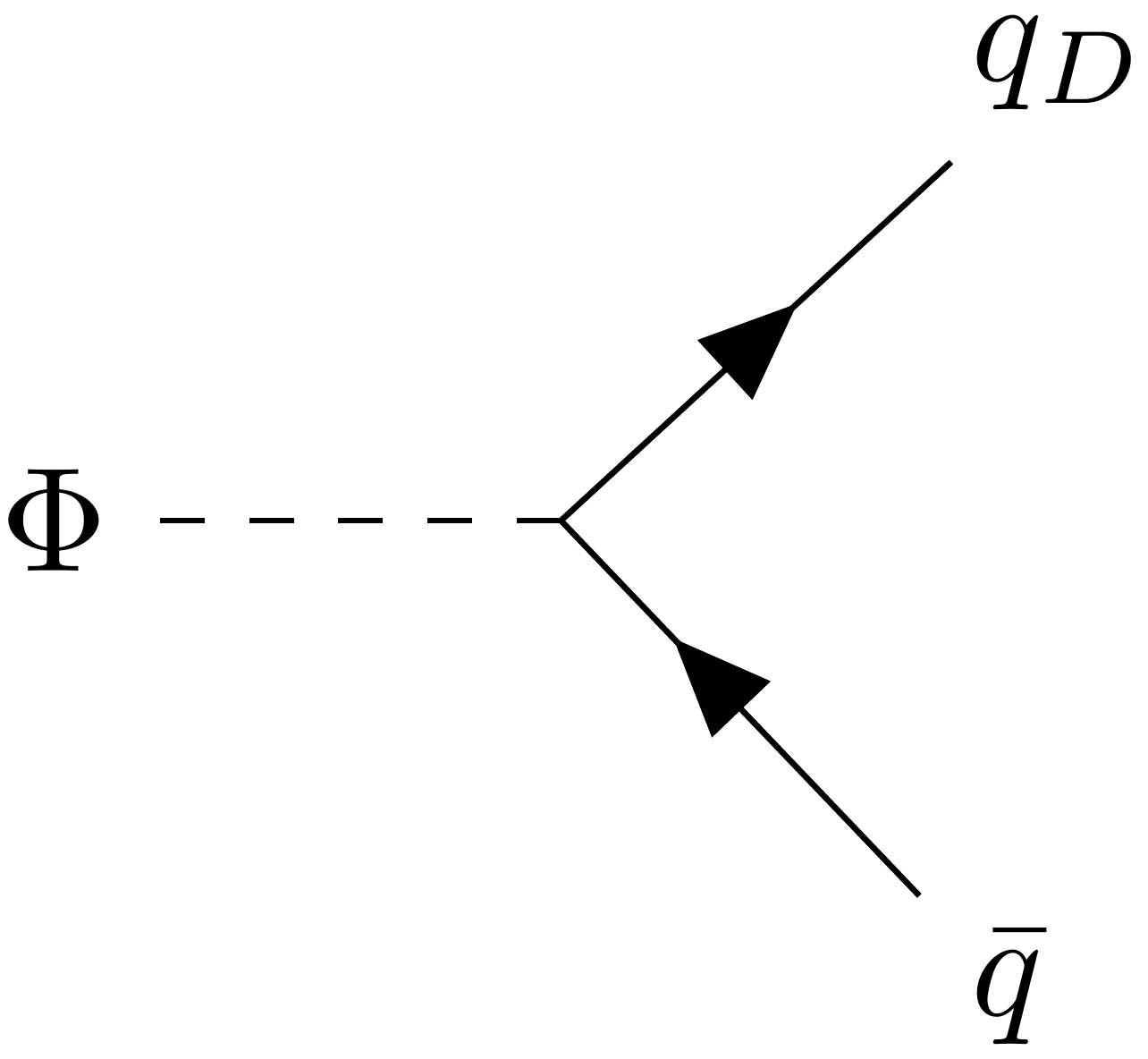}
        \caption{{\bf Yukawa:} a scalar $\Phi$ carrying SM gauge charges couples SM fermions (e.g.~quarks $q$) to $\qD$, which may be SM-neutral. This is often referred to as a ``$t$-channel mediator'' in a collider context or a ``fermion portal'' in other contexts.}
        \label{fig:tqq}
    \end{subfigure}
    \hfill
    \begin{subfigure}[t]{0.32\textwidth}
        \centering
        \includegraphics[width=0.6\linewidth]{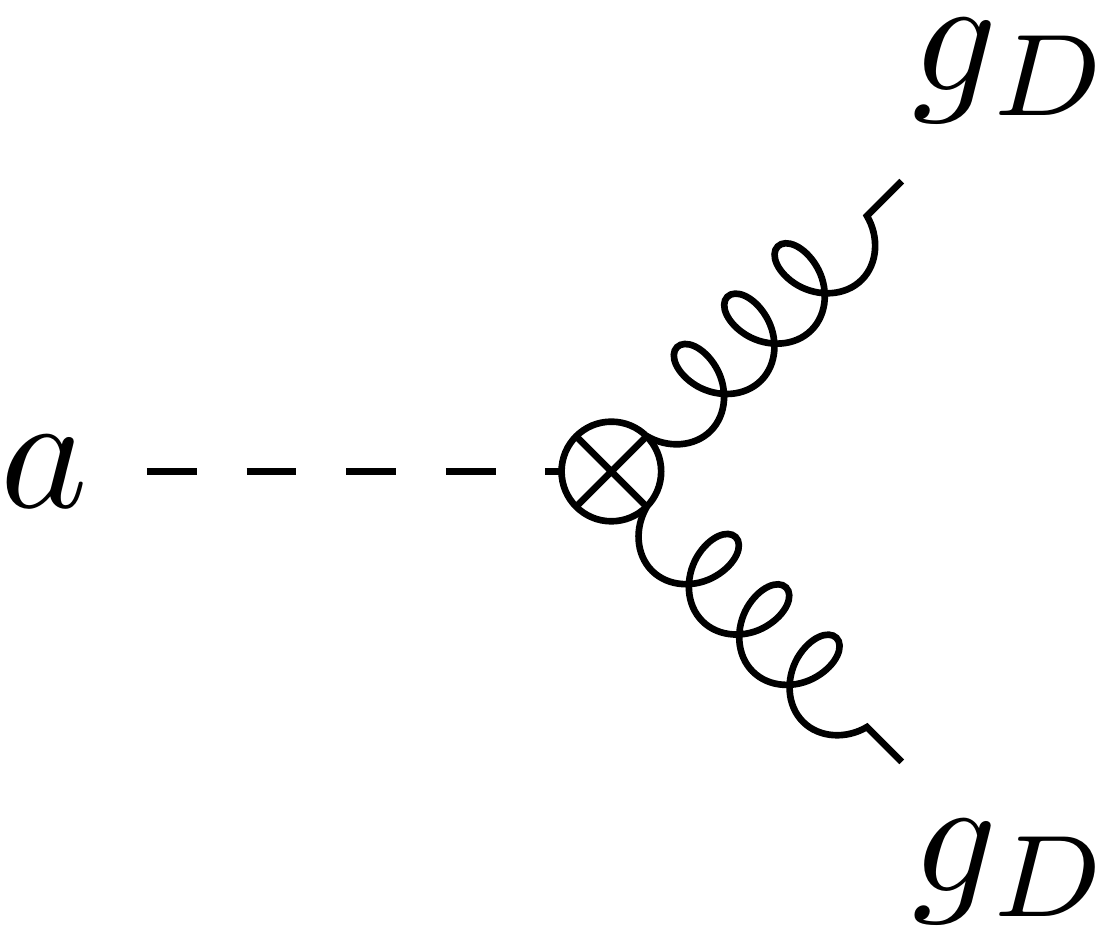}
        \caption{{\bf ALP:} a pseudoscalar $a$ couples to dark gluons $g_D$ through a dimension-5 operator and/or dark quarks, and $a$ may have various couplings to SM gauge and matter fields.}
        \label{fig:ggALP}
    \end{subfigure}

    \caption{Overview of some portal interactions that can connect the dark partons to the SM\@. Crossed dots denote where there may be higher-dimensional operators or mixings inserted. }
    \label{fig:partonPortal}
\end{figure}

Confining DM models can be classified according to various criteria, for example: 
\begin{itemize}

    \item \textbf{Dark Matter Candidate} --- Baryons, mesons, glueballs, or more exotic dark hadronic bound states can be the primary DM candidate. In some theories, there can be multiple DM candidates.

    \item \textbf{Abundance Mechanism} --- The DM abundance can be determined thermally or non-thermally, and the confining phase transition can play a significant role. 
    Strongly-coupled theories provide a natural home for an asymmetric dark matter abundance mechanism, since the symmetric abundance can often be depleted by the strong couplings.

    \item \textbf{Stabilizing Mechanism} --- A variety of mechanisms can stabilize dark hadrons, see e.g.~Refs.~\cite{Hambye:2009fg,Antipin:2015xia,Dondi:2019olm,Alfano:2025non} . This includes a conserved baryon number, other forms of flavor symmetry, accidental discrete symmetries, kinematics, or suppression of the dark hadron's decay rate by the mass of a heavy mediator or another UV scale. 

    \item \textbf{Portal to the SM} --- The dark sector can be connected to the SM via a multitude of different portal interactions, as depicted in \cref{fig:partonPortal}. Unlike non-confining DM models, dark quarks can 
    transform under the 
    SM gauge groups 
    while some dark hadrons still have suppressed SM interactions. Instead, or in addition to this, the two sectors could be coupled via other 
    renormalizable portals, as well as portals involving higher dimensional operators
    \cite{Strassler:2006im}. 

    \item \textbf{Mass Hierarchies } --- 
    The masses of the dark quarks relative to the dark confinement scale significantly affect phenomenology.
    As will be discussed in the upcoming sections, different regimes require different calculational techniques or tools to compute the IR spectra and interactions.

\end{itemize}

A recurring theme throughout this review is that these dark sectors resist a simple classification that maps between the UV model and phenomenology, as a given signal can arise from various models, and a given UV theory can realize a variety of signatures. 
For the remainder of this section, we focus on the DM candidate as the categorization principle, and each of the other aforementioned schemes is addressed in the following sections.

\subsection{Dark Baryons}
\label{subsec:model_baryon}

One of the few things we know about the 
nature of DM is that it is cosmologically stable.
Given that the
SM proton is stabilized by baryon number symmetry, it is only natural to consider that DM is stabilized by a counterpart dark baryon number.
This was originally proposed in the context of technicolor, where
techni-baryons 
were proposed as a dark matter candidate
\cite{Nussinov:1985xr}. 
Here, the prefix ``techni'' simply reflects the fact that such DM candidates arise as hadrons of a new confining force.
Interestingly, Ref.~\cite{Nussinov:1985xr} also proposed an explanation for the similarity between the present-day abundances of SM baryons and DM; this highlights the long-standing connection between confining DM models and attempts to address the abundance puzzle.
Models of techni-baryon DM were subsequently developed for a variety of confining gauge groups, and their phenomenology was studied in detail \cite{Bagnasco:1993st,Gudnason:2006ug,Gudnason:2006yj,Lewis:2011zb,Ryttov:2008xe}.
These models could also be embedded in technicolor solutions to the hierarchy problem, providing a unified solution for two of the major open questions in the field.

With the discovery of the Higgs boson and the accumulation of electroweak precision measurements favoring a SM-like Higgs sector, minimal technicolor solutions to the hierarchy problem are now stringently constrained. 
Nevertheless, alternative frameworks based on strong dynamics, such as composite Higgs models and warped extra-dimensional theories (related through AdS/CFT correspondence \cite{Maldacena:1997re,Witten:1998qj,Arkani-Hamed:2000ijo}), remain viable explanations.
It was soon realized that such frameworks can naturally accommodate viable DM candidates once supplemented with additional symmetries, see e.g.~Ref.~\cite{Agashe:2004ci}.

The connection between the origin of the electroweak scale and confining DM has also been explored outside the frameworks described above.
In one of the modern realizations of confining DM \cite{Hur:2007uz}, 
this connection was studied at the effective field theory level, where the dynamics of a dark confining sector contributes to the Higgs mass parameter and can trigger electroweak symmetry breaking.\footnote{See also Ref.~\cite{Hambye:2013dgv,Chang:2014ida} 
for other mechanisms in which the electroweak scale is dynamically generated in conjunction with a confining dark sector.}
In these models, both dark baryons and dark mesons can act as viable DM candidates.
While the stability of the baryon can arise from a conserved dark baryon number, the dark meson DM typically requires an additional stabilizing symmetry,as further discussed below.

Pursuing the connection between the weak scale and confining dark sectors, many authors have explored the possibility that SM gauge interactions themselves act as the portal between the visible and dark sectors. In this approach, the new strongly interacting sector contains matter fields charged under SM gauge groups, allowing interactions between the two sectors without introducing additional mediators. Examples include scenarios where the dark sector communicates with the SM through  the SU(2)$_L ~ \times ~$U(1)$_Y$ electroweak interactions \cite{Kribs:2009fy,Bai:2010qg,
LatticeStrongDynamicsLSD:2013elk,
LSD:2014obp,Antipin:2014qva,
Appelquist:2015zfa,Appelquist:2015yfa,
Huo:2015nwa,Mitridate:2017oky,Contino:2020god,Fleming:2024flc},\footnote{See also Refs.~\cite{Carvunis:2020exc,Blennow:2026psy} for models in which the portal is provided by a gauged flavor symmetry of the SM\@.} or through the SU(3)$_c$ gauge group \cite{Cline:2016nab,DeLuca:2018mzn}.

Another early study was carried out in Ref.~\cite{Antipin:2014qva}, where the portal interaction through the SU(2)$_L$ gauge group also plays a role in triggering electroweak symmetry breaking. In these models, additional symmetries can be imposed to stabilize some of the dark mesons, making them viable DM candidates.
Ref.~\cite{Huo:2015nwa}
also 
considered dark quarks charged under the SM electroweak gauge group, where they assumed
thermal freeze-out at the unitarity bound for the annihilation rate, giving the well-known prediction for the mass of the confined DM to be $\mathcal{O}(100)$~TeV\@.

In the heavy dark quark limit, where the constituent dark quarks are much heavier than the confinement scale, DM behaves as (dark) quarkonium, leading to unusual bound-state dynamics and collider signatures.
Ref.~\cite{Kribs:2009fy} developed a DM model inspired by this idea and explored its distinctive phenomenology across a variety of experiments. 
Similar to some of the models discussed above, the dark quarks in this framework are charged under the SM electroweak gauge group SU(2)$_L ~ \times ~$U(1)$_Y$, thereby allowing the SM gauge interactions to act as the portal between the two sectors. 
The model considers scalar dark quarks charged under a confining SU(2) gauge group, which form scalar dark baryons after confinement. These scalar dark baryons serve as the DM candidate and give rise to direct detection signatures that differ from those expected for fermionic dark baryons.

If the stability of the DM candidate relies on a global dark baryon number, quantum gravity effects are expected to violate this symmetry, whose effects are studied in Ref.~\cite{Huo:2015nwa}. 
Consequently, the dark baryon can undergo extremely rare decays into SM particles. In the specific framework studied there, such decays could potentially account for certain astrophysical anomalies (e.g.~the AMS-02 excess) and may be probed through $\gamma$-ray observations or constraints from the Cosmic Microwave Background (CMB).

While such portals help render the phenomenology of these models highly predictive, a detailed understanding of the properties of dark baryons is still required in order to make precise quantitative predictions. In particular, in the heavy quark limit, methods inspired by the quark model can be employed to study the properties of the dark hadrons.
In this regime, a variational approach was proposed in Ref.~\cite{Mitridate:2017oky} to estimate the masses and binding energies of dark baryons. These quantities enter directly into the calculation of dark baryon annihilation rates, e.g.~through rearrangement processes, and therefore play an important role in determining the relic abundance. Ref.~\cite{Mitridate:2017oky} also pointed out the possibility of long-lived dark hadrons that eventually decay into the DM candidate while injecting energy into the SM plasma.
More generally, effective field theories originally developed for QCD have proven useful in studying confining dark sectors in the heavy quark limit. For example, Ref.~\cite{Assi:2023cfo} applies techniques from potential nonrelativistic QCD (pNRQCD) to predict the spectrum of dark hadrons in this regime.

\subsection{Dark Mesons}
\label{subsec:model_meson}

Dark mesons, defined here as dark $(q\bar{q})$ bound states, can also serve as the DM candidate. 
Many studies, e.g.~Refs.~\cite{Cheng:2003ju,Agashe:2007jb,Essig:2009nc, Bhattacharya:2013kma,Cline:2013zca,Hochberg:2014kqa, Harigaya:2016rwr, Kopp:2016yji, Berlin:2018pwi, Beauchesne:2018myj, Bernreuther:2019pfb, Contino:2020god, Carmona:2024tkg, Cheng:2026bqf}, have explored 
discrete or continuous symmetries for obtaining stable 
dark meson DM 
candidates. As mentioned above, Ref.~\cite{Hur:2007uz} argued that a dark quark flavor symmetry can stabilize certain dark mesons. Similar ideas, as well as their possible role in triggering electroweak symmetry breaking, were further explored in Ref.~\cite{Hur:2011sv}. (For additional models in which dark mesons are connected to the electroweak hierarchy problem, see Ref.~\cite{Carmona:2015haa}.)

One of the critically important early observations is that dark sector  theories can contain a symmetry analogous to the SM $G$-parity that can stabilize one or more of the dark pions \cite{Bai:2010qg}. 
The original idea proposed dark fermions that transform in vectorlike representations of both the dark gauge group and the SU(2)$_L$ group, and it is this vector-like nature of the fermions that implies, after confinement, the dark sector contains an exact $G$-parity.\footnote{In the absence of additional higher-dimensional interactions that could violate $G$-parity.}  
$G$-parity implies the lightest $G$-odd dark pion is stable, which has important implications for colliders, DM, and cosmology. This symmetry was exploited in Ref.~\cite{Antipin:2014qva} to stabilize DM in a framework that also links the dark sector to the generation of the electroweak scale. This line of work was further developed for different confining gauge groups and matter content \cite{Antipin:2015xia}, including studies of scenarios in which such models arise as the infrared limit of certain Grand Unified Theories (GUTs).
Identifying the stable and unstable dark pions is essential for searches at colliders \cite{Schwaller:2015gea,Beauchesne:2018myj,Asadi:2025vfr}.
In cosmology, the leading annihilation rate of dark pions can be back into other dark pions.  The theory space divides into several distinct categories,
corresponding to the importance of  constraints from
indirect detection
which range from
severe to virtually
non-existent
\cite{Beauchesne:2019ato}.

In many confining dark sector models, the dark quarks that ultimately form the DM candidate transform in the fundamental representation of the confining gauge group. Nevertheless, scenarios involving dark quarks in higher representations have also been explored and can give rise to qualitatively different dynamics, e.g. different Sommerfeld enhancement factors in annihilation processes\cite{Mitridate:2017oky}, and mass spectrum \cite{Contino:2018crt,Buttazzo:2019mvl}.
For example, Ref.~\cite{Contino:2018crt} studied a setup in which the dark quarks transform in the adjoint representation of an SU($N$) confining gauge group and bound states of a single quark and a gluon (gluequark) can act as the DM candidate. 
More generally, Ref.~\cite{Buttazzo:2019mvl} explored dark quarks in non-fundamental representations for a variety of gauge groups and analyzed the resulting spectrum of dark hadrons and their viability as DM candidates.

Global anomalies, well known from QCD,  can play an important role in dark sector theories. For example, chiral anomalies can lead to the decay of some dark meson species, and it is not always obvious which species those are \cite{Asadi:2025vfr}.
The dark sector $\theta$ phase can also affect phenomenology, as explored in \cite{Berryman:2017twh}.

\subsection{Dark Glueballs}
\label{subsec:model_glueball}

Another class of DM candidates arising from confining dark sectors consists of glueballs, i.e.~color-neutral bound states composed purely of dark gluons. 
The possibility that glueballs could constitute DM was proposed in the contexts of string theory \cite{Faraggi:2000pv,Halverson:2016nfq,Halverson:2018olu} and self-interacting DM more generally due to the strong self-interactions they inherit from the UV \cite{Boddy:2014yra,Soni:2016gzf,Acharya:2017szw}.
In particular, Ref.~\cite{Boddy:2014yra} also proposed the possibility of glueballinos, i.e.~bound states involving gluinos in supersymmetric theories.

There are twelve glueball species distinguished by their spin $J$, charge conjugation $C$, and parity $P$ quantum numbers (each species is denoted by $J^{PC}$) that are stable in the absence of any portals \cite{Morningstar:1999rf}.\footnote{The $C$-odd glueballs are only expected to exist in the spectrum when there is a meaningful sense of charge conjugation within the pure Yang-Mills sector that cannot be written as a gauge transformation. This is not the case for SU(2), Sp(2$N$), or SO($2N+1$) gauge groups \cite{Juknevich:2009ji}. Other SU($N$) groups and SO($2N$) do have this property, but the glueball spectrum for SO($2N$) has additional subtleties \cite{Teper:2018gaw}.} For general discussions of glueball physics, see Refs.~\cite{Robson:1977pm,Donoghue:1980hw,Cornwall:1982zn,Jaffe:1985qp,Mathieu:2008me,Juknevich:2009ji,Juknevich:2009gg,Ochs:2013gi}, as well as Refs.~\cite{Berg:1982kp,Berg:1983qd,Michael:1988jr,Bali:1993fb,Morningstar:1997ff,Morningstar:1999rf,Gregory:2012hu} for lattice studies of the glueball spectrum. 
The lightest glueball state is expected to be the $0^{++}$, a scalar that is even under both parity and charge conjugation. Unlike baryons or flavor-charged mesons, this state's decays are not forbidden by any symmetry. 
Any couplings between the SM and dark sector must therefore be strongly suppressed for the $0^{++}$ to be a viable DM candidate. These interactions are induced by higher-dimensional operators of the effective theory coupling the dark gluons to the SM, and the leading decay rates are computed in Refs.~\cite{Juknevich:2009ji,Juknevich:2009gg}. 
These operators involve the Higgs field at dimension-6, as well as SM gauge bosons at dimension-8. Two of the twelve species, the $0^{-+}$ and $1^{+-}$, are not allowed to decay via the $CP$-even dimension-6 Higgs portal and are thus the longest-lived dark glueballs when the Higgs portal is dominant. In cases where the $0^{++}$ decays before Big Bang Nucleosynthesis (BBN), the $1^{+-}$ may be a particularly compelling DM candidate, as its dimension-8 decays can be naturally forbidden when charge conjugation acting only on the dark sector is a good symmetry of the UV \cite{Forestell:2017wov,McKeen:2024trt}. 

A precise calculation of the relic abundance of dark glueballs is a challenging task and has motivated a number of studies in the literature \cite{Forestell:2016qhc,Forestell:2017wov,Carenza:2022pjd,Carenza:2023eua,McKeen:2024trt,Asadi:2025btr}. 
One source of difficulty arises from the presence of strong number-changing processes, in particular $3\to 2$ cannibal interactions of the $0^{++}$, that cause the abundance to fall logarithmically with the cosmological scale factor \cite{Farina:2016llk,Dolgov:2017ujf}. 
These rates may be estimated by large-$N$ scaling \cite{Forestell:2016qhc,Forestell:2017wov,McKeen:2024trt} or expansion of the glueball potential in a Polyakov loop model \cite{Carenza:2022pjd,Carenza:2023eua}, but $\mathcal{O}(1)$ uncertainties remain that have led to discrepancies between some relic density estimates. 
Studies that incorporate glueball species beyond the $0^{++}$ find that the $0^{++}$ glueball is by far the most abundant before it decays, while the abundances of heavier states are Boltzmann-suppressed. 

Other sources of uncertainty stem from dynamics near the confining phase transition. Lattice calculations \cite{Borsanyi:2012ve,Caselle:2018kap} have been used to model the dark gluon/glueball equation of state and energy density near the critical temperature \cite{McKeen:2024trt}, but non-equilibrium dynamics such as bubble nucleation may also play at important role \cite{Carenza:2023eua}. The phase transition is only weakly first-order for SU(3) \cite{Brown:1988qe}, but it becomes 
increasingly strongly
first-order in the large-$N$ limit, where supercooling may cause the approximation of instantaneous conversion of gluons into glueballs at the critical temperature to lose validity. 
These uncertainties underscore the need for more precise and systematic studies of the glueball DM relic abundance.

\subsection{Large Dark Nuclei}

Large $A$ number composite dark nuclei states may form when the origin of the abundance of DM involves an \emph{asymmetric} mechanism leading to a preference for dark baryons (over dark anti-baryons).
Holding large number of constituents together requires either 
confining dark nuclear dynamics or an attractive Yukawa force mediated by a light scalar. 
In Ref. \cite{Krnjaic:2014xza}, efficient ``darkleosynthesis'' was shown to occur when a short-range nuclear attraction is supplemented by a light mediator that allows the binding energy to be radiated. The dark self-interaction bounds favor a confinement scale above $\Lambda_{\rm QCD}$, so the dark binding energies can be large and the repulsive Coulomb-like barriers comparatively weak, making dark nuclei easier to synthesize and harder to dissociate than ordinary nuclei.

The possibility of a more elaborate big bang dark nucleosynthesis has also been explored. In asymmetric DM scenarios, it has been argued that bound states containing a very large number of constituents---potentially as large as $10^8$---could form during this epoch \cite{Hardy:2014mqa,Hardy:2015boa}. Such scenarios can lead to striking direct detection and astrophysical signatures. Dark matter candidates composed of extremely large numbers of constituents are often referred to as dark ``nuggets'' \cite{Witten:1984rs}. 
Large $A$ nuggets arising from an asymmetric population of dark constituents into macroscopic or mesoscopic composites were considered in Refs.~\cite{Gresham:2017zqi,Gresham:2017cvl}. 
A conserved dark asymmetry leaves behind many dark fermions, and a sufficiently attractive finite-range force allows them to bind into large stable objects.
Ref.~\cite{Gresham:2017zqi}
computed the nugget structure, identifying nonrelativistic Coulomb-like, relativistic Coulomb-like, and saturation regimes.
Ref.~\cite{Gresham:2017cvl} followed early-universe nugget synthesis through fusion reactions and estimated the mass function after synthesis freeze-out. The late-time phenomenology of these asymmetric DM nuggets was studied in Ref.~ \cite{Gresham:2018anj}, which emphasized that large nuggets are a characteristic consequence of attractive asymmetric DM without a dark Coulomb barrier, and that exothermic dark fusion and self-interactions can affect galaxies and constrain the allowed nugget masses and couplings. A different but closely related formation route appears in Ref.~\cite{Bai:2018dxf}, where dark quark nuggets form during a first-order dark confinement transition provided there is a nonzero dark baryon asymmetry. 
This can trap enormous dark baryon number into macroscopic quark-matter nuggets, with possible residual free dark baryons giving more conventional direct detection signals.
Additional work impacting indirect detection and other cosmological implications
includes
Refs.~\cite{Fedderke:2024hfy,Kaplan:2024dsn,Bai:2024muo}, with formation mechanisms and stability critically discussed in Refs.~\cite{Bai:2024amm,Bai:2025zpm}.

\section{Abundance Mechanisms}
\label{sec:simplifiedUV}

There is a variety of mechanisms that have been considered to achieve the DM relic abundance.
In this section, we review some of these mechanisms and how confining dark sectors have been used as their UV completion.

\subsection{Symmetric Abundance}

Thermal dark matter is generally bounded to lie below $\mathcal{O}(100)$~TeV by partial-wave unitarity of the annihilation cross section \cite{Griest:1989wd}. Confining dark sectors provide a natural way to approach this limit: their strong dynamics can produce geometric annihilation cross sections \cite{Antipin:2014qva,Appelquist:2015yfa}. Consequently, thermal relics in such sectors are often driven toward the upper end of the allowed window, $m_{\rm DM}\sim \mathcal{O}(100)~\mathrm{TeV}$.

Several features of confining dark sectors can open up parameter space for significantly heavier DM candidates. 
First, if the DM is composite, its spatially extended wavefunction can modify the parametric scaling of annihilation cross sections. This effect can enhance annihilation rates relative to those expected for point-like particles, thereby allowing viable thermal relics at higher masses \cite{Smirnov:2019ngs}.
In addition, after the confining phase transition, dark baryons can undergo a second stage of annihilation as their velocities decrease. During this epoch, baryons may form bound states and experience recombination and rearrangement processes \cite{Mitridate:2017oky,Gross:2018zha}. These processes become more efficient at low velocities \cite{Contino:2018crt,Geller:2018biy}, leading to enhanced annihilation rates. Such late-time annihilation can significantly reduce the relic abundance and is one of the key mechanisms that allows these models to evade the conventional unitarity bound.

On the other hand, significantly smaller composite dark matter masses can be realized in a variety of ways.
One early idea \cite{Buckley:2012ky}
considers an SU(2) dark sector where the mesons and baryons both act as pseudo-Nambu-Goldstone bosons, allowing the dark baryon dark matter to be parametrically lighter than the confinement scale.
Another possibility is to exploit the observation that baryon-anti-baryon annihilation to mesons in the non-relativistic limit at fixed velocity is expected to be exponentially suppressed in the large $N$ limit due to the small probability of a quark finding an anti-quark scaling exponentially with $N$ 
\cite{Witten:1979kh}.
This was exploited in
Ref.~\cite{Morrison:2020yeg}, for the freeze-in abundance of dark baryons.  
This mechanism was also proposed in a separate study of light confining DM where the two sectors initially begin in thermal equilibrium \cite{Fleming:2024flc}.
In this case, Ref.~\cite{Fleming:2024flc} showed that an important constraint arises from the annihilation of DM into a long-lived flavor-singlet $\eta^\prime$-like  state, whose subsequent decays into SM particles can significantly affect cosmological observables.
More broadly, the large range of annihilation rates suggest a wide range of dark baryon dark matter masses is possible, and is worthy of continued investigation.

Another mechanism to achieve a symmetric abundance of light composite dark matter, in this case  as dark pions, utilizes the Strongly Interacting Massive Particle (SIMP) mechanism \cite{Hochberg:2014dra,Hochberg:2014kqa,Tsai:2020vpi}. 
In this framework, number-changing processes within the dark sector, such as $3\to 2$ interactions among dark pions, play a crucial role in determining the relic abundance.
Vector mesons of the same confining sector can participate in the thermal processes that control the relic density, leading to substantial changes in the predicted DM abundance within the SIMP paradigm \cite{Bernreuther:2023kcg}.

Cannibal DM \cite{Pappadopulo:2016pkp,Farina:2016llk} can also arise naturally in confining dark sectors, particularly in scenarios where glueballs constitute all or part of the DM abundance \cite{Acharya:2017szw,Buen-Abad:2018mas,McKeen:2024trt}, as discussed in \cref{subsec:model_glueball}. 
Number-changing $3\to 2$ interactions among glueballs induce a cannibal phase, in which the DM energy density redshifts non-trivially during the thermal evolution of the universe.

The densely-populated spectrum of hadronic states in confining dark sectors naturally realizes a variety of simplified DM scenarios that may otherwise appear fine-tuned or unnatural in the absence of a UV completion. One example is the forbidden DM framework \cite{DAgnolo:2015ujb}, in which the relic abundance is set by DM annihilations into particles that are slightly heavier than the DM itself. 
These final states may consist of heavier SM particles or new unstable states that subsequently decay into the SM\@. Such a mechanism can arise naturally in confining dark sectors, where the spectrum of hadrons typically contains many states with comparable masses \cite{Abe:2024mwa}.

Continuum DM models take the idea of a densely-populated spectrum in strongly interacting dark sectors to an extreme \cite{Brax:2019koq,Chaffey:2021tmj,Csaki:2021gfm,Csaki:2021xpy,Hong:2022gzo,Hong:2024zsn}. In these frameworks, the dark sector effectively exhibits a quasi-continuous spectrum of states rather than a discrete set of particles.
Such models can also realize mechanisms similar to forbidden DM, where annihilations into slightly heavier states determine the relic abundance \cite{Ferrante:2023bcz}.
While some initial studies have begun to investigate possible signatures \cite{Ferrante:2023fpx,Ferrante:2025ofe}, the range of potential signals arising from these models is still only beginning to be understood.

Nightmare DM scenarios---in which the dark sector interacts with the SM only through gravity---can also involve confining dark sectors \cite{Morrison:2020yeg,Garani:2021zrr}. In such models, the dark sector can be populated through gravitational freeze-in, and the relative temperature of the two sectors becomes an important parameter controlling the subsequent cosmological evolution. Even in these scenarios, observational constraints remain relevant: astrophysical bounds and limits on the effective number of relativistic degrees of freedom, $N_\mathrm{eff}$, can significantly restrict the allowed parameter space \cite{Garani:2021zrr}.

While such nightmare scenarios are logically consistent, the absence of non-gravitational interactions prevents them from addressing the observed similarity between the present-day abundances of DM and visible matter, i.e.~the abundance puzzle. As discussed earlier, providing a possible explanation for this puzzle is one of the motivations for studying confining DM models, which we will discuss in more detail in \cref{subsec:asymmetric}.

\subsection{Asymmetric Abundance}
\label{subsec:asymmetric}

Solving the abundance similarity puzzle requires a dynamical mechanism that predicts the DM and visible matter energy densities to be \textit{generically} of the same order today. Framed in this way, it becomes clear that asymmetric DM models (where the abundance is set by a matter-anti-matter asymmetry within the dark sector) alone do not constitute a complete solution to the problem. In such models \cite{Kaplan:2009ag}, one typically obtains comparable number densities for visible and DM\@. However, reproducing the observed DM abundance then requires the DM mass to lie in the $\mathcal{O}(5)\,$GeV range. 
Alternatively, there are models where the number density of dark matter can be suppressed relative to matter.  An early example 
was technibaryons \cite{Chivukula:1989qb} and later, quirky dark matter \cite{Kribs:2009fy}, where an asymmetric abundance of these composite dark matter candidates could be obtained with masses of $\mathcal{O}(1)\,$TeV\@. 
In each case, 
the dark number density was suppressed because the process that
re-distributed baryon number and dark baryon number 
had a Boltzmann suppression leading to $n_{\rm DM} \sim n_b \exp[-m_{\rm DM}/T_{\rm sph}]$ where $T_{\rm sph}$ is the temperature where 
electroweak sphalerons begin shutting off.
Nevertheless, in the absence of explaining both the matter number densities as well as the ratio $m_{\rm DM}/m_p$, asymmetric DM models by themselves do not fully explain the similarity between the dark and visible matter abundances.

A natural way to correlate the DM mass with the baryon mass scale is therefore to assume that the DM mass itself originates from a second confinement scale that is dynamically linked to the QCD scale \cite{Bai:2013xga,GarciaGarcia:2015pnn,Farina:2015uea,Farina:2016ndq,Lonsdale:2017mzg,Lonsdale:2018xwd,Ibe:2018juk,Ibe:2019yra,Murgui:2021eqf,Ritter:2022opo,Alonso-Alvarez:2023bat,Ritter:2024sqv,Cox:2025wxk}. 
After all, the visible matter mass density arises predominantly from the QCD confinement scale.
Different realizations of this idea employ a variety of mechanisms to correlate the running of the QCD and dark gauge couplings---and hence their confinement scales---as well as different portals between the two sectors and different mechanisms for generating the asymmetric number densities. Nevertheless, they share the common feature that they can generically produce comparable dark and visible matter energy densities, thereby providing a dynamical explanation of the abundance similarity puzzle.

The class of asymmetric composite DM models extends beyond DM masses of $\mathcal{O}(10)\,$GeV\@. For example, Ref.~\cite{Bottaro:2021aal} developed an asymmetric DM realization of the model proposed in Ref.~\cite{Antipin:2015xia}, in which the present-day DM abundance consists of asymmetric dark baryons. In this framework, the DM asymmetry is generated through perturbative breaking of the dark baryon number symmetry. Achieving a sufficiently large symmetry breaking to produce the observed DM abundance, while simultaneously ensuring that the DM remains sufficiently stable on cosmological timescales, constitutes the main challenge of this class of models and leads to interesting phenomenological implications.
Another approach to explaining the similarity between the dark and visible matter abundances is to correlate the origin of the asymmetries themselves. For instance, this can occur if both asymmetries are generated through the Hawking radiation of primordial black holes \cite{Kuwahara:2025eeo}.

It is worth emphasizing that confining asymmetric DM models---with DM mass fixed dynamically to be $\mathcal{O}(1)$~GeV---are not the most general solutions to the abundance similarity puzzle. In the frameworks discussed above, the similarity between the dark and visible matter abundances arises because the models separately generate comparable number densities and comparable mass scales for the two sectors, which together lead to comparable energy densities today.
A more general approach would instead involve dynamical mechanisms that produce comparable energy densities directly, without requiring the masses of the two sectors to be similar. While some attempts in this direction have been proposed (see e.g.~Refs.~\cite{Brzeminski:2023wza,Banerjee:2024xhn}), this line of investigation remains relatively unexplored and offers significant opportunities for further work.

A natural framework for studying asymmetric confining DM models and addressing the abundance similarity puzzle is provided by Twin Higgs models \cite{Chacko:2005pe}. In these models, a broken $\mathbb{Z}_2$ symmetry relates the visible sector to a twin sector. The twin sector generically contains its own confining gauge group, and the lightest twin baryon can naturally acquire a mass in the GeV range, making it a viable asymmetric DM candidate when the asymmetric number densities are comparable. 

Asymmetric twin (or mirror) DM models have been extensively studied in the literature \cite{Hodges:1993yb,GarciaGarcia:2015pnn,Farina:2015uea,Farina:2016ndq,Terning:2019hgj,Ibe:2019ena,Feng:2020urb,Easa:2022vcw,Bodas:2024idn}. Owing to the rich particle spectrum of twin sectors, different realizations of this idea lead to a wide variety of phenomenological signatures across cosmological, astrophysical, and experimental probes. 
In the context of twin or mirror models, even small deviations from the SM parameters can significantly alter the spectrum of stable bound states and therefore the nature of the dominant DM component \cite{Barbieri:2017opf,Easa:2022vcw}. 
Various cosmological, astrophysical, and direct detection signatures of these scenarios have been explored in Refs.~\cite{Chacko:2018vss,Chacko:2021vin,Curtin:2021spx,Bittar:2023kdl,Das:2024tmx}. 
An important constraint on these models arises from limits on additional relativistic degrees of freedom, typically expressed in terms of $\Delta N_\mathrm{eff}$. In many twin Higgs DM scenarios, satisfying these bounds requires asymmetric reheating of the visible and twin sectors so that the twin sector is reheated to a lower temperature \cite{Curtin:2021alk,Alonso-Alvarez:2023bat}. 

The possibility that DM resides in a sector resembling the SM has also motivated speculation that, much like visible matter today, DM could predominantly exist in the form of larger composite objects, analogous to nuclei composed of many constituents.
Lattice studies have established the existence of stable nuclear states in simple confining gauge theories \cite{Detmold:2014qqa,Detmold:2014kba}. In particular, Ref.~\cite{Detmold:2014qqa} examined the indirect detection signatures of such nuclear DM scenarios in both symmetric and asymmetric relic abundance frameworks. Effective field theory techniques developed for nuclear physics, such as pionless EFT \cite{Bethe:1949yr,Bedaque:2002mn}, have also been applied to study the properties and phenomenology of nuclear DM \cite{Redi:2018muu}.

\section{Dark Sectors in the Early Universe}
\label{sec:cosmo}

In this section, we consider some of the more exotic phenomenology of confining dark sectors in the early universe. 
This includes novel effects of dark confining phase transitions, potential explanations of different astrophysical and cosmological anomalies, and exotic effects on various cosmological observables such as the matter power spectrum.

In \cref{sec:models}, we discussed models and processes that can deplete the DM number density \textit{after} the confining phase transition. 
The dynamics \textit{during} the phase transition itself can also have a dramatic impact on the final DM abundance. It was first pointed out by Witten \cite{Witten:1984rs} that a first-order confining phase transition in the SM could potentially play a non-trivial role in determining the present-day DM abundance. However, lattice results indicate that the QCD phase transition in the SM is a smooth crossover. In contrast, a first-order confining phase transition can occur in many confining dark sector models. Indeed, lattice studies show that confinement transitions are expected to be first order in the limits of both very light and very heavy quark masses \cite{Svetitsky:1982gs,Alexandrou:1998wv,Kaczmarek:1999mm,Lucini:2005vg,Aoki:2006we,Saito:2011fs,LatticeStrongDynamics:2020jwi}.

\begin{figure}
\begin{center}
\resizebox{0.8\columnwidth}{!}{
\includegraphics[scale=1]{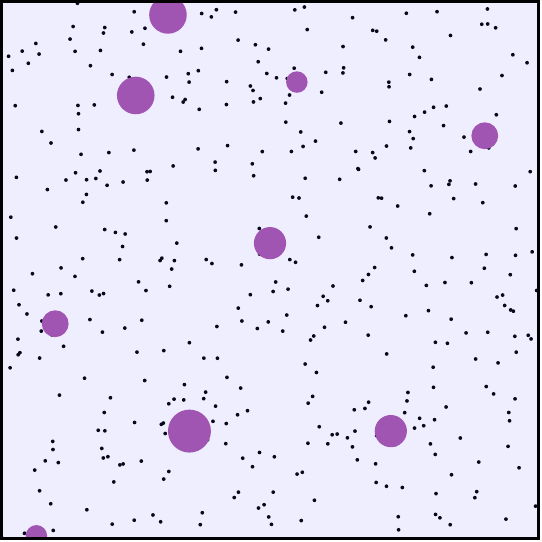}
\includegraphics[scale=1]{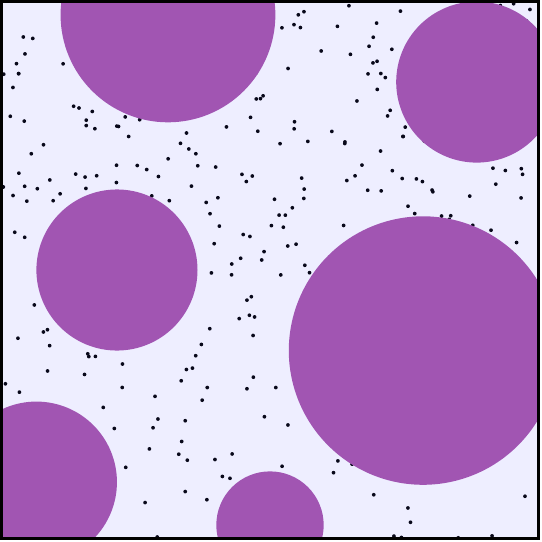}
\includegraphics[scale=1]{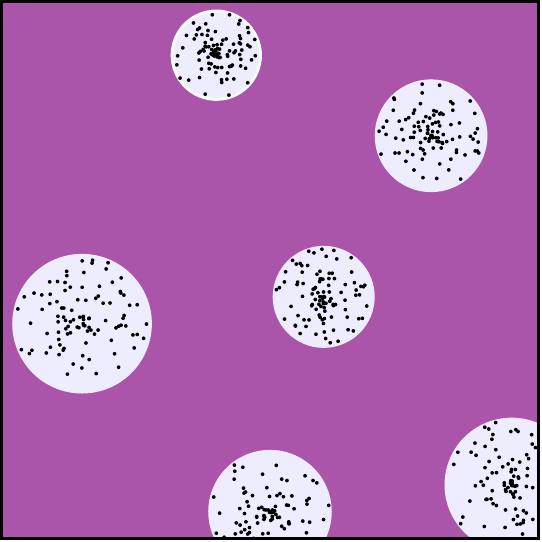}
}\\
\resizebox{0.8\columnwidth}{!}{
\includegraphics[scale=1]{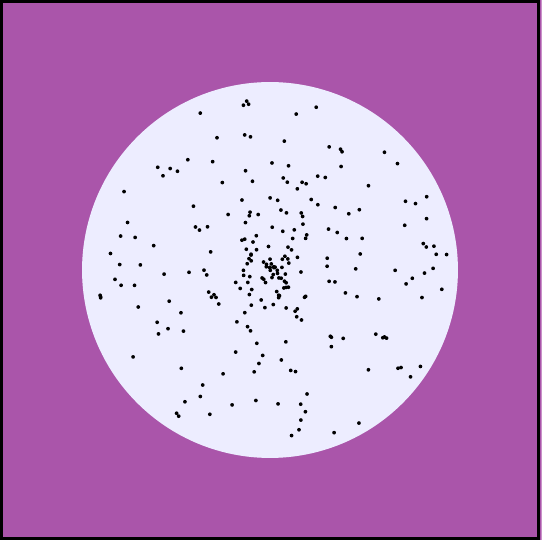}
\includegraphics[scale=1]{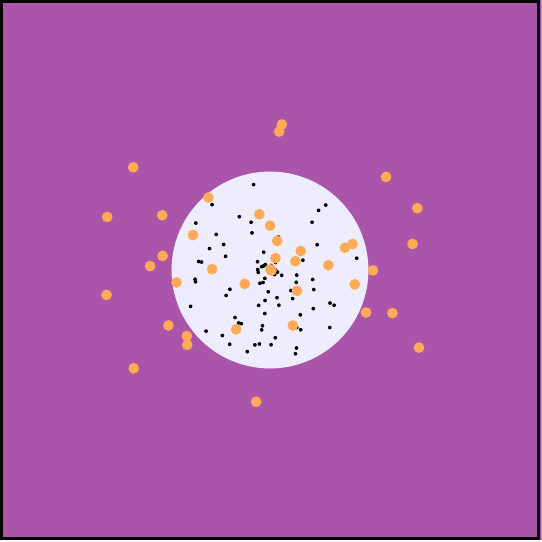}
\includegraphics[scale=1]{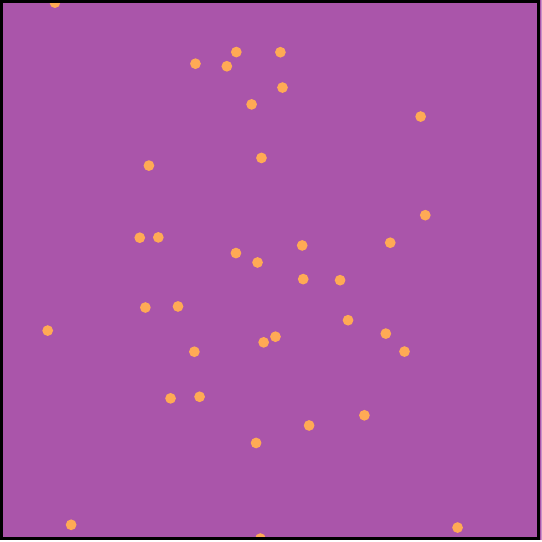}
}
\end{center}
\caption{Different phases of heavy dark quarks (black dots) being trapped in pockets of the deconfined phase (light purple) as the phase transition to the confined phase (dark purple) proceeds. In every single pocket (second row), dark quarks eventually recouple and the pocket shrinks, go through both annihilation and dark baryon DM (orange) formation. Eventually, only a small fraction of quarks in each pocket survives this process, highlighting the drastic impact of first order confining phase transition on DM abundance in such models. Figure from Ref.~\cite{Asadi:2021pwo}. }
\label{fig:squeeze}
\end{figure}

In the light quark limit, the collision of quarks with bubble walls of the confining phase transition instigates string formation and fragmentation, giving rise to parton showers \cite{Baldes:2020kam}. 
These processes can substantially reduce the DM relic abundance during the phase transition, thereby requiring larger DM masses (than the usual thermal freeze-out scenario) to achieve the observed DM abundance today \cite{Baldes:2021aph}.
In the opposite regime of very heavy dark quarks, string breaking is strongly suppressed. As a result, the diffuse gas of colored dark quarks becomes trapped inside contracting pockets of the deconfined phase during the phase transition \cite{Asadi:2021yml,Asadi:2021pwo}, see \cref{fig:squeeze}. As these pockets shrink, the quarks are brought back into contact with one another, leading to an efficient stage of annihilation that can significantly reduce their number density and allow for viable DM candidates at higher masses. 
In this regime, the large hierarchy between the confinement scale and the dark quark mass can also lead to the production of long-lived glueballs. The subsequent decay of these glueballs into SM particles injects entropy into the visible sector, further diluting the DM abundance and opening additional parameter space for heavy DM candidates \cite{Asadi:2022vkc}. 
Precise predictions are difficult given 
the strong dynamics uncertainties 
in these calculations,  motivating further work~\cite{Gouttenoire:2023roe} on the effect of inhomogeneities in contracting deconfined pockets, see e.g.~Ref.~\cite{Gouttenoire:2023roe}. 
A first-order confining phase transition has also been argued to be able to give rise to quark nuggets \cite{Witten:1984rs,Bai:2018dxf}. 
If they exist, these exotic objects can give rise to unique phenomenology \cite{Bai:2024muo}. 
Another possible consequence of a first-order phase transition is observable curvature perturbations due to bubble nucleation \cite{Ho:2026xas}.

Models featuring a first-order confining phase transition could also give rise to gravitational wave signals \cite{Schwaller:2015tja,Helmboldt:2019pan}. At first sight, one might expect that the strongly-interacting nature of these phase transitions would produce sizable gravitational wave backgrounds. 
However, it has been argued that the typically small amount of supercooling in confining phase transitions significantly suppresses the resulting signal, rendering it unlikely to be detectable with foreseeable gravitational wave experiments \cite{Asadi:2021pwo,Agrawal:2025xul,Ayyar:2026wht}.
As with many theoretical studies of gravitational wave production, the current analyses capture only the leading-order dynamics of these phase transitions. A more complete understanding of the resulting signals will require further detailed investigations of the underlying non-equilibrium dynamics.

Aside from harboring different DM candidates such as dark mesons or baryons, 
confining dark sectors can facilitate the formation of another DM candidate, namely primordial black holes (PBHs). 
The phase transition, as well as other aspects of confining dark sectors, can potentially give rise to large enough density overfluctuations that seed gravitational collapse and formation of PBHs \cite{Baker:2021nyl,Dvali:2021byy,Baker:2021sno,Gonin:2026xhe}. 
Such mechanisms remain one of the major open questions in models of PBH DM\@.

Confining dark sectors can also give rise to an early matter domination epoch. 
For instance, in models with no light quarks, the dark glueballs can be long-lived, which in turn allows them to dominate the energy density of the bath and initiate a matter domination epoch, see e.g.~Ref.~\cite{Acharya:2017szw,Asadi:2022vkc,Foster:2022ajl}. This can give rise to signatures such as early formation of DM halos and enhancements of small-scale structures, microlensing signals, Doppler-shift in the frequency of pulsars, or even modification to potential primordial gravitational wave signals, see e.g.~Refs.~\cite{Blinov:2021axd,Erickcek:2021fsu,Pearce:2023kxp}. 

These sectors may also play an important role during inflation. Although inflationary model building extends far beyond confining dynamics, several works have explored how composite sectors can give rise to novel inflationary mechanisms and signatures, see e.g.~Refs.~\cite{Channuie:2011rq,Konstandin:2011dr,Bezrukov:2011mv,Evans:2012jx,Channuie:2014ysa,Samart:2022pza,Cacciapaglia:2023kat,Cacciapaglia:2025xqd}. 
In natural and warm-inflation models, an axion-like field can act as the inflaton, with its potential generated by non-perturbative effects in an underlying confining theory \cite{Freese:1990rb,Kim:2004rp,Berghaus:2019whh,DeRocco:2021rzv,Kamali:2023lzq,Biondini:2024cpf,Berghaus:2025dqi}. 
In these scenarios, the same confining dynamics that shape the inflationary potential can also lead to distinctive early-universe phenomena. 
For example, it has been argued that reheating may proceed through collisions of bubbles produced during a 
confining phase transition \cite{Konstandin:2011dr}. 
Strongly-coupled sectors during inflation can also be probed through cosmological-collider signatures \cite{Hubisz:2024xnj,Kumar:2025anx}.

\section{Detecting Confining Dark Matter}
\label{sec:detection}

Strongly coupled theories admit a variety of composite dark matter candidates, many of which can be probed through direct detection searches that probe the scattering off detector materials, as well as indirectly through annihilation into astrophysically observable final states. 
In this section, direct detection is addressed first, naturally splitting into two qualitatively distinct possibilities:  elastic scattering, where the incoming and outgoing states remain the same, and inelastic scattering, where DM either upscatters into an excited state (endothermic) or an excited state of DM downscatters (exothermic) into DM during its interaction with the scattering material.
Then, we also discuss collective excitations
that can arise from extended composites. 
Finally, we discuss indirect detection and astrophysical signals.

\subsection{Direct Detection through 
Elastic Scattering}
\label{subsec:elastic_DD}

Dark matter 
may be detectable by 
searching for elastic scattering of DM
with target particles of detector on Earth.
In this section, we focus on composite DM that appears ``particle-like,'' that can be detected on Earth at non-relativistic velocities characteristic of the DM distribution in the galaxy. 
So long as the momentum exchange is small compared to the effective size of the composite, DM will generally appear ``point-like'' with
respect to elastic scattering interactions with nuclei (or electrons). 

This allows the formulation of an
effective field theory for the 
effective interactions of strongly-coupled composite DM\@.   
The EFT is characterized by 
the non-relativistic expansion of all possible interactions of the DM with SM fields, 
and if present, with dark sector fields.  
This IR description of the effective theory 
is then matched onto the actual UV theory, see e.g.~Ref.~\cite{Hill:2014yka}.  In many cases, some operators or classes of operators may be suppressed or absent due to the symmetries of the underlying UV theory.

\subsubsection*{Electromagnetic moments}

\begin{figure}
\centering
\begin{subfigure}[t]{0.32\textwidth}
\vspace{0pt}
\includegraphics[width=\textwidth]{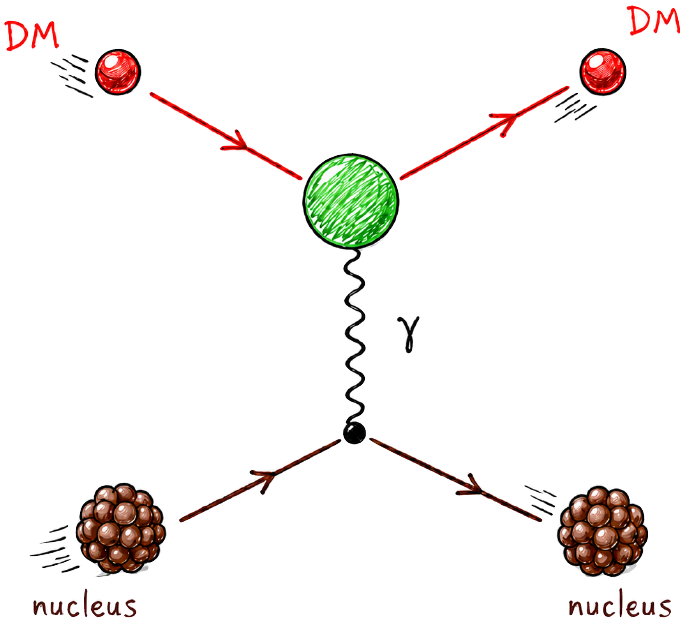}
\caption{Scattering through one-photon electromagnetic moments.
The green blob denotes the particular moment.}
\label{fig:dda}
\end{subfigure}
\hfill
\begin{subfigure}[t]{0.32\textwidth}
\vspace{0pt}
\includegraphics[width=\textwidth]{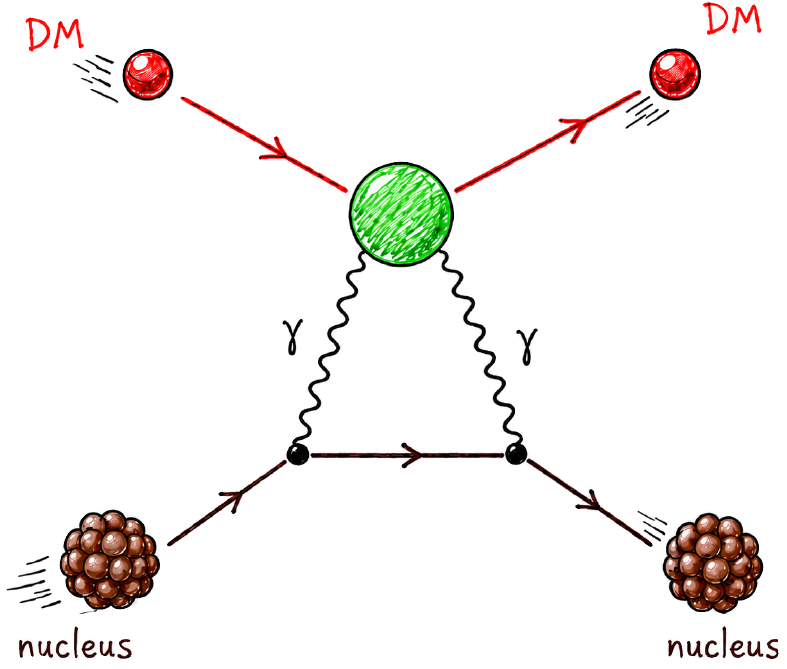}
\caption{Scattering through two-photon exchange electromagnetic form factors (polarizability).}
\label{fig:ddb}
\end{subfigure}
\hfill
\begin{subfigure}[t]{0.32\textwidth}
\vspace{0pt}
\includegraphics[width=\textwidth]{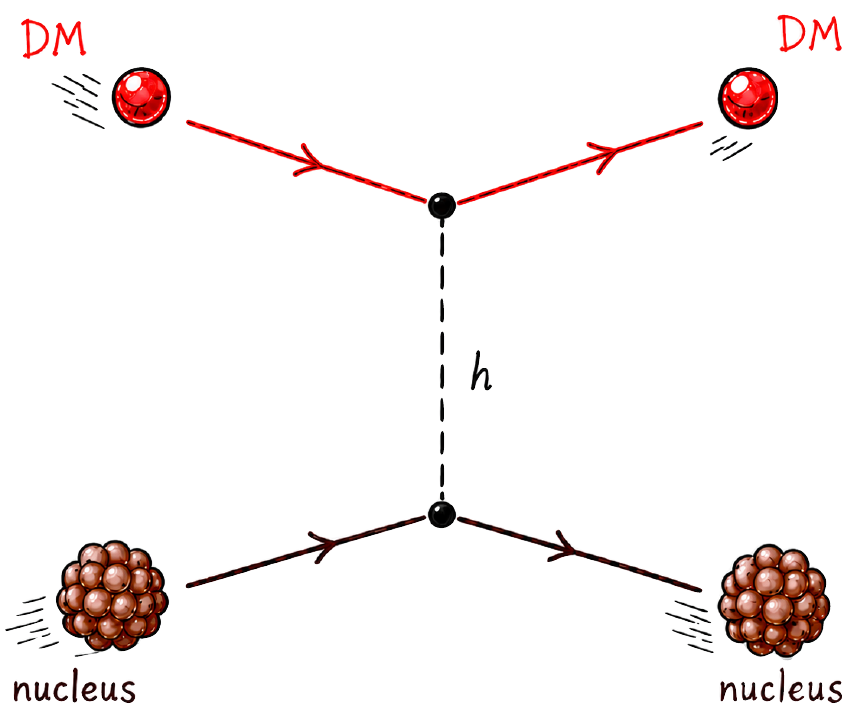} 
\caption{Scattering through Higgs exchange, when the Higgs boson directly couples to the fields in the dark sector.}
\label{fig:ddc}
\end{subfigure}
\\ 
\vspace{1em}
\begin{subfigure}[t]{0.32\textwidth}
\vspace{0pt}
\includegraphics[width=\textwidth]{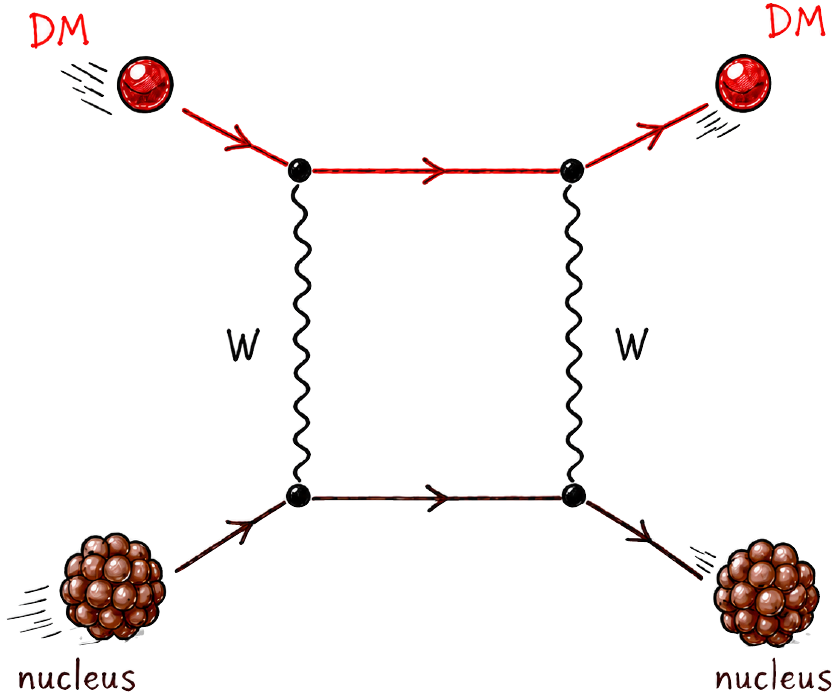}
\caption{Scattering through electroweak gauge boson exchange, when DM is a neutral component of an SU(2)$_L$ multiplet.
(Only one diagram is shown.)}
\label{fig:ddd}
\end{subfigure}
\hfill
\begin{subfigure}[t]{0.32\textwidth}
\vspace{0pt}
\includegraphics[width=\textwidth]{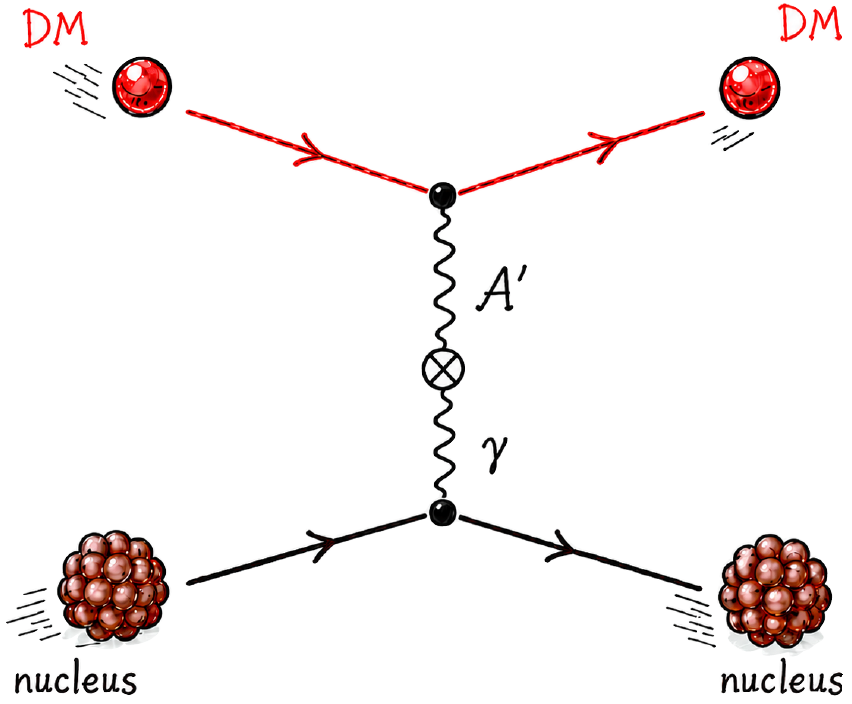} 
\caption{Scattering through dark photon exchange, when the dark sector couples to a new vector boson mediator that mixes with the SM\@.}
\label{fig:dde}
\end{subfigure}
\hfill
\begin{subfigure}[t]{0.32\textwidth}
\vspace{0pt}
\includegraphics[width=\textwidth]{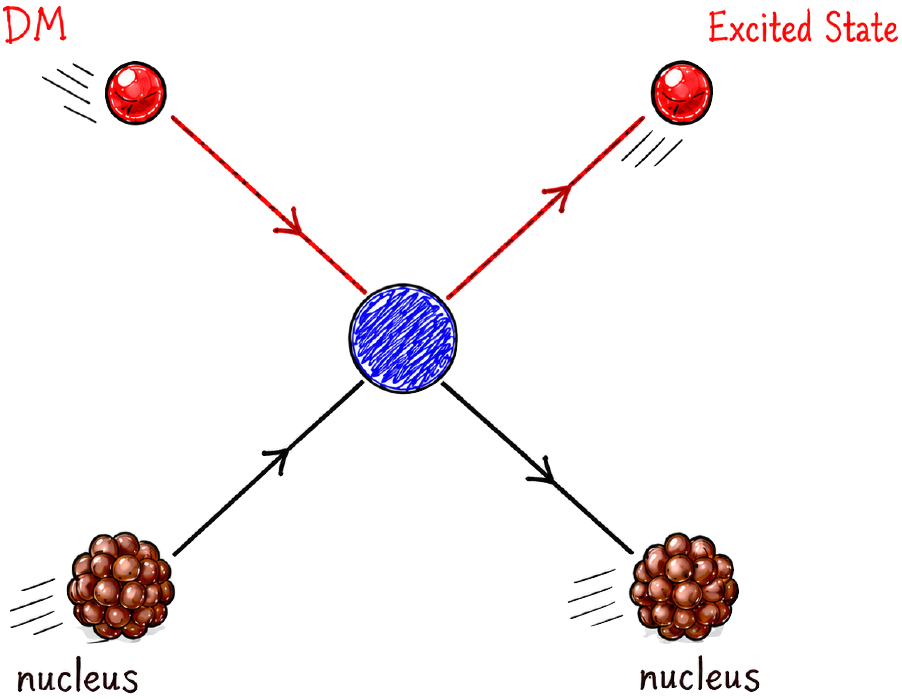} 
\caption{Inelastic scattering of dark matter to an excited state.  The blue blob denotes the specific interaction the enables the inelastic transition.}
\label{fig:ddf}
\end{subfigure}
\caption{Schematic examples of common mediators enabling confined DM to scatter off SM nuclei.}
\label{fig:dd}
\end{figure}

Electrically neutral composite DM that is itself composed of electrically charged constituents is expected to have electromagnetic moments
that enable scattering off nuclei in the SM, see \cref{fig:dda,fig:ddb}.
Technicolor theories that proposed a dynamical origin for electroweak symmetry breaking  provided the first set of examples of such composite DM with electromagnetic moments. Neutral technibaryons were recognized to have observable elastic scattering
\cite{Bagnasco:1993st}, 
where fermionic technibaryon DM could scatter through  a magnetic dipole interaction, while 
scalar technibaryon DM first occurs through the charge-radius interaction.  A general low-energy framework for constraining electromagnetic form factors of DM was developed in Ref.~\cite{Pospelov:2000bq}. This analysis provided phenomenological estimates for the electromagnetic form factors for magnetic and electric dipoles, anapole moments, charge-radius operators, and polarizabilities.
A subsequent study of electric and magnetic dipole DM showed how direct detection, indirect detection, and astrophysical constraints could set bounds on 
electromagnetic moments of neutral DM \cite{Sigurdson:2004zp}.

A more detailed analysis of the differential scattering cross sections for DM with electromagnetic moments was carried out in Refs.~\cite{Barger:2010gv,Banks:2010eh}. These papers derived recoil spectra for DM with an electric dipole, magnetic dipole, and a charge-radius interaction.  
The differential rate spectrum 
and velocity dependence 
of these electromagnetic multipoles 
could help disentangle the fundamental scattering process
\cite{Fitzpatrick:2010br},\footnote{
Ref.~\cite{Fitzpatrick:2010br} called these ``dark moments,'' but we reserve this term to refer to moments under a dark gauge group, not  electromagnetism.} 
making these operators natural probes of neutral composite states with charged constituents.
Ref.~\cite{Banks:2010eh} emphasized that dark baryons in gauge-mediated or technicolor-like sectors are natural candidates for such interactions.
An analysis of light DM with a magnetic moment in direct detection searches was done in Ref.~\cite{DelNobile:2012tx}. This work clarified how magnetic dipole scattering differs from conventional spin-independent scattering, including its dependence on nuclear charge, magnetic moments, and detector thresholds.  
Ref.~\cite{Aranda:2015jis} provided further studies of 
magnetic dipole moments for composite DM in models where neutral bound states are built from constituents charged under a new strong interaction.
Lattice form factors for composite DM with dark neutron-like states in a strongly-coupled gauge theory were computed in Ref.~\cite{LatticeStrongDynamicsLSD:2013elk}. The calculation showed how dark baryon magnetic moments and charge radii can be extracted nonperturbatively and then used to predict elastic scattering rates in direct detection experiments.
Finally, a broad survey of accidental composite DM in confining SU(N) and SO(N) gauge theories was presented in Ref.~\cite{Antipin:2015xia}. The paper identified when stable composite states acquire electric or magnetic dipoles, spin-dependent interactions, Higgs-mediated scattering, or inelastic channels.

If the one-photon electromagnetic moments (magnetic moment, charge radius, etc.) are suppressed or absent, the leading interaction involves two photons,
see \cref{fig:ddb}.  This is known as ``polarizability,'' which is well-known to be the effective interaction for the scattering of light off neutral atoms, known as Rayleigh scattering.\footnote{The similarity to light scattering off atoms led 
\cite{Weiner:2012cb}
to coin the term ``Rayleigh DM,'' however we will restrict our nomenclature to the DM's induced multipole response.}
Ref.~\cite{Weiner:2012cb}
emphasized that 
in theories where the (composite) DM particle is a neutral scalar or Majorana bound states, these candidates often lack one-photon form factors, making polarizability the leading elastic electromagnetic interaction.
Further investigations of UV completions where polarizability 
(as well as transition dipole operators)
could emerge from charged mediator sectors was studied in Refs.~\cite{Weiner:2012gm,
Barducci:2025twe}.
Direct detection of DM polarizability 
probes different nuclear response functions as compared with conventional spin-independent scattering, analyzed in detail in Ref.~\cite{Ovanesyan:2014fha}.
Polarizability-induced elastic scattering was shown to depend on nuclear two-body charge distributions.
Lattice simulations were employed to calculate the polarizability of ``stealth DM'' \cite{Appelquist:2015zfa}, a strongly-coupled composite DM candidate that has no single-photon electromagnetic moments. In this specific strongly-coupled theory, the polarizability of the scalar baryon DM candidate was found to be suppressed (compared with a na\"ive rescaling of polarizability of the neutron to the scale of the composite DM candidate).

A systematic classification of DM-photon effective operators was given in Ref.~\cite{Kavanagh:2018xeh}. The resulting operator basis organizes the low-energy electromagnetic interactions---dipoles, anapoles, charge radii, and polarizabilities---that can arise from dark-sector compositeness.
A recent analysis of direct detection through DM electromagnetic properties was presented in Ref.~\cite{Ibarra:2024mpq}. The paper treated charge, charge radius, electric dipole, magnetic dipole, and anapole interactions, including updated nuclear recoil and Migdal-effect phenomenology.

Vectorlike confining dark sectors can possess a symmetry 
($H$-parity) that suppresses the one-photon electromagnetic moments of dark baryons \cite{Asadi:2024bbq}. This result explains how apparently generic composite dark baryons can evade elastic direct detection bounds, leaving polarizability, electroweak loop effects, or electromagnetic dipole transition moments as the leading probes.

\subsubsection*{Higgs portal}

In addition to electromagnetic moments, composite DM may also be detected when the underlying UV theory itself interacts with the SM Higgs field,
see Fig.~\ref{fig:ddc}.
The first strongly-coupled theory that contained a composite DM candidate with a large Higgs interaction arose in  ``quirky composite DM'' \cite{Kribs:2009fy}.  
In this theory, DM is a scalar baryonic bound state of a confining SU(2) group where the dark fermions are chiral with respect to the electroweak group.  
This led to large interactions with the Higgs boson, that were allowed by direct detection constraints at the time of the paper (2009), but were eventually  ruled out by the substantial improvements in direct detection searches.   
The DM candidate did not have a charge radius interaction (due to a parity symmetry), and so its leading electromagnetic moment was polarizability. 
The chiral nature of this theory enabled the possibility to obtain the relic abundance through an asymmetric mechanism tied to electroweak sphalerons.  
A vectorlike confining SU(2) theory was also explored in 
Ref.~\cite{Pasechnik:2014ida},
where a phenomenological coupling between the confined states and the Higgs field was introduced to further explore direct detection.

Higgs interactions can be suppressed if the dark fermion field content permits vectorlike masses as well as Higgs interactions.  A theory
with these characteristics with four flavors of dark quarks transforming under a dark SU(4) confining group,
called Stealth Dark Matter, 
was constructed in Ref.~\cite{LSD:2014obp,Appelquist:2015yfa}.  Lattice simulations were used to calculate the effective Higgs interaction, and hence the Higgs-mediated spin-independent elastic scattering cross section.  A nonzero Higgs interaction was required so that the lightest dark meson had a mechanism to decay, leaving the lightest scalar dark baryon as the sole DM candidate.   

Higgs exchange was also explored in Ref.~\cite{Mitridate:2017oky} where dark fermion singlets and electroweak doublets permitted Yukawa couplings with the Higgs field.  After electroweak symmetry breaking, the light constituent acquires a Higgs coupling through mixing, and after confinement, a dark baryon Higgs coupling is generated.  This was explored for a variety of confining theories in the heavy dark fermion limit, 
where the dark baryon is well approximated as a weakly bound state of heavy constituents. The Higgs coupling follows, to leading order, from the Higgs dependence of the constituent fermion masses. 
In the presence of additional scalar mediators, vectorlike dark fermions can have interactions with Higgs field.  
Ref.~\cite{Gouttenoire:2023roe}  showed that very heavy composite dark baryons can remain cosmologically viable while still being probed by direct detection experiments through these interactions.

\subsubsection*{Electroweak portal}

Direct detection of composite electroweak multiplet dark baryons has also been studied, opening up an ``electroweak portal,'' see Fig.~\ref{fig:ddd}, that is distinct from both Higgs exchange and electromagnetic multiple moments.  
Ref.~\cite{Antipin:2015xia}
recognized that 
$Y=0$ triplet or quintuplet dark baryons inherit loop-level direct-detection rates analogous to Minimal Dark Matter \cite{Cirelli:2005uq}, while Majorana technibaryons can avoid vector $Z$-exchange and electromagnetic dipole moments, leaving only suppressed spin-dependent $Z$-exchange or Higgs-mediated scattering. A more detailed treatment was given in Ref.~\cite{Mitridate:2017oky}, where $Y=0$ dark baryons scatter through $W$ loop diagrams, while additional electroweak-charged constituents can induce suppressed $Z$- and Higgs-mediated interactions through mixing. The realization that the
neutral components of composite odd-dimensional SU(2)$_L$ multiplets can have their leading electromagnetic moments forbidden by the discovery of $H$-parity \cite{Asadi:2024bbq} implies electroweak loop-induced scattering is the dominant direct-detection channel unless the lightest baryon is an electroweak singlet or only weakly mixed with non-singlet states
\cite{Asadi:2024tpu}. 
Another example where the lightest dark baryon state is nearly a pure singlet, and thus has highly suppressed $Z$-exchange is  
``Hyper Stealth DM'' \cite{Fleming:2024flc}. 
The direct detection rate is controlled by a small non-singlet admixture, that is also correlated to the lifetimes of long-lived dark mesons that accompany the model.

\subsubsection*{Dark photon portal}

If the dark sector transforms under a dark U(1) gauge symmetry, 
composite DM candidates may be probed in direct detection experiments via their interactions with a dark photon that kinetically mixes with the photon, see Fig.~\ref{fig:dde}. This has been studied in the contexts of a chiral dark sector \cite{Co:2016akw}, asymmetric DM \cite{Ibe:2018juk, Ibe:2018tex}, and self-interacting dark baryons \cite{Cline:2022leq}. 
In general, the kinetic mixing that enables scattering with the SM an additional free parameter of the theory, and so the direct detection rates are not determined by the strongly-coupled sector itself.  However, Ref.~\cite{Ibe:2018tex} did consider generating the kinetic mixing with a specific UV completion.
Ref.~\cite{Alonso-Alvarez:2023rjq} studied the case of ``non-Abelian kinetic mixing,'' where the photon has an effective interaction with the dark gluon and a scalar adjoint of the dark sector. In the IR, this leads to an effective mixing of the photon with a bound state that couples to the DM candidate.

\subsubsection*{Other higher-dimensional operators}

Direct detection of dark glueball DM has been considered in several contexts \cite{Soni:2016gzf,McKeen:2024trt,Li:2026nse}. 
A common element is that dark glueball scattering of nuclei is often strongly suppressed by the high dimension of the operators that couple the glueballs from the dark sector to the SM\@.

Alternatively, new mediators can connect the dark sector to the SM\@.  
In Ref.~\cite{Carmona:2024tkg},
stable dark pions scatter off SM fermions via a contact interaction that arises from the deconfined-phase portal interaction depicted in \cref{fig:tqq}.
One of the interesting consequences of this model is that 
the scattering cross section for direct detection is directly tied to the lifetime of other unstable dark pions that can produce emerging/semi-visible jet signatures, see \cref{sec:darkShowers}.

Dark matter that is a bound state of new heavy quarks transforming under QCD was considered in \cite{DeLuca:2018mzn}.
In this work, 
the leading scattering cross section with the SM occurs through
the chromo-electric polarizability of the   bound state.

\subsubsection*{Direct detection of dark nuclei}

Strongly-coupled confining dark sectors can contain DM candidates that are \emph{dark nuclei}, made up of a few to possibly a huge number of nucleons, depending on the specific model and parameters.

In an SU(2) confining dark sector theory with two flavors, 
Refs.~\cite{Detmold:2014qqa,Detmold:2014kba}
recognized that dark nuclei could form, and the light states were stable.  The cosmology of this dark sector is novel, containing both dark nucleons and dark nuclei, where dark nucleosynthesis can populate low $A$ nuclear states \cite{Detmold:2014qqa}.
The spectrum including the masses, form factors, etc., was  established by 
lattice spectroscopy \cite{Detmold:2014kba}.
Although Ref. \cite{Detmold:2014qqa} focused mainly on cosmology and indirect detection, the lattice calculations included the $\sigma$-terms of the lowest dark baryon states, which control scalar-current couplings to the visible sector through Higgs- or scalar-mediated nuclear scattering.

Subsequent work on the formation of low $A$ number nuclei arising from strongly-coupled sectors utilized the pionless effective theory for nucleon-nucleon interactions \cite{Redi:2018muu}.
This analysis considered 
certain choices of the strong gauge group with fermions transforming as triplets (or singlets) of SU(2)$_L$, and assuming the DM abundance arose from an asymmetric mechanism. Ref.~\cite{Redi:2018muu} found that for nucleon masses in the TeV range, baryonic DM made of electroweak constituents can form a significant fraction of dark deuterium, but a much smaller fraction of dark tritium.  Beyond the implication of having multi-component DM, the implications for direct detection remain an open question.

Refs.~\cite{Wise:2014jva,Wise:2014ola}, considered a simple Dirac-fermion plus light-scalar Yukawa theory, showing that stable nuggets with large fermion number  can form.  They compute their structure: 
as the fermion number is increased (but for moderate values), the bound state becomes more compact, while once the core becomes relativistic the radius begins to grow again, leading to a minimum-size configuration.  This leads to a calculable elastic form factor relevant for direct detection. 

Investigations of large $A$ number composites arising from an asymmetric DM scenario were considered in Ref.~\cite{Hardy:2014mqa}. They found that big bang dark nucleosynthesis can produce very large dark nuclei, often with $A \ge 10^8$ or larger, depending on the relative efficiency of the fusion reactions.  The consequences for direct detection \cite{Hardy:2015boa} are that elastic scattering on SM nuclei can be coherently enhanced by as much as $A^2$, but the finite spatial size of the dark nucleus introduces a dark form factor that distorts the recoil spectrum, potentially producing peaks and troughs rather than a simple WIMP-like falling spectrum.  
Ref.~\cite{Butcher:2016hic} investigated 
the novel features in the recoil spectra that arise from nuclear-sized composite DM scattering, and considered the sensitivity of direct detection experiments to these unusual features. Low-energy collective excitations of the large composite may provide additional inelastic scattering processes.

The direct detection implications of large $A$ ``nuggets'' 
were developed in several complementary directions
\cite{Witten:1984rs,Hardy:2014mqa,Hardy:2015boa,Gresham:2017cvl,Gresham:2017zqi,Gresham:2018anj,Bai:2018dxf,Fedderke:2024hfy,Kaplan:2024dsn,Bai:2024muo,Bai:2024amm,Bai:2025zpm}. 
Ref.~\cite{Grabowska:2018lnd} abstracted the problem to a ``dark blob'': if strong dark-sector self-interactions coalesce DM into large composites, their low number density suppresses ordinary single-scatter searches, but their enhanced interaction with the SM motivates transit-style searches using scintillation, calorimetry, accelerometers, strain detectors, and spin-precession experiments. Ref.~\cite{Coskuner:2018are}  specialized to large asymmetric DM bound states with $10^4$
or more dark nucleons and showed that direct detection can proceed through nuclear recoils, phonon or collective-mode excitation, and electronic excitation.  Large keV-threshold detectors are most useful when the composite is effectively pointlike, while low-threshold targets such as superfluid helium, polar materials, and superconductors become more powerful for extended composites or light mediators. Ref.~\cite{Acevedo:2024lyr} then emphasized a more diffuse, loosely bound regime, closer to dark nuclei or dark molecules than saturated nuggets, in which the binding energy per constituent is below the constituent mass. Such objects can give very large nuclear scattering rates scaling like $A^4$, but individual recoils may have much less than a keV while the total energy deposited in one detector passage is much larger than a keV\@.

\subsubsection*{Paleodetectors and supermassive DM}

The paleodetection of DM utilizes ancient minerals as passive, gigayear-exposure nuclear-recoil detectors, where DM  leaves nanoscale tracks that can later be read out with microscopy 
\cite{Snowden-Ifft:1995zgn,
Snowden-Ifft:1997vmx,Baltz:1997dw,Baum:2018tfw}.
Macroscopic, approximately nuclear-density DM could leave long, straight melted cylinders in ordinary rock slabs, giving a geological fossil-track search for supermassive DM with large geometric cross sections
\cite{SinghSidhu:2019znk}.
Ref.~\cite{Ebadi:2021cte} 
suggested that dark sector self-interactions could clump DM into heavy composite states with low number density, proposing to use old quartz that is read out by electron microscopy to search for long, straight-damage tracks. Ref.~\cite{Acevedo:2021tbl} recast ancient muscovite mica data as limits on high-mass DM-nucleus scattering, emphasizing the gigayear-scale exposure of excavated minerals and including the important attenuation/overburden effects for strongly interacting particles. Quite recently,
Ref.~\cite{Boukhtouchen:2026rfz} re-considered muscovite mica with mineral-melt modeling and X-ray readout, targeting large composites with radii from nanometers to microns where ordinary etch-based methods may be inefficient. 
Paleodetectors are thus well matched to large $A$ number strongly-coupled composite DM: the same large size and strong coupling that suppress conventional detector acceptance can generate macroscopic or semi-macroscopic tracks that remain stored in minerals over geological time.

\subsection{Inelastic scattering of composite DM}

One of the hallmarks of composite DM is an ensemble of excited states of the DM that could be accessible through inelastic scattering and direct detection, see Fig.~\ref{fig:ddf}.
There are variety of microscopic processes that can lead to the excitations, including:
hyperfine splittings, transition dipoles, dark atomic levels, radiative capture processes
and more exotic possibilities
that qualitatively differ from ordinary elastic nuclear recoils.

\subsubsection*{Hyperfine inelastic composite DM}

An early foray into composite inelastic DM 
\cite{Alves:2009nf}
introduced a scenario in which DM is a meson-like bound state of a new confining force 
with two widely separated mass scales for the constituents.
They observed that spin-spin interactions naturally generate hyperfine splittings of order $100$~keV, allowing nuclear scattering to proceed dominantly through inelastic transitions between nearby composite states.
Parity violation in composite inelastic DM models was studied in Ref.~\cite{Lisanti:2009am}. 
They showed that parity-violating higher-dimensional operators can mix elastic and inelastic channels, modifying the direct-detection phenomenology relative to the simplest purely inelastic composite scenario.
The cosmology of this meson-like composite inelastic DM was analyzed in Ref.~\cite{SpierMoreiraAlves:2010err}. Here, the hyperfine splitting was connected to dark meson and dark baryon formation in the early universe, showing how the same confining dynamics controls relic abundance, spectroscopy, as well as inelastic scattering.

\subsubsection*{Transition dipoles, de-excitation photons, and luminous signals}

Magnetic inelastic DM was proposed as a scenario in which scattering proceeds through an electromagnetic transition dipole between two nearly degenerate dark states \cite{Chang:2010en}. 
The original proposal involved elementary fermions, though this work motivated the investigation of composite DM because transition moments can naturally appear when dipole moments are forbidden or suppressed by symmetries of the UV theory.
Magnetic ``fluffy'' DM was developed as a framework with a tower of nearby excited states and inelastic magnetic transitions \cite{Kumar:2011iy}. This construction is conceptually close to composite DM because a dense spectrum of excitations is a generic feature of extended or bound-state systems.
The ``inelastic frontier'' of direct detection was surveyed in Ref.~\cite{Bramante:2016rdh}. The analysis emphasized that splittings of order hundreds of keV or larger remain experimentally interesting and are broadly applicable to any type of DM with dominantly inelastic transitions when scattering off nuclei.

One of the distinct signals of inelastic DM is possibility that the nuclear recoil signal is also accompanied by a prompt de-excitation photon \cite{Chang:2010en}. Given the vastly improved constraints on the the nuclear recoil signal, the de-excitation likely occurs well outside the detector. Indeed, a related idea, ``Luminous DM,'' proposed that DM upscatters in the Earth and then decays radiatively inside a detector \cite{Feldstein:2010su}. While their idea was not intrinsically a confining model, 
and although the original model is now highly constrained by improved nuclear recoil constraints,
the mechanism is a natural phenomenological template for composite DM with metastable excited states and transition dipoles.
More recently, luminous DM with low masses ($\sim 0.1$~--~$15$~GeV range) and splittings ($\sim 0.1$~--~$5$~keV) 
was considered in Ref.~\cite{Bell:2022dbf},
where
endothermic scattering in the detector or surrounding Earth produces an excited state whose decay photon gives an electron-recoil line.

Heavy inelastic DM, 
masses of a few hundred GeV or above,
can acquire a sidereal-daily modulation signal in the excited state decay to a photon 
\cite{Eby:2019mgs}. 
Large inelastic splittings,
$300$--$600$~keV, where xenon 
nuclear-recoil searches lose kinematic reach, provide a clear opportunity when the DM 
upscattering occurs off heavy trace elements, such as Pb, in the Earth.
Large underground detectors can be sensitive to these photons, including both Borexino and large gaseous detectors such as CYGNUS. 

Magnetic inelastic DM 
itself was revisited in
Ref.~\cite{Eby:2023wem} where
DM upscatters in the Earth, 
through a magnetic transition dipole, and the excited state later decays inside a large underground detector, giving a monoenergetic photon line signal.  In this work, the same magnetic dipole transition operator controls both the Earth upscattering rate and the excited-state decay length, so the photon flux and detector acceptance are tightly correlated.  The crucial experimental signal is a sidereal-daily modulation: because the DM wind points roughly from constellation Cygnus, the amount of Earth material upstream of the detector changes as the Earth rotates, causing a predicted daily modulation in the photon-line rate. 

In theories with strongly-coupled composite DM, 
Ref.~\cite{Asadi:2024bbq}
identified symmetries of the UV theory that suppress one-photon  electromagnetic moments while permitting transition dipole moments.  In the specific theory that was studied, the mass splitting was not expected to be small enough to permit inelastic scattering off nuclei in direct detection.  Nevertheless, composite theories can provide an elegant mechanism to enable inelastic scattering while suppressing elastic scattering.

\subsection{Exotic detection possibilities}

Large $A$ composite DM 
\cite{Hardy:2015boa} can have
a nontrivial recoil spectrum from the extended spatial structure of the dark composite, 
as we already mentioned. 
In addition, 
collective excitations of the composite can produce coherently enhanced inelastic scattering, although this is generally subdominant to elastic scattering in their benchmarks.

Ref. \cite{Coskuner:2018are} also considered direct detection of large composite asymmetric DM nuggets by comparing nuclear recoil experiments with low-threshold targets sensitive to nuclear recoils, phonons, and electronic excitations. The main result is that conventional xenon-like searches are optimal for compact nuggets and heavy mediators, whereas low-threshold semiconductor, superfluid-helium, superconducting, and polar-material detectors can dominate for extended nuggets or light mediators because they preserve coherence at lower 
momentum transfer.

Ref.~\cite{Acevedo:2020avd} proposed a qualitatively different direct-detection signature for very large asymmetric composite DM: rather than coherent elastic recoils, nuclei that penetrate the composite are accelerated by the internal scalar potential, producing ionization, bremsstrahlung radiation, and in some cases nuclear fusion. The most promising terrestrial searches are therefore large-area, large-volume detectors such as IceCube and liquid scintillator neutrino experiments, which could detect the intense optical/UV/X-ray radiation or fusion byproducts from a rare transiting ultraheavy composite. Conventional nucleus scattering is subdominant, and overburden stopping, Earth heating, white-dwarf ignition, and possible isotope anomalies provide important complementary constraints or discovery channels.

Resonant scattering between DM and baryons was analyzed in Ref.~\cite{Xu:2020qjk}. Motivated in part by strongly interacting and composite candidates such as sexaquark DM, the paper showed that attractive interactions can generate nonperturbative, velocity-dependent, and target-dependent scattering behavior.

While somewhat out of the main topic of this review, we remark that there is a variety of interesting models and signals that arise in theories with ``dark atoms,'' typically bound by a weakly coupled U(1)$_D$, reviewed in Ref.~\cite{Cline:2021itd}.
Another interesting possibility, self-destructing DM
\cite{Grossman:2017qzw},
considers scattering on terrestrial matter converts a cosmologically metastable dark bound state into a short-lived state that annihilates or decays visibly, releasing energy of order the DM mass rather than a keV-scale recoil. The most distinctive direct-detection targets are large neutrino detectors such as Super-K, Borexino, SNO+, and DUNE, where the signal can be one or more lepton/jet pairs with fixed invariant mass, unusual multiplicity, and nontrivial directionality.

\subsection{Astrophysical and Cosmological Signals}
\label{subsec:indirect}

Compared with minimal freeze-out DM, confining dark sectors do not generically predict a simple one-to-one relation between the relic abundance and present-day indirect detection signals. 
In the vanilla WIMP picture, both are largely controlled by the same annihilation cross section, which motivates the canonical thermal target of $\sigma v \sim 10^{-26}$~cm$^3$/s for indirect searches.
In a confining dark sector, by contrast, a host of non-perturbative effects (the confining phase transition itself, bound-state formation, rearrangement processes, and the presence of a rich hadron spectrum) can all modify the relic abundance calculation. 
At the same time, this same structure opens qualitatively new annihilation, de-excitation, and transition channels today, each of which giving rise to novel signals. 
Indirect detection in confining sectors therefore probes a broader range of processes than in elementary DM models.

For symmetric DM, the signals depend sensitively on the spectroscopy of the confining sector. 
In the light dark quark regime, the dense spectrum of hadrons allows DM to be excited through scattering into unstable dark hadrons, which subsequently decays to SM \cite{Carmona:2024tkg}. 
In the heavy quark regime, dark hadronic DM can instead annihilate into dark glueballs, which then decay to the SM and generate indirect-detection signals \cite{Curtin:2022oec}. A quantitative prediction for these spectra requires theoretical control over hadronization in pure-glue sectors, an ingredient that has only recently started to become available \cite{Curtin:2022tou,Batz:2023zef}. 
If the dark states arising from DM up-scattering or annihilation are long-lived enough, they can also give rise to signals in CMB or BBN.

Confinement can also qualitatively change the viable DM mass range relevant for indirect searches. 
In particular, it can open parameter space at masses above those usually emphasized in perturbative thermal relic models \cite{Mitridate:2017oky,Baldes:2020kam,Baldes:2021aph,Asadi:2021pwo,Asadi:2021yml}. Related non-perturbative processes, such as rearrangement \cite{Geller:2018biy,DeLuca:2018mzn}, can lead to geometric cross sections and, when they control freeze-out, allow thermal relic DM above the usual unitarity limit. 
Similar phenomena can arise in theories where the dark quarks are not in the fundamental representation of the confining group \cite{Contino:2018crt}. In such cases, DM can still annihilate efficiently in halos today and produce indirect detection signals, but in a different mass and energy regime from the standard WIMP window.

A particularly interesting feature of confining sectors is that they can remain visible to indirect searches even when the present-day DM abundance is predominantly asymmetric, see e.g.~Ref.~\cite{Mahbubani:2020knq}. 
In elementary asymmetric DM models, conventional indirect signals are typically absent because the symmetric component has been efficiently depleted.\footnote{An exception arises if DM and anti-DM can oscillate into one another. In that case, a residual anti-DM population can survive or be regenerated at late times, allowing DM--anti-DM annihilations today and thereby producing indirect-detection signals, e.g. see Ref.~\cite{Cai:2009ia,Cirelli:2011ac,Tulin:2012re,Hardy:2014dea,Ibe:2019yra}.} 
In composite theories, however, the rich spectrum of stable and metastable bound states opens additional possibilities. 
In particular, \textit{dark nucleosynthesis} can produce observable signals even in asymmetric scenarios \cite{Detmold:2014qqa}. 
The underlying reason is that once more than one stable bound state exists in the spectrum, the relevant observable is set by binding-energy release and level splittings during inelastic capture and transition processes, rather than by particle-antiparticle annihilation alone.
As emphasized in Ref.~\cite{Mahbubani:2019pij}, the formation of heavier dark bound states, such as dark nuclei, can be accompanied by an emission of visible radiation. 
This can yield monochromatic photons or other sharp features in indirect-detection data \cite{Redi:2018muu}, and has even been invoked in explanations of the Galactic-center excess \cite{Detmold:2014qqa} (with limited success). 
Models in which DM is an extended dark hadron, such as puffy DM \cite{Chu:2018faw}, can give rise to similar up-scattering and de-excitation signals. The same extended structure leads to non-trivial velocity-dependent self-interactions, which have been used to address small-scale structure anomalies \cite{Chu:2018faw}.

Confining spectra also motivate signals with no close analog in minimal DM models. For instance, if the DM possesses a transition dipole moment to a nearly degenerate excited hadron, then background photons can be resonantly absorbed, leading to line-like absorption features in astrophysical spectra \cite{Kribs:2009fy,Ganguly:2023rzm}. 
As a second example, the confinement transition, in some models, may produce macroscopic dark nuggets \cite{Bai:2018dxf}. Their subsequent collision and annihilation can give rise to unusual transient signals, including very brief bursts and other non-standard astrophysical signatures \cite{Fedderke:2024hfy,Kaplan:2024dsn,Bai:2024muo}. 
Models with SUEP-like collider signals can also produce rich, exotic annihilation cascades, leading to distinctive spectra of antinuclei final states \cite{DiMauro:2026owr}.

Additionally, both the DM and other composite states can decay to SM, giving rise to indirect detection signals, see e.g.~Refs.~\cite{Falkowski:2009yz,Boddy:2014qxa,Cohen:2016uyg,Dondi:2019olm}. 
The decay can take place to different SM channels. In particular, even the possibility of decay to neutrinos and a signal at neutrino detectors such as Icecube have been considered as well \cite{Falkowski:2009yz,Das:2024tmx}. 
If these states decay to new sterile neutrinos which subsequently decay to SM particles during the recombination epoch, the injected energy can be detectable at CMB, as well as at a host of different searches \cite{Ahmed:2023vdb}. 
The dark glueballs (or glueballinos) can also naturally have very long lifetime and give rise to decaying DM signal \cite{Boddy:2014qxa,Cohen:2016uyg}.

Overall, the broad lesson is that while confinement weakens the standard link between relic abundance and present-day annihilation, the richness of dynamics and the spectrum in the dark sector enlarges the space of possible indirect and astrophysical signals. 
Gamma rays, cosmic rays, absorption features, dark nuclear transitions, and transient macroscopic events can all arise in well-motivated regions of parameter space, including scenarios with asymmetric DM\@. 
A rich phenomenology is awaiting exploration in such models.

Confining dark sectors may also address cosmological anomalies such as the Hubble tension \cite{Buen-Abad:2015ova,Allali:2021azp,Fujikura:2023lkn}. For sufficiently low confinement scales, dark gluons can behave as interacting dark radiation around the CMB epoch, affecting both $H_0$ and $S_8$; see also Ref.~\cite{Zu:2023rmc} for related twin-Higgs realizations. The same sectors may also undergo first-order phase transitions capable of generating a stochastic gravitational-wave background in the NANOGrav band \cite{Fujikura:2023lkn,Chen:2023bms,Balan:2025uke,NANOGrav:2023gor}.

Models with richer dark spectra, such as twin sectors exhibiting dark acoustic oscillations, can leave additional imprints on the matter power spectrum. In these scenarios, twin nuclei may remain coupled to dark radiation until dark recombination, delaying their contribution to structure growth. If DM consists of several twin-nuclear components with different recombination epochs, each component can begin clustering at a different time, leading to nontrivial effects on the matter power spectrum that depend on their relative abundances and recombination histories \cite{Chacko:2018vss,Barron:2025dys}. Small-scale structure anomalies, such as the core-cusp and too-big-to-fail problems, can also be addressed by self-interacting DM models, which arise naturally in confining dark sectors \cite{Cline:2013zca}.

\section{Collider Phenomenology}
\label{sec:pheno}

\begin{figure}[t]
    \centering

    \begin{subfigure}[t]{0.32\textwidth}
        \centering
        \includegraphics[width=0.6\linewidth]{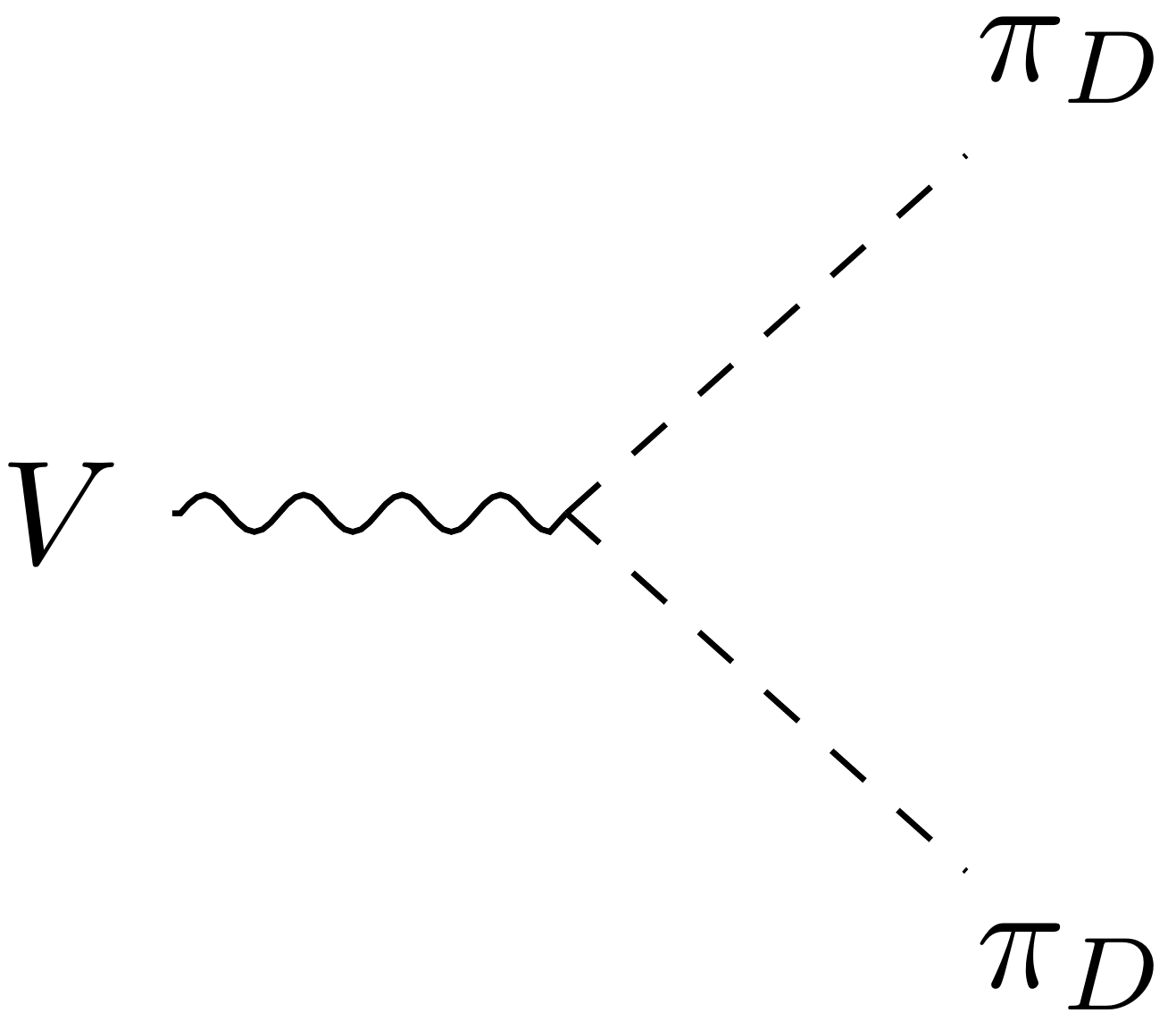}
        \caption{{\bf Gauge boson:} dark hadrons couple  directly to SM gauge bosons or a new gauge boson $V$. The dark hadrons may be in a variety of (potentially exotic) gauge representations.}
        \label{fig:pipiV}
    \end{subfigure}
    \hfill
    \begin{subfigure}[t]{0.32\textwidth}
        \centering
        \includegraphics[width=0.6\linewidth]{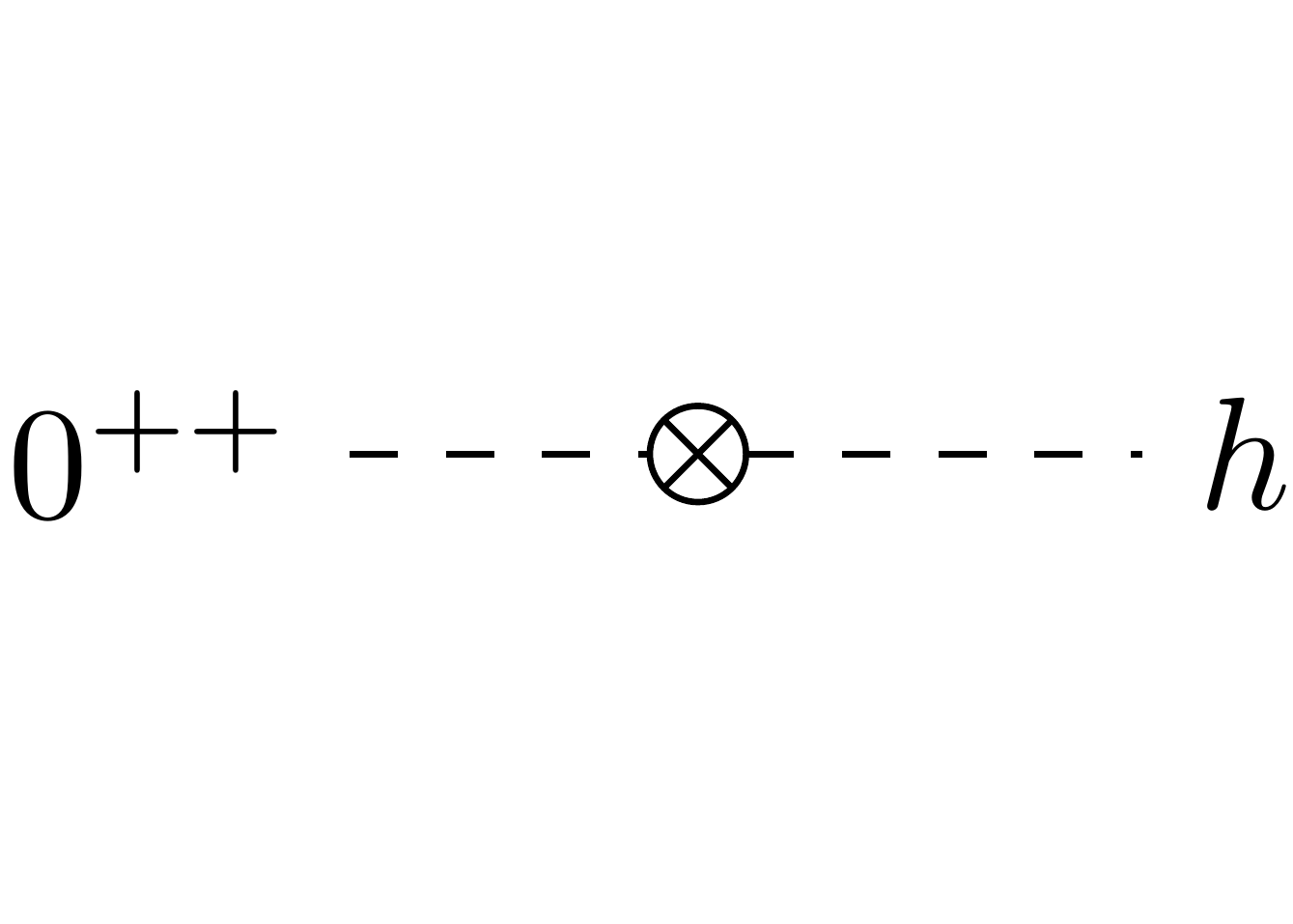}
        \caption{{\bf Scalar mixing:} 
        a scalar such as the $0^{++}$ glueball mixes with the Higgs boson. The $\piD$ could also mix with the Higgs if the dark sector violates $CP$. A pseudoscalar such as the $0^{-+}$ glueball could also mix with an ALP.
        }
        \label{fig:SSmix}
    \end{subfigure}
    \hfill
    \begin{subfigure}[t]{0.32\textwidth}
        \centering
        \includegraphics[width=0.6\linewidth]{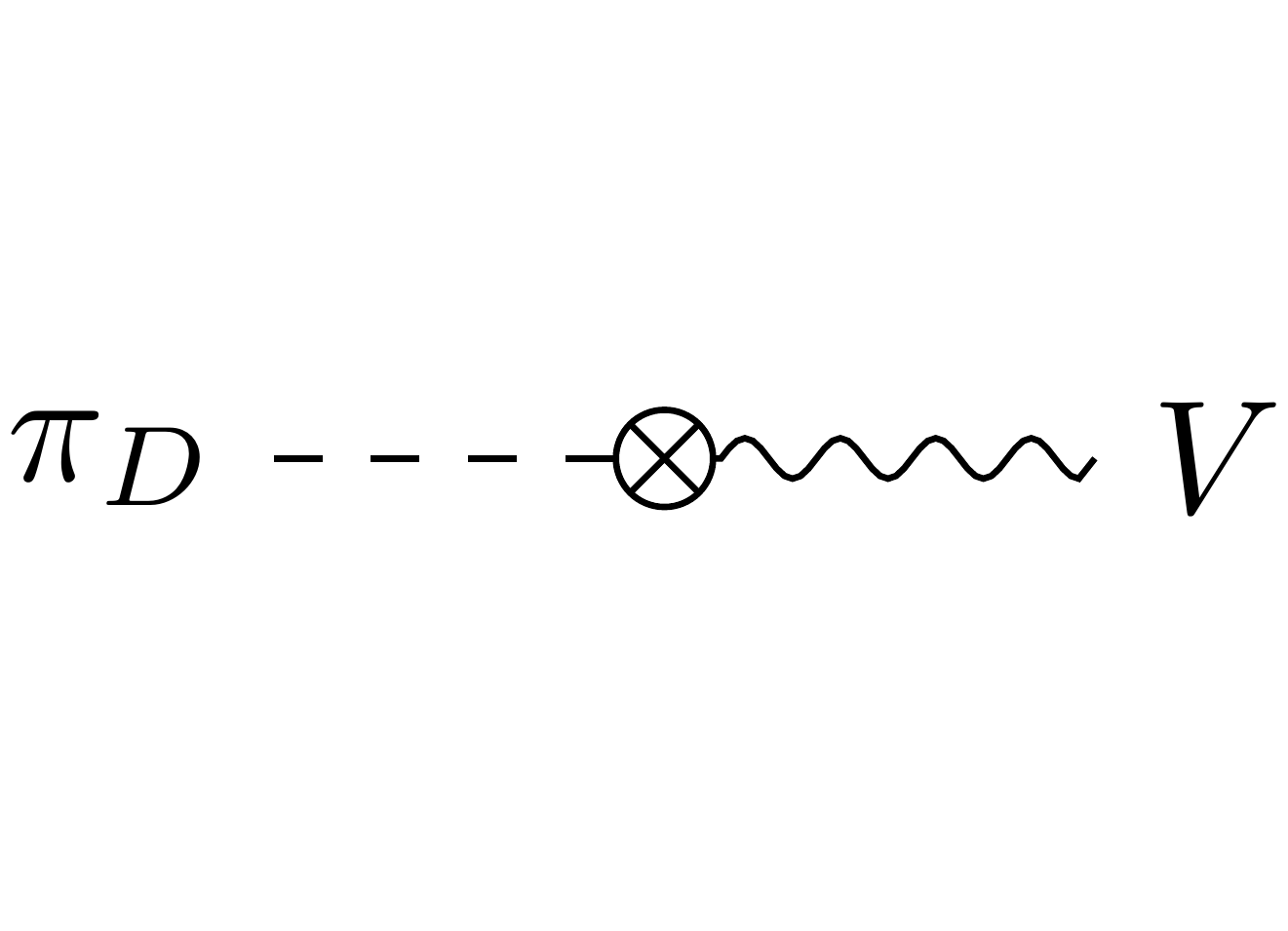}
        \caption{{\bf Goldstone mixing:} 
        $\piD$ mixes with the longitudinal mode of a massive vector boson $V$ (or equivalently, with a Goldstone of the SM Higgs or a dark Higgs multiplet). This requires the dark quarks to have parity-violating gauge couplings. 
        }
        \label{fig:PSVmix}
    \end{subfigure}

    \begin{subfigure}[t]{0.32\textwidth}
        \centering

        \includegraphics[width=0.6\linewidth]{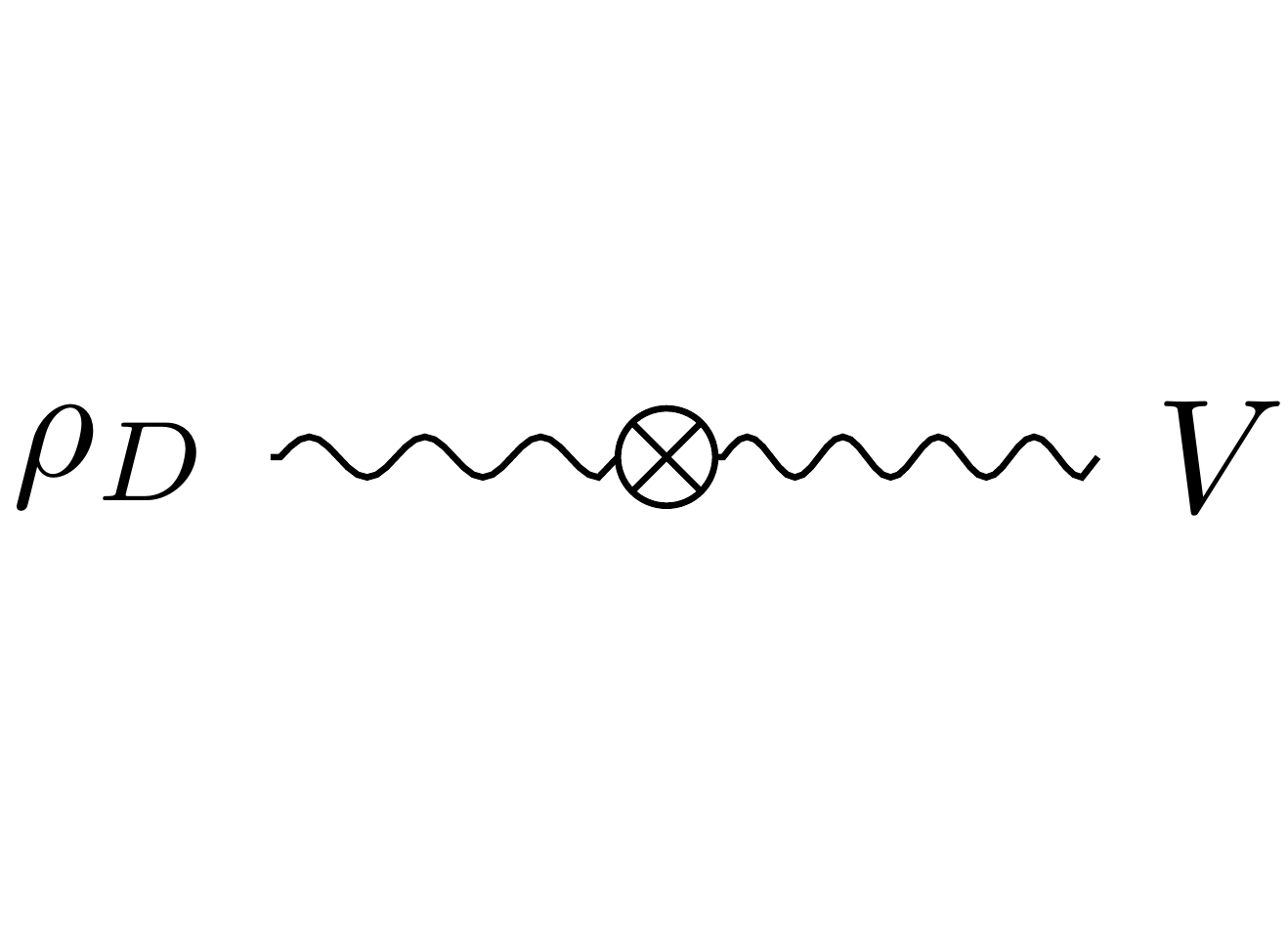}
        \caption{{\bf Vector mixing:} a dark vector hadron such as the $\rhoD$ mixes with some other vector boson $V$. This can be understood in the language of Vector Meson Dominance \cite{alma991020825129706011,Schildknecht:2005xr}.} 
        \label{fig:VVmix}
    \end{subfigure}    
    \hfill
    \begin{subfigure}[t]{0.32\textwidth}
        \centering
        \includegraphics[width=0.6\linewidth]{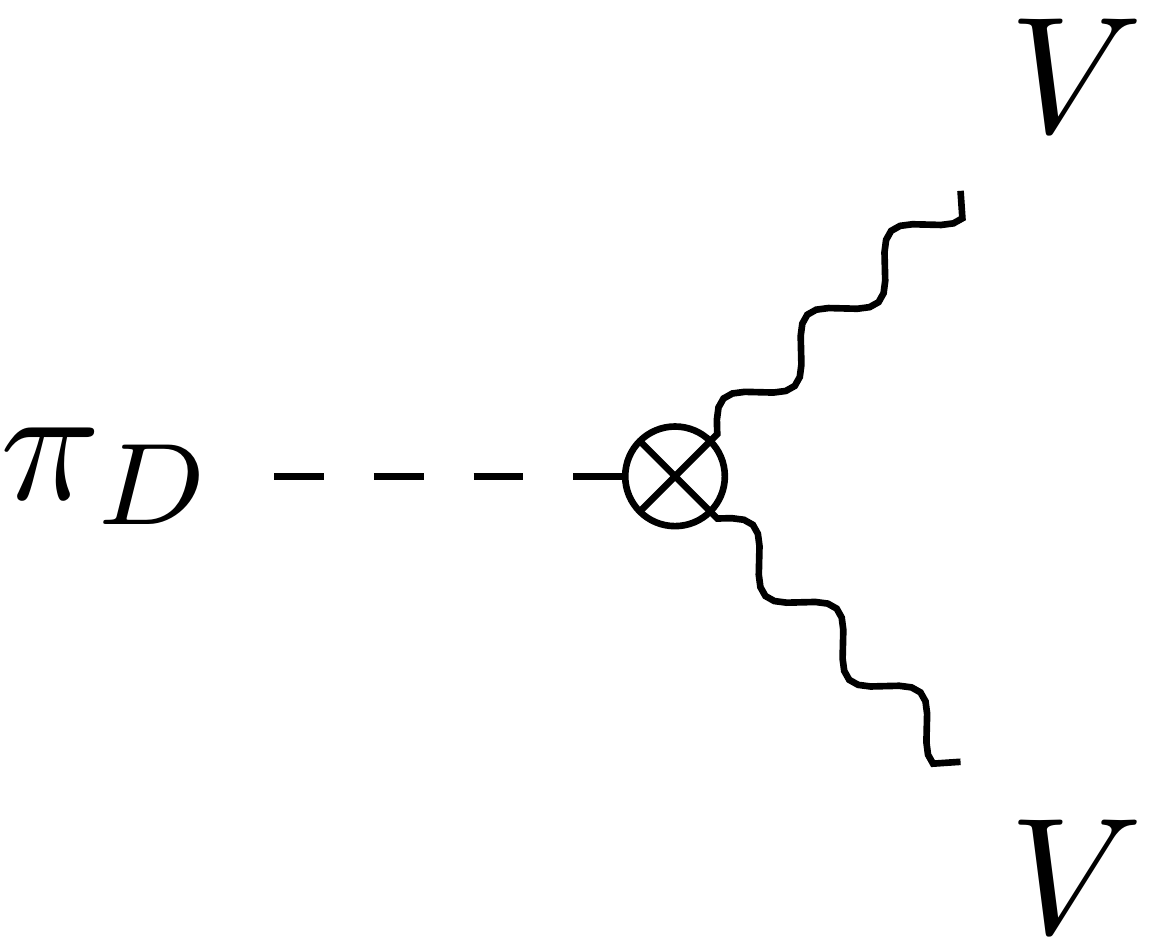}
        \caption{{\bf Chiral anomaly:} $\piD$ has an effective dimension-5 interaction with gauge bosons $V$. These may be SM gauge bosons if the dark quarks carry SM charge. This portal may arise even in vector-like dark sectors.}
        \label{fig:piVV}
    \end{subfigure}
    \hfill
    \begin{subfigure}[t]{0.32\textwidth}
        \centering
        \includegraphics[width=0.6\linewidth]{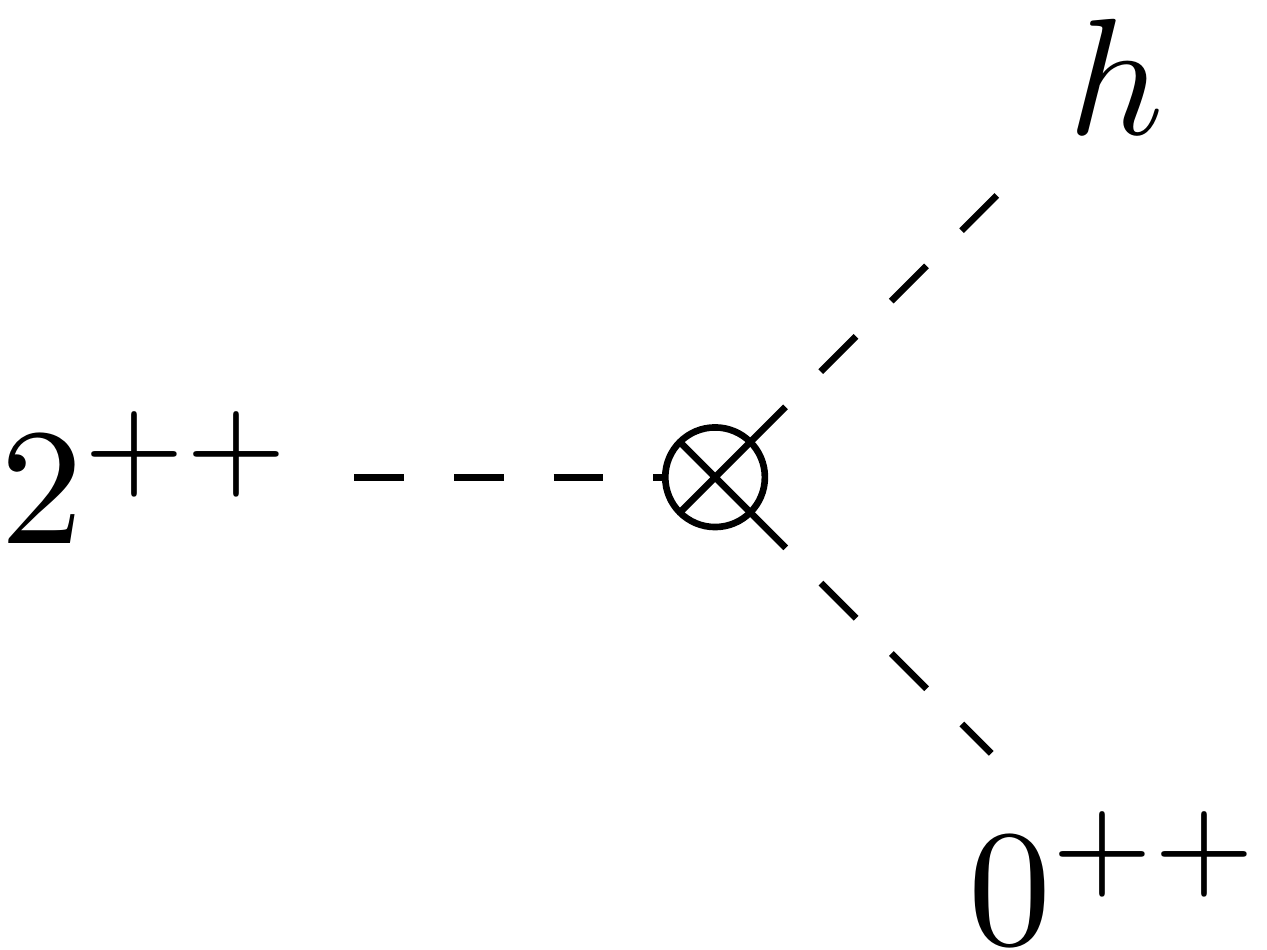}
        \caption{{\bf Radiative transition:} couplings of the partons to the SM induce transitions involving excited dark hadron states. For example, a tensor ($2^{++}$) glueball can transition to the scalar ($0^{++}$) by radiating a Higgs boson.}
        \label{fig:radDecay}
    \end{subfigure}
    \caption{Overview of some portal interactions that can connect dark hadrons either directly or indirectly to the SM, focusing on those relevant to collider phenomenology. Crossed dots denote where there may be higher-dimensional operators or mixings inserted.
    } 
    \label{fig:hadronPortal}
\end{figure}

Confining dark sectors can realize a broad variety of collider signatures depending on the flavor structure, hierarchies of mass scales both in the dark sector and the SM, and the (potentially multiple) portals to the SM\@. In \cref{fig:hadronPortal}, we sketch a non-comprehensive taxonomy of hadron-level portal interactions studied in the collider context. The benchmark models used in collider studies may or may not feature a viable DM candidate.

\begin{figure}[t]
    \centering
    \includegraphics[width=0.8\linewidth]{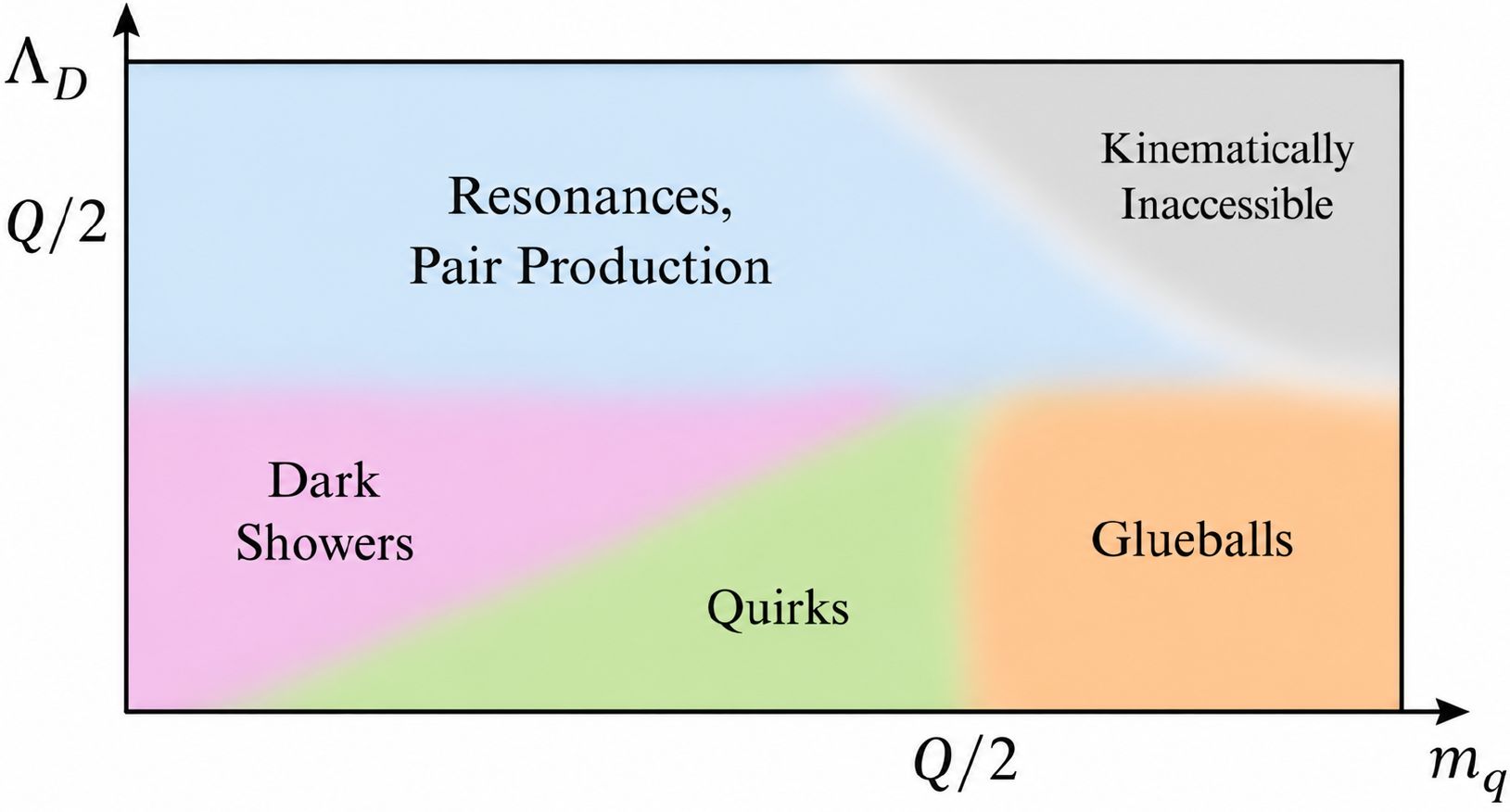}
    \caption{Sketch of mass scale regimes for collider phenomenology in terms of the dark quark mass $m_q$ and the dark sector confinement scale $\LD$. The characteristic scale of the hard interaction $Q$ is held fixed in the plot, though it can vary event-by-event due to parton distribution functions. The boundaries between regions are blurry and carry some implicit dependence on details such as flavor structure. When $\mq$ and $\LD$ are too large, no dark hadrons can be produced.}
    \label{fig:colliderSketch}
\end{figure}

Collider signals are partially dictated by the hierarchies of a few mass scales: the quark mass $\mq$, the confinement scale $\LD$, and the characteristic scale $Q$ of the hard interactions that produce the dark sector states, e.g.~the mass of a resonance that decays to dark quarks. The rough regimes of interest are depicted in \cref{fig:colliderSketch} and further described below. There are other important factors, such as the number of dark quark flavors and their relative mass hierarchies determining whether or not a light pion-like state exists, but $\LD$, $\mq$, and $Q$ fix the overall landscape.

The portal interactions determine the dark hadron lifetimes and observable final states. As seen below, common features of confining dark sectors, including (quasi-)stability of dark hadrons, lead to potentially striking signals such as displaced vertices (DVs) and missing transverse energy (MET). Many features of the collider phenomenology were first pointed out within the original Hidden Valley paradigm \cite{Strassler:2006im, Strassler:2006qa}, where light states that are charged under the confining group are accessed at high energies via heavy mediators, though we take a broader point of view in this section. 

We focus here on experiments at the Large Hadron Collider (LHC), but other existing and historical colliders provide relevant constraints, and future colliders pushing the energy and precision frontiers may expand sensitivity. 
We also omit a detailed discussion of flavor and electroweak precision tests (see e.g.~Refs.~\cite{Cai:2008au,Renner:2018fhh,Cheng:2019yai,Cheng:2024hvq,Blennow:2026psy}), as these are more model-specific.

\subsection{Resonances and Pair Production}
\label{sec:resonances}

Consider the regime where the dark sector is in the confined phase at the hard interaction scale $Q$, so the appropriate effective field theory is always that of dark hadrons, only a small multiplicity of which are produced in collider events. Then, some dark hadrons can be resonantly produced e.g.~via $\rhoD$ mixing with SM gauge bosons \cite{Kilic:2009mi,Kilic:2010et,Sayre:2011ed,Bai:2011mr,Carmona:2015haa,Antipin:2015jia,Beylin:2016kga,Cline:2017aed,Kribs:2018oad,Kribs:2018ilo,Bottaro:2021srh,Palmisano:2024mxj} or a $\piD$ with a chiral anomaly being produced via gluon-gluon or vector boson fusion \cite{Antipin:2015xia,Draper:2018tmh,Palmisano:2024mxj,Asadi:2025vfr}. 
A chiral anomaly can also produce a single $\piD$ in association with a gauge boson \cite{Kilic:2009mi,Antipin:2015xia,Bottaro:2021srh,Abe:2024mwa}. Dark hadron pair production can be mediated by, e.g.,~exotic Higgs decays \cite{Curtin:2014pda,Alipour-Fard:2018lsf,Batz:2023zef} or direct interactions with SM gauge bosons \cite{Kilic:2010et,Bai:2010mn,Freitas:2010ht,Sayre:2011ed,Bai:2011mr,Pasechnik:2014ida,Appelquist:2015yfa,Antipin:2015xia,Carmona:2015haa,Antipin:2015jia,Arkani-Hamed:2016kpz,Englert:2016ktc,Beylin:2016kga,Cline:2017aed,Barducci:2018yer,Ahmed:2023vdb,Palmisano:2024mxj,Asadi:2025vfr,Cacciapaglia:2026jlv}.

Some common features of confining dark sectors lead to characteristic signals in the low-multiplicity regime. Resonances can be constrained by searches for QCD dijets \cite{ATLAS:2019fgd,CMS:2019gwf}, dilepton final states \cite{ATLAS:2019erb,CMS:2021ctt}, and pairs of electroweak bosons \cite{ATLAS:2018iui,ATLAS:2020tlo,ATLAS:2020fry,CMS:2021wlt,CMS:2021klu,ATLAS:2021uiz,ATLAS:2023sua,CMS:2024nht,CMS:2024vps}. Since various $\piD$ species are often long-lived, individual DVs \cite{Curtin:2014pda,Alipour-Fard:2018lsf} (the targets of several dark sector searches \cite{Maravin:2007vdd,D0:2009mtx,CMS:2011xlr,CDF:2011dnt,CMS:2012axw,LHCb:2014jgs,ATLAS:2015mlf,ATLAS:2022gbw,CMS:2024bvl,CMS:2025fnr}) or disappearing tracks signatures \cite{Asadi:2025vfr} (the targets of recent SUSY searches \cite{CMS:2020atg,ATLAS:2022rme,CMS:2023mny,ATLAS:2024umc,ATLAS:2025lhc,ATLAS:2026hnb}) appear in various UV models. For either prompt or displaced decays, branching ratios of pseudoscalar $\piD$ decays to heavy fermions are enhanced by helicity suppression (and similarly for a scalar dark glueball decaying through a Higgs portal \cite{Juknevich:2009gg}), so dark sector searches such as Ref.~\cite{ATLAS:2024xbu} can leverage signatures specific to $b$ and $t$ quarks. 

When the produced dark hadron species are not the lightest in the spectrum, they may undergo ``cascade'' decays \cite{Csaki:2021gfm,Asadi:2025vfr}. In this case, the heavy species transitions down to a lighter species by emitting soft SM decay products. Then, the lighter species may itself transition to an even lighter species, and so on until the lightest available state is produced. 
That lightest species may leave a long-lived particle (LLP) signature described above or escape the detector and result in missing energy. Correlating an LLP or MET signal with the soft radiation from decays of heavier states in the cascade may aid in matching the signature to a UV theory.

\subsection{Dark Showers}
\label{sec:darkShowers}

In the dark showers regime, SM-singlet dark quarks and gluons produced at the high-energy scale $Q$ undergo a perturbative parton shower, iteratively splitting into lower-energy pairs of partons. As the shower energy scale decreases toward $\LD\ll Q$, the dark sector coupling $\aD$ becomes larger, until the partons eventually bind into a large multiplicity of collimated hadrons in the non-perturbative process of hadronization. The decay of these dark hadrons back to the SM creates a signal that can resemble a QCD jet. Distinguishing these jets from the enormous QCD background is the characteristic challenge of searching for dark showers. 

Some common features of dark showers aid in this effort. For example, some hadron species often are (quasi-)stable due to symmetries or large mediator masses, which leads to DVs or MET\@. The SM decay products also tend to be less collimated than a QCD jet due to the multi-stage process of dark sector showering, then decay, then a QCD shower from quark/gluon decay products \cite{Han:2007ae}. Moreover, unless the portal is leptophobic, the jets tend to contain more leptons than QCD jets, including muons that can be triggered on and tagged \cite{Han:2007ae,Born:2023vll,Niedziela:2024khw,Lu:2025cty}. 

Qualitatively different signatures arise depending on how many hadrons from the dark shower decay promptly, decay displaced within the detector, or are collider-stable. The ubiquity of large lifetime hierarchies enables each of these limits to manifest from many UV theories. There are two broadly-studied cases:
\begin{itemize}

    \item Semivisible jets: the dark shower produces some prompt and some collider-stable dark hadrons \cite{Cohen:2015toa}. The characteristic feature is missing momentum aligned with a jet.

    \item Emerging jets: the dark shower dominantly produces displaced dark hadrons \cite{Schwaller:2015gea}. The characteristic feature is several DVs within a jet cone.
    
\end{itemize}
There is a continuum between these limits, and thus a combination of insights may be useful \cite{Carrasco:2025bct}, but we discuss each separately below. 

\subsubsection{Semivisible Jets}

\noindent Between events, different dark showers will produce different multiplicities of dark hadrons that are collider-stable or prompt. The quantity ``$r$-invisible''
\begin{equation}
    r_{\text{inv}} \equiv \left\langle
    \frac{\text{stable multiplicity}}{\text{total multiplicity}}
    \right\rangle
\end{equation}
describes the average proportion of collider-stable dark hadrons within a semivisible jet \cite{Cohen:2015toa}, which determines how a typical semivisible jet's transverse momentum $p_T$ is distributed between visible $p_T$ and contributions to MET\@. From a UV perspective, $r_{\text{inv}}$ is a rough proxy for the fraction of the theory's light dark hadron species (assuming approximate mass-degeneracy) that have lifetimes beyond the collider scale. Fluctuations in stable vs.~unstable dark hadrons from jet to jet can cause an asymmetry in the contributions to missing momentum from jets within the same event. The measured total missing transverse momentum vector $\vec{\cancel{p}}_{T}$ can thus have a large magnitude and be somewhat closely aligned with one jet. In traditional searches for multijet+MET signals targeting, e.g., partially-invisible decays of SUSY-like states, events with MET aligned with the jet are assumed to originate from jet energy mis-measurement \cite{CMS:2014tzs,ATLAS:2014jxt}, so a semivisible jet signal would be cut from the analysis. 

The portal used in several studies, e.g.~Refs.~\cite{Cohen:2015toa,Bernreuther:2019pfb,Cohen:2020afv,Kar:2020bws,Liu:2024rbe,Buckley:2025hty} and some searches \cite{CMS:2021dzg,ATLAS:2025kuz}, is a leptophobic $s$-channel $Z^\prime$ mediator,\footnote{Realizing this scenario in a UV theory is not necessarily straightforward. One must take care to cancel gauge anomalies \cite{FileviezPerez:2010gw,FileviezPerez:2011pt,Duerr:2013dza}, or the SM fermions may acquire effective couplings to the $Z^\prime$ through higher-dimensional operators after integrating out heavier states \cite{Fox:2011qd}, and the necessary addition of heavy states with SM charge may come with its own constraints in either case. The $Z^\prime$ may also need to preferentially decay to the dark sector to evade QCD dijet searches for heavy resonances \cite{ATLAS:2019fgd,CMS:2019gwf}.} as shown in \cref{fig:qqVp}. The flavor-diagonal mesons (analogs of the SM $\pi^0$ and $\rho^0$) are neutral under the new U(1) and decay by vector mixing as in \cref{fig:VVmix} (the $\rhoD\to\piD\piD$ decay is often taken to be kinematically forbidden) or pseudoscalar-vector mixing as in \cref{fig:PSVmix}. The flavor-off-diagonal mesons (analogs of the SM $\pi^\pm$ and $\rho^\pm$) are charged under the new U(1) and taken to be stable (though this story can change depending on how the U(1) and dark flavor symmetries are broken). The flavor-diagonal $\piD$ decay (which prefers heavy fermionic decay products due to helicity suppression) is often suppressed compared to that of the $\rhoD$, so it is sometimes taken to be collider-stable. In this case of resonant $\Zp$ production, a powerful discriminating observable is the transverse mass $M_T$, defined via 
\begin{equation}
    M_T^2=M_{jj}^2 + 2\left(|\vec{\cancel{p}}_T|\sqrt{M_{jj}^2+|\vec{p}_{Tjj}|^2}\,
    -
    \vec{\cancel{p}}_T\cdot \vec{p}_{Tjj}
    \right),
\end{equation}
where $M_{jj}$ and $\vec{p}_{Tjj}$ are the total invariant mass and transverse momentum of the two hardest jets, respectively. The distribution of $M_T$ is closer to that of the true total invariant mass of the dark hadrons than the  $M_{jj}$ distribution itself \cite{Cohen:2015toa}.

Other portals generating semivisible jets have also been considered. Some studies \cite{Cohen:2017pzm,Mies:2020mzw,Kar:2020bws,Kar:2022hxn,Buckley:2022zry,Carmona:2024tkg} and an ATLAS search \cite{ATLAS:2023swa} have analyzed the $t$-channel portal shown in \cref{fig:tqq}, which has the additional challenge compared to the $s$-channel portal of lacking a resonant $M_T$ distribution. Ref.~\cite{Knapen:2021eip} classified several portals in terms of effective operators coupling the dark sector to the SM Higgs or gauge bosons, including mixing between the photon and a dark photon that can lead to visible decays of the $\piD$ from a chiral anomaly as shown in \cref{fig:piVV}. Ref.~\cite{Knapen:2021eip} also emphasized that there are trade-offs between lightness, multiplicity, and promptness: as the dark hadrons are taken lighter, they are produced in greater quantities but tend to become more displaced as the phase space for their decays decreases and more hadronic decay channels become closed. 
ATLAS \cite{ATLAS:2012wib} and preliminary CMS \cite{CMS:2025caz} searches have exploited the fact that portals that do not lead to dark hadrons decaying exclusively to SM hadrons will cause the semivisible jets to be enriched with leptons (though they may be too soft to tag efficiently) \cite{Han:2007ae,Cazzaniga:2022hxl,Beauchesne:2022phk,CMS:2025caz}. Other portals, such as the $\piD$ mixing with an ALP, may lead to photon-enriched jets \cite{Knapen:2021eip,Cazzaniga:2024mmv}. A Higgs portal is an attractive option \cite{Strassler:2006qa,Craig:2015pha,Pierce:2017taw,Cheng:2021kjg,Batz:2023zef}, in part due to modest constraints on the Higgs-to-invisible branching ratio of $\sim\!11\%$ \cite{ATLAS:2023tkt}, though jets coming from Higgs decays tend to be too soft to pass $p_T>\mathcal{O}(100)\,$GeV trigger thresholds used in existing searches \cite{CMS:2021dzg,ATLAS:2025kuz} unless the Higgs recoils off of hard initial state radiation.

\subsubsection{Emerging Jets} 

The prototypical emerging jet looks like collimated SM hadrons appearing inside the tracker with tracks that do not originate from the primary collision vertex \cite{Schwaller:2015gea}. Traditional searches for displaced dijets or DVs assume that a significant portion of the final state's momentum comes from one individual DV, and thus jets would be cut from the analysis if they emerge from an ensemble of DVs that are each relatively soft \cite{CMS:2014wda,LHCb:2014jgs,ATLAS:2014fzk,ATLAS:2015mlf}. Emerging jets are nonetheless a spectacular signal and are the target of multiple ATLAS \cite{ATLAS:2025kuz,ATLAS:2025bsz,ATLAS:2025lfx} and CMS \cite{CMS:2018bvr,CMS:2024gxp} searches. Starting in Run 3 of the LHC, ATLAS even has a dedicated emerging jets trigger using the fraction of a jet's momentum coming from displaced vs.~prompt tracks \cite{ATLAS:2024xna}. 

A commonly-studied emerging jet portal is the $t$-channel scalar shown in \cref{fig:tqq} \cite{Schwaller:2015gea,Renner:2018fhh,Mies:2020mzw,Carrasco:2023loy}. In addition to the $t$-channel production mechanism, the scalar itself may be pair-produced, in which case each produced scalar decays to one SM QCD jet and one emerging jet. Other studies have used a portal where both production and decay are mediated by a $Z^\prime$ that mixes with U(1)$_Y$, and the $\rhoD$ decays as shown in \cref{fig:VVmix} \cite{Bernreuther:2022jlj,Cheng:2024hvq,Bernreuther:2025xqk,Liu:2025bbc}. Combinations of vector boson mixing and Higgs portals have also been considered \cite{Born:2023vll,Cheng:2024aco}. These particular portals highlight the strengths of LHCb, $e^+e^-$ colliders, and beam-dump experiments to probe light displaced resonances. In the limit where the dark hadron lifetimes exceed the tracker scale, proposed dedicated LLP experiments such as AL3X, ANUBIS, and MATHUSLA gain sensitivity \cite{Archer-Smith:2021ntx,Liebersbach:2024kzc}.

\subsubsection{Challenges of Model Dependence} 

There is no one-to-one bottom-up map between dark shower signals and UV theories. One must make a choice of specific models to simulate in Monte Carlo generators to derive constraints, but the mapping between constraints on different UV models that give rise to the same qualitative signature is highly non-trivial. Dark showers are thus incompatible with the usual Simplified Model framework used in SUSY and WIMP contexts \cite{LHCNewPhysicsWorkingGroup:2011mji, Abdallah:2014hon, Abdallah:2015ter}, where observables and Lagrangian parameters have a sharper correspondence \cite{Cohen:2017pzm} and reinterpreting constraints between models with similar signals is relatively straightforward. Likewise, the complicated IR dynamics of dark showers impedes analysis within the SM Effective Field Theory framework \cite{Grzadkowski:2010es,Pomarol:2013zra,Falkowski:2015fla}, where UV physics is integrated out, and its imprint on the IR can be mapped onto coefficients of non-renormalizable operators. 

This issue motivated the ``simplified parametrization'' described in Ref.~\cite{Cohen:2017pzm}, where one specifies a minimal number of parameters (such as $\LD$, the mass scale of the dark hadrons, the mass scale of the portal, and $r_{\text{inv}}$) and designs an inclusive search program to target any given UV theory that maps onto those parameters. However, the true minimal set of parameters is unclear. For example, $\Nc$ and $\Nf$ can have a strong impact on the multiplicity of hadrons produced by a shower \cite{Mueller:1982cq}, which limits the reliability of reinterpretations of searches that fix these parameters. 

Monte Carlo modeling is another concern. The Hidden Valley Module \cite{Carloni:2010tw,Carloni:2011kk} of \textsc{Pythia} \cite{Bierlich:2022pfr} has enabled dark shower event generation based on the technology of \textsc{Pythia}'s SM perturbative shower and hadronization algorithms. Recently, similar capabilities have been added to \textsf{Herwig} \cite{Bahr:2008pv,Bellm:2015jjp} as well \cite{Kulkarni:2024okx}. While these tools are enormously useful for performing studies, they have many phenomenological parameters that cannot be tuned to match data, unlike for SM QCD. Hadronization in particular is a non-perturbative process that cannot be computed from first principles with any technique currently in use. An example of why this is important is the Hidden Valley Module setting \texttt{HiddenValley:probVector}, which determines relative production rates of $\rhoD$ and $\piD$ species. It is set to 0.75 by default, reflecting a three-to-one spin degree-of-freedom enhancement for the $\rhoD$, but we know this value is incorrect for the SM, where $\rho$ production is mass-suppressed compared to the $\pi$ (a generic feature of the light-quark limit). The $\rhoD$ and $\piD$ often have different decays and lifetimes, so their relative production rates can impact signals such as MET.

These challenges lead to difficulties in determining appropriate observables and analysis techniques for characterizing and searching for dark showers. For example, the fact that the showers can produce invisible states as well as states that decay to multiple SM hadrons is suggestive that jet substructure would be a powerful probe \cite{Cohen:2020afv}. Infrared and collinear (IRC) safe substructure observables have the advantage that they are perturbatively calculable with corrections suppressed by powers of $\LD^2/Q^2$ (although these power corrections are sometimes unintuitively large \cite{Manohar:1994kq,Bright-Thonney:2023gdl}). However, important aspects of dark showers (such as relative production rates of stable vs.~unstable dark hadron species) are inherently IRC unsafe, which obscures the robustness of this approach. Another proposed tool to mitigate systematic uncertainties is the Lund Jet Plane \cite{Dreyer:2018nbf}, which aids in separating jet phase space into regions that are more or less sensitive to IR effects \cite{Cohen:2023mya}. See Ref.~\cite{Li:2026jop} for discussion of finite mass and gauge group effects on the Lund Jet Plane. 

A multitude of machine learning (ML) strategies have been suggested or applied to dark showers \cite{Heimel:2018mkt,Dreyer:2020brq,Lim:2020igi,Bernreuther:2020vhm,Canelli:2021aps,Barron:2021btf,Beauchesne:2021qrw,Finke:2022lsu,Faucett:2022zie,Anzalone:2023ugq,Bai:2023yyy,Pedro:2023sdp,Favaro:2023xdl,Lu:2023gjk,Chhibra:2023tyf,Bhardwaj:2024djv,CMS:2025lmn,Cazzaniga:2025piw} (a description of which is beyond the scope of this work) with varying degrees of interpretability and IRC safety of inputs. ML is a powerful tool for optimally exploiting observables, but the robustness of some architectures against variations in model parameters and non-perturbative systematic uncertainties is unknown. Since highly distinct UV models can map onto qualitatively similar signals with under-explored variations in the IR, the black-box nature of some ML implementations impedes the reinterpretability of experimental constraints. As further discussed in \cref{sec:outlook}, these challenges can be seen as exciting opportunities.

\subsection{Exotic Signatures}
\label{sec:exoticCollider}

Some regimes of confining dark sectors can produce collider signals unlike anything that SM hard processes can achieve. For example, there is a particular limit of the dark showers regime where the dark sector has a large 't Hooft coupling ($4\pi\aD\Nc\gg1$) and is quasi-conformal (i.e.~there are sufficiently many flavors that the coupling runs very slowly over a large energy range). In this case, wide-angle parton emissions carry an $\mathcal{O}(1)$ fraction of the dark shower's energy. When the dark hadrons also decay promptly, this leads to a high-multiplicity, quasi-isotropic, and soft ($\mathcal{O}(100)\,$MeV) set of SM decay products known as a Soft Unclustered Energy Pattern (SUEP). This signal was originally dubbed a ``soft bomb'' \cite{Knapen:2016hky} and is a particular manifestation of the ``unparticle'' scenario \cite{Georgi:2007ek,Strassler:2008bv}. SUEPs can arise from, for example, certain models with extra dimensions \cite{Brax:2019koq,Costantino:2020msc}.

SUEP production can be simulated by assuming the dark hadron phase space is approximately a thermal distribution with a ``temperature'' near the confinement scale \cite{suep_generator}.\footnote{The quasi-conformal shower itself should not be simulated using current implementations of \textsc{Pythia}'s parton shower algorithm due to errors introduced by approximations of the $\beta$ function \cite{Kulkarni:2025rsl}.}
Triggering on SUEPs is characteristically challenging because they resemble pileup contamination, but various strategies have been proposed, including MET, low-$p_T$ track multiplicity, and exotic ring-shaped density patterns of tracker hits dubbed ``belts of fire'' \cite{Knapen:2016hky,DiPetrillo:2022qsb}. Measures of event isotropy have also been suggested as discriminating observables \cite{Cesarotti:2020hwb,Cesarotti:2020uod,Cesarotti:2020ngq}. SUEPs are the target of a recent CMS search \cite{CMS:2024nca}, an upcoming ATLAS search \cite{Lory:2022upc}, and a preliminary CMS search \cite{CMS:2026adz} that employed data scouting.

\begin{figure}[t]
    \centering
    \includegraphics[width=0.6\linewidth]{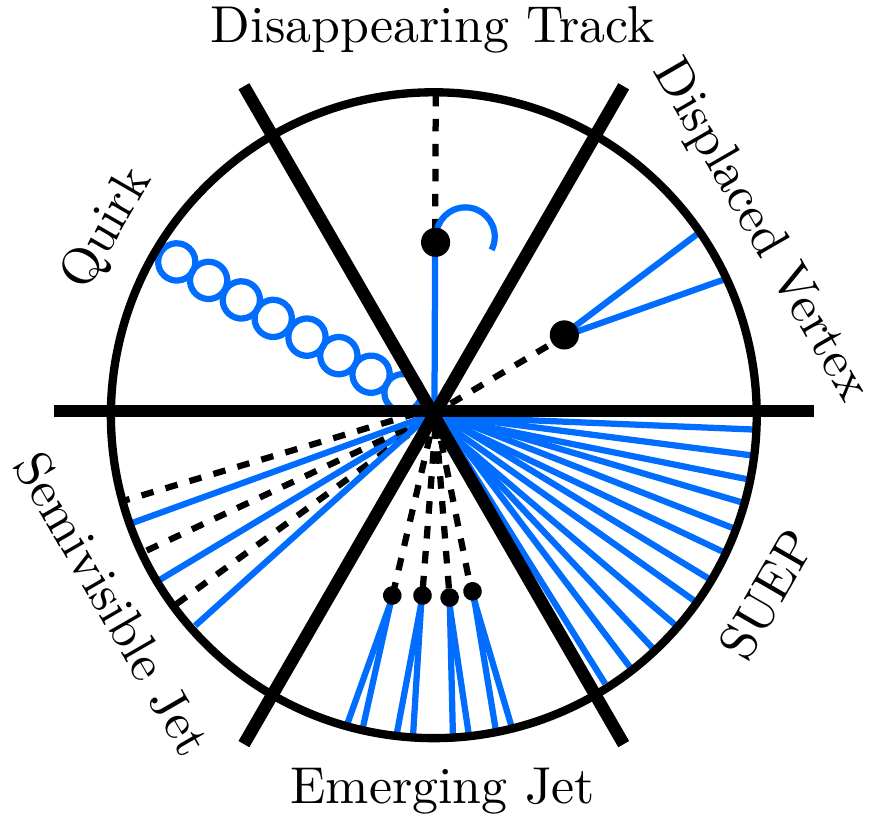}
    \caption{Sketches of how various exotic/long-lived particle collider signatures would look inside the detector, similar to those in Refs.~\cite{Alimena:2019zri,Darwish:2023jix}. Solid blue lines denote visible particles (which in some cases must be charged when tracking is essential), dashed black lines denote invisible particles, and dots denote decay vertices.}
    \label{fig:colliderSignals}
\end{figure}

Another signal topology that is not realized in the SM is the ``quirk'' scenario \cite{Kang:2008ea}, which arises when $\LD \ll \mq \lesssim Q/2$. Specific UV realizations have been discussed in the contexts of folded SUSY \cite{Burdman:2008ek,Cheng:2018gvu}, little Higgs \cite{Cai:2008au}, composite DM \cite{Kribs:2009fy}, baryogenesis \cite{Craig:2010au}, gauge unification \cite{Martin:2010kk}, and dark quarks in various electroweak representations \cite{Fok:2011yc}. In a collider, a $q \bar{q}$ pair is produced that is joined by a dark color string, and breaking that string by producing another $q \bar{q}$ pair from the vacuum is exponentially suppressed. 
The separation between $q$ and $\bar{q}$ oscillates, and they cross each other some number of times before annihilating. If the quark-anti-quark pair does not annihilate promptly, their trajectory in the lab frame looks like two bodies whirling around and oscillating about their center of mass. A qualitative picture of this, as well as a summary of other signals described in this section, is illustrated in \cref{fig:colliderSignals}.
There is a broad range of possible phenomenology (a thorough description of which is beyond the scope of this work) depending on the dark quark charge assignments and the length of the string, which scales as $\mq/\LD^2$. Modeling quirk dynamics is a challenge. There have been efforts to model energy loss due to radiation \cite{Harnik:2011mv} or analytically solving the quirk equation of motion \cite{Li:2020aoq}, but certain aspects such as de-excitation via dark glueball radiation are non-perturbative and thus come with non-trivial uncertainties.

Various quirk search strategies have been proposed, including using anomalies in the underlying event \cite{Harnik:2008ax}, emissions of electroweak bosons and QCD jets \cite{Harnik:2011mv}, coplanar non-helical tracks \cite{Farina:2017cts,Knapen:2017kly,Sha:2024hzq,Condren:2025czc,CidVidal:2026ist}, monojets \cite{Farina:2017cts}, ionization energy loss \cite{Evans:2018jmd,Li:2019wce}, resonances decaying to di-Higgs \cite{Barela:2023exp}, far-displaced signals in MATHUSLA \cite{Li:2023jrt}, decays of radiated dark glueballs \cite{Forsyth:2025wks}, and SUEP-like isotropic emissions of SM pions \cite{Curtin:2025ngf}. 
Quirks were targeted in one Tevatron search \cite{D0:2010kkd}, but there are no dedicated LHC searches. 
Given the distinctive nature of the signal, its generic appearance in certain limits of confining theories, and the fact that macroscopic quirk discovery would imply a stringent upper bound on the reheating temperature of the early Universe \cite{Asadi:2025btr}, developing dedicated LHC searches for quirks is well-motivated.

\section{Calculating at Strong Coupling}
\label{sec:calc_tools}


One of the major difficulties in theories involving a strongly-coupled sector, whether or not it contains a DM candidate, is the calculation of the properties of the resulting strongly-coupled states. This is a well-known challenge in QCD, and many of the tools and techniques developed to study QCD mesons and baryons have also been applied to strongly-coupled dark sectors. In this section, we briefly review the techniques that have been most commonly used in this context.

\subsection{Chiral Lagrangian for Dark Mesons}

As the strongly-coupled theory confines, the proper EFT changes from a theory of quarks and gluons to one of hadrons. 
In theories with $N_f$ flavors of fermions transforming as fundamentals under the dark group, SU($N$), a chiral Lagrangian can be constructed when
$\mq \ll \LD$.
Just like with QCD, the global flavor symmetry
is spontaneously broken to the diagaonal
\begin{eqnarray}
    \mathrm{U}(N_f)_L \times \mathrm{U}(N_f)_R  &\rightarrow&  \mathrm{U}(N_f)_V
\end{eqnarray}
with $N_f^2 -1$ pseudo Nambu-Goldstone bosons (pNGBs) $\pi^a$---the pions---and one $\eta'$ that acquires mass from the U(1)$_A$ anomaly.
The chiral Lagrangian is constructed analogously to QCD, utilizing a non-linear sigma model with 
\begin{equation}
\Sigma(x) = \exp\left[ \frac{2 i}{f_\pi} \pi^a T^a \right],
\end{equation}
where the $T^a \in $~$\mathfrak{su}$($N_f$)
are the broken generators and $f_\pi$ is dark pion decay constant. 
The EFT
\begin{equation} \label{eq:chiralLagrangian}
    \mathcal{L} = \frac{f_\pi^2}{4}\rm{Tr}(\partial_\mu \Sigma^\dagger\partial^\mu \Sigma)
\end{equation}
then allows us to make perturbative predictions involving the $\pi^a$ fields up to a scale of $\sim\!4\pi f_\pi$.\footnote{The interactions of vector mesons analogous to the $\rho$ can also be described in the formalism of Hidden Local Symmetry \cite{Bando:1987br,Harada:2003jx}.} See Ref.~\cite{Scherer:2002tk} for a pedagogical review.

The paradigm of vector-like confinement \cite{Kilic:2009mi,Kilic:2010et} emphasized the utility of the chiral Lagrangian for characterizing the spectrum and properties of dark pions.  Dark pions that carry SM charges are 
straightforwardly handled by promoting the partial derivatives in \cref{eq:chiralLagrangian} to covariant derivatives, just as occurs for the charged pions of QCD that transform under electromagnetism.  The spectrum, couplings, 
and phenomenology
of dark pions has been
studied in a variety of different theories
(see \cref{subsec:model_meson} for details).

\subsection{Heavy quarks and NREFT}

In theories with heavy quarks, $m_f \gg \LD$, the tools of non-relativistic effective field theory (NREFT) can be utilized to calculate  properties of the heavy composite states, see Refs.~\cite{Brambilla:1999xf,Petrov:2016azi} for reviews of these techniques.
In the meson sector, heavy quarkonium states were considered long ago \cite{Barger:1987xg}, 
with calculations that focused on dark sector ``quirkonium'' states with fermions in vector-like \cite{Cheung:2008ke} 
and chiral or mixed representations \cite{Fok:2011yc}.
Dark quarkonium formation in the early universe was further clarified in 
Refs.~\cite{Alves:2009nf,SpierMoreiraAlves:2010err,Geller:2018biy},
which showed that the rearrangement process of a heavy-light meson interactions has a cross section of geometric size, with no heavy quark mass suppression.

Considerable work has explored the baryonic sectors of confining SU($N$)
dark sector theories.  
While the dark ``baryonic'' states in a confining SU(2) are comparatively  simple, just two valence quark bound states, the heavy quark limit utilized in Ref.~\cite{Kribs:2009fy}
allowed for the calculation of the spectrum and splittings among the lightest states.  By examining the interactions of the baryonic states with matter, Ref.~\cite{Kribs:2009fy} showed that the leading interactions of ``quirky DM'' were suppressed compared to na\"ive expectations based solely on power counting of the EFT\@.

More extensive discussions of dark baryonic DM in the heavy quark regime appeared in Ref.~\cite{Mitridate:2017oky}, where a variational method was utilized to obtain the spatial wavefunctions and thus the spectra of the theory.  
In the context of dark quarks transforming in arbitrary irreducible representations of SU(2)$_L$ of the SM, Ref.~\cite{Asadi:2024tpu} described how to find the SU(2)$_L$ representations of the dark baryons, as well as computing the baryon mass spectrum in the non-relativistic quark limit. The dark baryons are commonly admixtures of states in trivial and non-trivial electroweak representations, suggesting that composite DM with heavy constituents is generically a superposition of electrically neutral states.
Systematic application of pNRQCD for generic heavy hadrons in SU($N$) confining dark sectors was carried out in Ref.~\cite{Assi:2023cfo}. 

Finally, a pionless EFT \cite{Bethe:1949yr,Bedaque:2002mn} has been utilized to enable the calculations of the possible binding of dark baryons into dark nuclei 
\cite{Redi:2018muu}.  It was shown that dark deuterium could form when there is another field, such as a photon, to enable the dark baryon fusion reaction.  The main results focused on asymmetric dark matter, though the basic reactions may also be able to occur if there is a symmetric abundance of dark baryons and anti-baryons.  Signals of dark deuterium were considered in subsequent work
\cite{Mahbubani:2019pij,Mahbubani:2020knq}.

\subsection{Lattice Simulations of Dark Sectors}

In the regime where a confined SU($N$) dark sector has fermions with masses $\mq \sim \LD$, non-perturbative methods are often necessary to make progress.  While some insight can be gained in theories that closely match QCD, the fundamental difficulty is that QCD contains both light, intermediate, and heavy quarks that complicates a clean extrapolation.  In several cases, one can ``re-purpose'' existing lattice studies and apply the results to confining dark sectors \cite{DeGrand:2019vbx}. 
In particular, existing lattice studies of the spectrum and form factors for glueballs in SU(3) \cite{Morningstar:1997ff,Morningstar:1999rf,Chen:2005mg,Athenodorou:2020ani}
and glueballs in SU($N$) theories 
\cite{Gregory:2012hu,Athenodorou:2021qvs}
have been extensively used 
(see 
\cref{subsec:model_glueball}
for details).
In this section, we summarize a variety of work that has employed lattice simulations for specific strongly-coupled dark sectors.

The Lattice Strong Dynamics (LSD) collaboration has carried out a sequence of first-principles lattice studies of dark baryon dark matter. In an initial SU(3) study, they simulated QCD-like theories with two and six degenerate fundamental fermions and computed electromagnetic form factors of electroweak-neutral dark baryons, extracting the  charge radius and anomalous magnetic moment that determine one-photon mediated direct detection rates 
\cite{LatticeStrongDynamicsLSD:2013elk}.
LSD subsequently considered SU(4), where quenched Wilson-fermion simulations were used to calculate the bosonic baryon and meson spectrum, compare SU(3) and SU(4) spectroscopy, and determine the effective Higgs coupling of the scalar dark baryon from the fermion-mass dependence of its mass 
\cite{LSD:2014obp}. These lattice inputs were used in Higgs-portal phenomenology in a fully developed model,
``Stealth Dark Matter,'' that used 4 flavors of dark quarks with an SU(4) dark color group to calculate spin-independent direct detection through Higgs exchange 
\cite{Appelquist:2015yfa}. A companion background-field calculation computed the electromagnetic polarizability of the lightest SU(3) and SU(4) baryons and translated it into a calculation of
the leading electromagnetic moment in the theory (polarizability) and used this to estimate the cross section for the irreducible direct detection signal \cite{Appelquist:2015zfa}. More recent finite-temperature SU(4) simulations studied the confinement phase transition of  Stealth Dark Matter, using Polyakov-loop observables to determine how heavy the dark fermions must be for the transition to become first order, thereby identifying the regime in which the early-universe transition can potentially source a stochastic gravitational-wave background 
\cite{LatticeStrongDynamics:2020jwi}.
The SU(4) spectroscopy program was subsequently refined with Laplacian-Heaviside smearing and octahedral-group irreducible representations, reducing excited-state contamination in the single-baryon channel, improving low-lying meson and baryon spectra, presenting odd-parity baryons, and laying groundwork for future calculations of baryon-baryon self-interactions 
\cite{LatticeStrongDynamicsLSD:2023vtk}. Finally, in a one-flavor SU(4) theory, dynamical M\"obius domain-wall fermion simulations were used to reconstruct the Polyakov-loop effective potential, determine first-order finite-temperature thermodynamics, and calculate the resulting gravitational wave spectrum, with sea-quark effects found to reduce the interface tension and hence lower the gravitational wave amplitude below the reach of planned future searches
\cite{Ayyar:2026wht}.

For a theory with a confining SU(2) dark sector, 
Ref.~\cite{Lewis:2011zb}
considered a theory with 
two massless Dirac flavors.
The goal was to simultaneously provide a technicolor mechanism for electroweak symmetry breaking, while leaving the uneaten pNGBs as a complex scalar DM candidate.  
Lattice calculations 
were used to confirm the expected chiral symmetry-breaking pattern by determining the Nambu-Goldstone spectrum.
They  
were also used to 
compute the electromagnetic form factor and hence estimate the charge radius of the composite Goldstone DM candidate \cite{Hietanen:2013fya}. 
Interestingly, the form factor was found to be consistent with vector-meson dominance even though this arose in a two-color theory.

An SO(4) gauge theory was considered in 
Ref.~\cite{Hietanen:2012qd}
where lattice simulations with two Dirac Wilson fermions in the vector representation were carried out. The results included the phase diagram including identifying the bulk transition, and measured pseudoscalar and vector meson masses.  The resulting DM candidate was found to be an isotriplet composite state.  One-flavor theories involving SU(2) were considered in 
Refs.~\cite{Francis:2016bzf,Francis:2018xjd}, where the lattice studies of the hadron spectrum
found that baryon number symmetry is enlarged to SU(2)$_B$, with the lightest baryon identified as a vector boson, providing a composite realization of hidden-vector DM\@.  
An SU(2) gauge theory was also explored in 
Ref.~\cite{Detmold:2014qqa}, with 
lattice simulations 
carried out in Ref.~\cite{Detmold:2014kba}
focusing on nuclear spectroscopy in multi-baryon sectors. One key result was evidence that $J=1$ systems with baryon number $B=2,3$ are bound, along with mixed meson-baryon states.  

Further detailed studies of SU(2) lattice simulations
include
Ref.~\cite{Hietanen:2014xca},
that focused on composite Higgs sectors, and 
Refs.~\cite{Arthur:2016dir,Arthur:2016ozw} that built on the earlier analyses with lighter quark masses and a variety of other improvements, to then
extract the continuum spectrum, including the scalar and pseudoscalar singlets. Finite temperature studies of SU(2) were also considered in Ref.~\cite{Lee:2017uvl}.

Lattice simulations were also performed for an Sp(4) dark sector theory with two flavors, leading to a set of five pNGBs and providing a UV completion for SIMPs \cite{Zierler:2022uez}.  This work built on earlier lattice simulations of Sp(4) that focused on the spectrum, decay constants and scattering length \cite{Maas:2021gbf,Zierler:2022qfq}.
In Ref.~\cite{Dengler:2024maq},
lattice simulations were used to compute the dark-pion scattering phase shift to fit an effective-range expansion, obtaining velocity-weighted self-interaction cross sections and thus a lower limit on the dark-matter mass from astrophysical constraints.
Further studies of dark sector Sp(4) theories appeared in Refs.~\cite{Bennett:2024bhy,TELOS:2026mim}.

\subsection{Approximately Conformal Sectors}

Confining dark sectors with enough flavors can
flow to an IR fixed point, exhibiting approximate conformal invariance
characteristic of a
conformal field theory (CFT), see e.g.~Ref.~\cite{Ryttov:2007sr}.
If some of these flavors have mass, the conformal sector becomes a confining gauge theory below the scale of these heavy fields, leading to hidden sector phenomenology akin to hidden valley models 
\cite{Strassler:2008bv}. 
This opens the possibility of using the AdS/CFT correspondence to reformulate the calculation in a perturbative dual description.

In the context of a warped extra dimensions model, Ref.~\cite{Gherghetta:2010cq} used AdS/CFT to interpret the dark photon and DM as composite states of a new strong sector. 
Ref.~\cite{Bunk:2010gb}
used Randall-Sundrum geometry to make hidden gauge sectors more phenomenologically accessible.
Ref.~\cite{vonHarling:2012sz} explicitly coupled secluded DM to a confining hidden CFT, used the dual warped picture to generate the mediator scale naturally, and worked out how the CFT changes freeze-out, annihilation channels and other dark sector phenomenology.
In Ref.~\cite{Hong:2019nwd},
the dark sector was assumed to be a conformal sector when it was populated by freeze-in interactions.
This allowed for the use of CFT operator scaling involving interactions with genuinely large anomalous dimensions.  
A stable composite DM candidate appears only after interactions between the conformal dark sector with the SM induce the breaking of conformal symmetry.

In the gravitational production framework,
dark sector production rates are written in terms of stress-tensor correlators and the CFT central charge, which makes it possible to compute abundances without ever resolving the microscopic hadron spectrum \cite{Redi:2020ffc}. Thermalization, confinement, and the emergence of dark glueball or dark dilaton DM after the conformal era ends were also discussed in Ref.~\cite{Redi:2020ffc}.
Related holographic methods appear in warped or approximately AdS descriptions, where a nearly conformal dark/mediator sector is represented by a 5D dual, giving calculable access to spectra, couplings, and long-range interactions in regimes where a purely four-dimensional strong-coupling treatment would be difficult \cite{Brax:2019koq}. More recent work applies the same 5D logic to the breaking of conformality itself, using radion/dilaton dynamics to study whether the confining phase transition completes and how that transition affects the dark matter abundance \cite{Redi:2022myr}.

\subsection{Large \texorpdfstring{$N$}{N}}

Large $N$ is well known as a powerful technique to probe strongly-coupled theories, originating in the classic papers of~'t Hooft and Witten \cite{tHooft:1973alw,Witten:1979kh,Witten:1980sp,Witten:1983tx} (for reviews, see e.g.\ Refs~\cite{Manohar:1998xv,Lucini:2012gg}).
At large $N$, mesons are  weakly interacting states with decay constants $f \propto \sqrt{N}$, and thus amplitudes that scale as $1/N$, suppressing decay widths and meson scattering rates.
In addition, the $\eta'$ becomes a pNGB, as the chiral anomaly is suppressed by large $N$ \cite{Witten:1979vv,Veneziano:1979ec}.
Baryons are qualitatively different: they contain $N$ quarks, so their masses scale as $m_B \propto N \LD$, 
with a structure that is intrinsically many-body, and their couplings to mesons are generically of order $g_{B\pi\pi} \propto \sqrt{N}$. Large-$N$ baryons behave like heavy semiclassical solitonic objects rather than weakly interacting few-body bound states \cite{Witten:1979kh,Jenkins:1998wy}.

In dark sector theories, large $N$ is often used
to obtain the
parametric behavior of the sector's hadron phenomenology. In Ref.~\cite{Kilic:2009mi}, vectorlike confinement was formulated utilizing large-$N$ scaling to estimate the couplings and widths of vector resonances, the sizes of pseudo-Goldstone interactions, and the various DM and collider-stable states. Similar uses can be found in
broader model-building \cite{Antipin:2014qva,Antipin:2015xia}, where large-$N$ NDA was used to relate $m_\rho$, $f_\pi$, and baryon masses. 
In \cite{Morrison:2020yeg},
a one-flavor SU($N$) dark sector completed decoupled from the SM was shown to be
surprisingly predictive, with a dark $\eta'$ meson and an $N$-quark baryon that provided two calculable DM candidates.  A one-flavor SU($N$) dark sector was also considered in \cite{Fleming:2024flc}, where large $N$ was again used to get a handle on the $\eta'$ and baryon mass spectrum, where now the $\eta'$ could decay through higher-dimensional operators mediated by heavy flavors that transform under the SM\@.

Large $N$ has also been used  as a guide to the thermodynamics and relic abundance of heavy baryons in confining SU($N$) theories. Ref.~\cite{Mitridate:2017oky} showed that a two-stage cosmology is generic, with perturbative freeze-out followed by confinement and recombination into dark baryons.  
The large $N$ expectation that pure or nearly pure SU($N$) confinement is strongly first order was exploited as a calculational framework for ``squeezeout'' \cite{Asadi:2021pwo,Asadi:2021yml}, demonstrating that such theories can accommodate dark baryon masses far above the conventional thermal unitarity bound.
Further use of large $N$ appeared in Refs.~\cite{Gouttenoire:2023roe,Sanchez-Garitaonandia:2023zqz}
where the thermodynamic hierarchy between the deconfined plasma phase of gluons carries energy and entropy densities of order $O(N^2)$ compared with the confined glueball phase that carries only $O(1)$.
Dark baryons in confining SU($N$) were also speculated to collapse into Planckian black-hole relics that could constitute the DM, where large $N$ was critical to central mechanism \cite{Profumo:2025var}.

Finally, large $N$ has also been used for pure-glue (pure Yang-Mills) dark sectors. In Ref.~\cite{Soni:2016gzf}, dark glueball DM was already shown to have qualitatively new behavior in the large $N$ regime, including DM self-interactions, warm DM, and the possibility of the formation of compact-object.  In Ref.~\cite{McKeen:2024trt},
although the main focus is on cosmological evolution, the same glue-dominated regime is followed through confinement and glueball freeze-out to show how remnant dark glueballs can dominate or substantially modify the final relic abundance, see \cref{subsec:model_glueball}. 
Ref.~\cite{Yamada:2023thl} showed that moderately large $N$ in pure SO($2N$) theories produces a stable ``baryonic glueball'' DM candidate.  In this theory, large $N$ 
suppresses the reconnection probability of cosmic strings (interpreted as macroscopic color flux tubes) formed after confinement in the dark sector.

\subsection{Supersymmetry}

Supersymmetric gauge theories are well known to be calculable at strong coupling (for reviews, see e.g.~Refs.~\cite{Intriligator:1995au,Shadmi:1999jy,Intriligator:2007cp,Dine:2010cv}). 
Early work that considered DM candidates in the hidden or messenger sectors included Refs.~\cite{Banks:2005hc,
Hamaguchi:2007rb,Hamaguchi:2008rv,Hamaguchi:2009db,Mardon:2009gw}.  
Deformed supersymmetric QCD theories with weakly-coupled magnetic duals have metastable SUSY-breaking vacua with pseudomoduli or pNGBs charged under hidden global symmetries can be stable composite DM candidates \cite{Fan:2010is}. 
One example that explicitly uses the Seiberg-Witten metholds \cite{DelZotto:2016fju}
calculated kinetic mixing between a visible U(1) and a strongly-coupled hidden sector with an approximate $\mathcal{N}=2$ supersymmetry.
In Ref.~\cite{Csaki:2022xmu}, a model of self-interacting DM based on the low energy effective theory of supersymmetric QCD
was shown, where the structure of the theory ensures a resonant enhancement of the self-interactions of the low energy mesons.
Ref.~\cite{Cacciapaglia:2023syp}
considered an exact $\mathcal{N}=1$ supersymmetric hidden sector as a new paradigm for (supersymmetric) dark sectors, with the feature that many properties of the theory can be fully computed using Veneziano-Yankielowicz-type effective descriptions \cite{Veneziano:1982ah} and lattice input.
Ref.~\cite{Csaki:2024plt} provided a nice example of using the $\mathcal{N}=2$
Seiberg-Witten SU(2) theory as a fully calculable strongly-coupled axion model with light monopoles (or dyons) and exact duality-invariant axion-photon amplitudes.

Recent work in the past few years introduced the idea of using supersymmetric QCD deformed by anomaly-mediated supersymmetry breaking (AMSB) \cite{Randall:1998uk,Giudice:1998xp} as a controlled proxy for ordinary QCD-like dynamics \cite{Murayama:2021xfj,Csaki:2022cyg}.
When the AMSB scale is small compared to the strong scale, holomorphy, duality, and supersymmetric strong dynamics make the theory calculable, while the non-supersymmetric spectrum has the same massless degrees of freedom expected in the QCD-like theory.
Ref.~\cite{Murayama:2021xfj} used this to argue for chiral symmetry breaking for $N_f < 3 N_c/2$ and conformal behavior for larger $N_f$.
A systematic guide toward AMSB QCD was given \cite{Csaki:2022cyg},  carefully analyzing chiral-symmetry-breaking vacua, baryonic directions, and possible runaways in SU($N_c$) with $N_f$ flavors, finding that loop effects stabilize many na\"ive baryonic runaways, while the $N_f = N_c$ quantum-modified case remains less calculable because of incalculable K\"ahler terms. 
Some applications include 
a confining Nelson-Barr-style solution to strong $CP$ \cite{Csaki:2025ikr}, where spontaneous $CP$ violation comes from a dark chiral condensate and the dark pions can serve as freeze-in DM candidates. 
Another application \cite{Gherghetta:2025kff}
uses supersymmetric chiral dynamics perturbed by AMSB to construct a high-quality composite QCD axion
using a calculable strongly coupled supersymmetric chiral gauge theory also deformed by AMSB.

\section{Paths Forward}
\label{sec:outlook}

Composite dark sectors provide well-motivated frameworks with rich phenomenology and a wide range of exotic signals across experiments. They also offer versatile extensions of the SM that can address several of its outstanding shortcomings. 
In this review, we discussed the motivations for these models, dark hadron DM candidates, embeddings of simplified DM abundance mechanisms in confining sectors, tools for calculation in the non-perturbative regime, and the novel early-universe behavior of such sectors. 
We also surveyed their phenomenology across collider searches, direct and indirect detection, and cosmological and astrophysical probes.

We close by highlighting some open questions that motivate future work. First, from the perspective of the early universe, the confining phase transition is a rich, under-explored, and unifying thread across many open questions. It can dramatically reshape the relic abundance of dark glueballs, baryons, mesons, and nuclei. In first order phase transitions, the bubble-wall velocity simultaneously influences both these abundance calculations and gravitational wave predictions, but is still not well-understood. 
Beyond the transition itself, uncertainties from non-perturbative processes make precise abundance calculations challenging across all these candidates, motivating systematic early-universe studies of confining dark sectors to further refine predictions. See \cref{sec:cosmo,subsec:model_glueball}.

There are some specific scenarios in which direct detection, indirect detection, and astrophysical phenomenology should be further developed:
\begin{itemize}
    \item The macroscopic quirk regime, where DM can be an extended object.

    \item DM candidates for which polarizability is the leading direct detection channel. Nuclear form factors are a particular challenge, see \cref{subsec:elastic_DD}.

    \item DM candidates that are a WIMP-like admixture of states in different electroweak representations. 
    \item Multi-component (e.g.~multiple stable dark nuclei) composite DM.

    \item Dissipative forces causing DM to exhibit substructure, see \cref{subsec:indirect}.

    \item Exotic DM annihilations involving emission of SM states, rearrangement into dark mesons, and dark glueball emission.

    \item Glueball DM with an ALP portal.
\end{itemize}
Correlations between these types of signals, early universe implications, and collider signatures are a ubiquitous feature of confining dark sectors. Further mapping out these connections would aid in maximizing information extracted from a future discovery.

From a collider perspective, with the ongoing LHC searches and High-Luminosity LHC on horizon, dark showers are a well-motivated and timely research direction. As discussed in \cref{sec:pheno}, further study is needed to understand systematic uncertainties associated with non-perturbative processes such as hadronization, optimal strategies and triggers to tag exotic signatures, and mappings between signal and theory space. These unsolved problems represent an exciting opportunity to develop experimental methods and phenomenology for the modern LHC era. 
Special care should be taken to explicitly discuss parameters used in both benchmark UV models and phenomenological models such as hadronization in the Hidden Valley module. When evaluating the performance of a candidate strategy, its robustness under variations of these parameters should be explored. UV benchmarks should be self-consistent, and if signals could arise that are targeted by conventional searches, e.g.~dilepton resonances, then the competing constraints should be considered. Reinterpretation efforts would benefit from model-inclusive search strategies and public availability of code frameworks that can be applied to simulated signal. Further investigation of the robustness of various ML architectures may permit optimal exploitation of observables while controlling systematic uncertainties. There are other open questions regarding simulation of exotic signatures such as quirks as well.

From the perspective of exploring the general space of theories, several avenues may lead to novel observations:
\begin{itemize}
    \item Constructing novel solutions to SM puzzles such as the hierarchy problem, the strong $CP$ problem, the Yukawa hierarchy, inflation, and baryogenesis. Novel solutions to the \textit{abundance similarity puzzle} (i.e.~the similarity of visible and dark matter energy densities) in particular may lead to qualitatively different phenomenology, e.g.~DM masses outside the usual $\mathcal{O}(10)\,$GeV mass window in existing solutions.

    \item Looking for new symmetries analogous to $H$- and $G$-parity that either stabilize dark hadrons or suppress detection signals. This may reveal other models that are naturally elusive to existing analyses but exhibit some alternative discovery mechanism.

    \item More investigations of theories with non-Abelian gauge groups beyond SU($N$) and matter representations beyond the fundamental.
    
\end{itemize}
Developing new calculational techniques, beyond those discussed in \cref{sec:calc_tools}, will aid in making predictions for these models, e.g.~those inspired by quantum information theory \cite{Asadi:2022vbl,Asadi:2023bat} or real-time lattice simulations on quantum computers \cite{Martinez:2016yna, Funcke:2023jbq}.

The breadth and depth of confining dark sectors, as well as their applicability to open questions across the field, strongly motivate their further exploration. An understanding of their general features reinforces one's ability to model-build effectively, drawing inspiration from different corners of the theory space and consistently analyzing phenomenology across different experiments. This review serves to emphasize these connections, aiming to present a unified compilation of a diverse literature that is nonetheless strongly coupled to itself. 

\section*{Acknowledgments}

We thank Prateek Agrawal, Reuven Balkin, David Curtin, Stefania Gori, Rashmish Mishra, and Hitoshi Murayama for illuminating discussions. 
We also thank Timothy Cohen and Simon Knapen for feedback on parts of the manuscript. 
We thank Ethan Neil for the invitation to write this review. 
The research of PA is supported in part by the U.S.
Department of Energy grant number DE-SC0010107.
The work of AB is supported by the U.S. Department of Energy, Office of Science Graduate Research (SCGSR) program, which is administered by the Oak Ridge Institute for Science and Education under contract number DE-SC0014664. The work of GK was supported in part by the U.S. Department of Energy under grant number DE-SC0011640.

\end{spacing}

\clearpage

\bibliography{ref}

@article{Randall:1998uk,
    author = "Randall, Lisa and Sundrum, Raman",
    title = "{Out of this world supersymmetry breaking}",
    eprint = "hep-th/9810155",
    archivePrefix = "arXiv",
    reportNumber = "MIT-CTP-2788, PUPT-1815, BUHEP-98-26",
    doi = "10.1016/S0550-3213(99)00359-4",
    journal = "Nucl. Phys. B",
    volume = "557",
    pages = "79--118",
    year = "1999"
}

@article{LatticeStrongDynamicsLSD:2023vtk,
    author = "Brower, R. C. and others",
    collaboration = "Lattice Strong Dynamics (LSD)",
    title = "{Stealth dark matter spectrum using Laplacian Heaviside smearing and irreducible representations}",
    eprint = "2312.07836",
    archivePrefix = "arXiv",
    primaryClass = "hep-lat",
    reportNumber = "FERMILAB-PUB-23-808-T, RIKEN-iTHEMS-Report-23, IPPP/23/71,
  LLNL-JRNL-858123",
    doi = "10.1103/PhysRevD.110.095001",
    journal = "Phys. Rev. D",
    volume = "110",
    number = "9",
    pages = "095001",
    year = "2024"
}

@article{Brambilla:1999xf,
    author = "Brambilla, Nora and Pineda, Antonio and Soto, Joan and Vairo, Antonio",
    title = "{Potential NRQCD: An Effective theory for heavy quarkonium}",
    eprint = "hep-ph/9907240",
    archivePrefix = "arXiv",
    reportNumber = "CERN-TH-99-199, HEPHY-PUB-716-99, UB-ECM-PF-99-06, UWTHPH-1999-34, UB-ECM-PF-99-13",
    doi = "10.1016/S0550-3213(99)00693-8",
    journal = "Nucl. Phys. B",
    volume = "566",
    pages = "275",
    year = "2000"
}

@book{Petrov:2016azi,
    author = "Petrov, Alexey A. and Blechman, Andrew E.",
    title = "{Effective Field Theories}",
    doi = "10.1142/8619",
    isbn = "978-981-4434-92-8, 978-981-4434-94-2",
    publisher = "WSP",
    year = "2016"
}

@article{Giudice:1998xp,
    author = "Giudice, Gian F. and Luty, Markus A. and Murayama, Hitoshi and Rattazzi, Riccardo",
    title = "{Gaugino mass without singlets}",
    eprint = "hep-ph/9810442",
    archivePrefix = "arXiv",
    reportNumber = "CERN-TH-98-337, LBNL-42419, LBL-42419, UCB-PTH-98-50, UMD-PP-99-037",
    doi = "10.1088/1126-6708/1998/12/027",
    journal = "JHEP",
    volume = "12",
    pages = "027",
    year = "1998"
}

@article{Kaplan:2009de,
    author = "Kaplan, David E. and Krnjaic, Gordan Z. and Rehermann, Keith R. and Wells, Christopher M.",
    title = "{Atomic Dark Matter}",
    eprint = "0909.0753",
    archivePrefix = "arXiv",
    primaryClass = "hep-ph",
    doi = "10.1088/1475-7516/2010/05/021",
    journal = "JCAP",
    volume = "05",
    pages = "021",
    year = "2010"
}

@article{Bernreuther:2019pfb,
    author = {Bernreuther, Elias and Kahlhoefer, Felix and Kr\"amer, Michael and Tunney, Patrick},
    title = "{Strongly interacting dark sectors in the early Universe and at the LHC through a simplified portal}",
    eprint = "1907.04346",
    archivePrefix = "arXiv",
    primaryClass = "hep-ph",
    reportNumber = "TTK-19-25, P3H-19-019",
    doi = "10.1007/JHEP01(2020)162",
    journal = "JHEP",
    volume = "01",
    pages = "162",
    year = "2020"
}

@article{Griest:1989wd,
    author = "Griest, Kim and Kamionkowski, Marc",
    title = "{Unitarity Limits on the Mass and Radius of Dark Matter Particles}",
    reportNumber = "CFPA-TH-89-013, FERMILAB-PUB-89-205-A",
    doi = "10.1103/PhysRevLett.64.615",
    journal = "Phys. Rev. Lett.",
    volume = "64",
    pages = "615",
    year = "1990"
}

@article{Smirnov:2019ngs,
    author = "Smirnov, Juri and Beacom, John F.",
    title = "{TeV-Scale Thermal WIMPs: Unitarity and its Consequences}",
    eprint = "1904.11503",
    archivePrefix = "arXiv",
    primaryClass = "hep-ph",
    doi = "10.1103/PhysRevD.100.043029",
    journal = "Phys. Rev. D",
    volume = "100",
    number = "4",
    pages = "043029",
    year = "2019"
}

@article{Witten:1979kh,
    author = "Witten, Edward",
    title = "{Baryons in the 1/n Expansion}",
    reportNumber = "HUTP-79-A007",
    doi = "10.1016/0550-3213(79)90232-3",
    journal = "Nucl. Phys. B",
    volume = "160",
    pages = "57--115",
    year = "1979"
}

@article{Aranda:2015jis,
    author = "Aranda, Alfredo and Barajas, Luis and Cembranos, Jose A. R.",
    title = "{Magnetic dipole moments for composite dark matter}",
    eprint = "1511.02805",
    archivePrefix = "arXiv",
    primaryClass = "hep-ph",
    doi = "10.1088/1475-7516/2016/03/034",
    journal = "JCAP",
    volume = "03",
    pages = "034",
    year = "2016"
}

@article{Cline:2016nab,
    author = "Cline, James M. and Huang, Weicong and Moore, Guy D.",
    title = "{Challenges for models with composite states}",
    eprint = "1607.07865",
    archivePrefix = "arXiv",
    primaryClass = "hep-ph",
    doi = "10.1103/PhysRevD.94.055029",
    journal = "Phys. Rev. D",
    volume = "94",
    number = "5",
    pages = "055029",
    year = "2016"
}

@article{Cline:2021itd,
    author = "Cline, James M.",
    title = "{Dark atoms and composite dark matter}",
    eprint = "2108.10314",
    archivePrefix = "arXiv",
    primaryClass = "hep-ph",
    doi = "10.21468/SciPostPhysLectNotes.52",
    journal = "SciPost Phys. Lect. Notes",
    volume = "52",
    pages = "1",
    year = "2022"
}

@article{Kavanagh:2018xeh,
    author = "Kavanagh, Bradley J. and Panci, Paolo and Ziegler, Robert",
    title = "{Faint Light from Dark Matter: Classifying and Constraining Dark Matter-Photon Effective Operators}",
    eprint = "1810.00033",
    archivePrefix = "arXiv",
    primaryClass = "hep-ph",
    reportNumber = "CERN-TH-2018-200",
    doi = "10.1007/JHEP04(2019)089",
    journal = "JHEP",
    volume = "04",
    pages = "089",
    year = "2019"
}

@article{Antipin:2014qva,
    author = "Antipin, Oleg and Redi, Michele and Strumia, Alessandro",
    title = "{Dynamical generation of the weak and Dark Matter scales from strong interactions}",
    eprint = "1410.1817",
    archivePrefix = "arXiv",
    primaryClass = "hep-ph",
    doi = "10.1007/JHEP01(2015)157",
    journal = "JHEP",
    volume = "01",
    pages = "157",
    year = "2015"
}

@article{Antipin:2015xia,
    author = "Antipin, Oleg and Redi, Michele and Strumia, Alessandro and Vigiani, Elena",
    title = "{Accidental Composite Dark Matter}",
    eprint = "1503.08749",
    archivePrefix = "arXiv",
    primaryClass = "hep-ph",
    reportNumber = "IFTP-TH-2015",
    doi = "10.1007/JHEP07(2015)039",
    journal = "JHEP",
    volume = "07",
    pages = "039",
    year = "2015"
}

@article{Mitridate:2017oky,
    author = "Mitridate, Andrea and Redi, Michele and Smirnov, Juri and Strumia, Alessandro",
    title = "{Dark Matter as a weakly coupled Dark Baryon}",
    eprint = "1707.05380",
    archivePrefix = "arXiv",
    primaryClass = "hep-ph",
    reportNumber = "CERN-TH-2017-151",
    doi = "10.1007/JHEP10(2017)210",
    journal = "JHEP",
    volume = "10",
    pages = "210",
    year = "2017"
}

@article{Sigurdson:2004zp,
    author = "Sigurdson, Kris and Doran, Michael and Kurylov, Andriy and Caldwell, Robert R. and Kamionkowski, Marc",
    title = "{Dark-matter electric and magnetic dipole moments}",
    eprint = "astro-ph/0406355",
    archivePrefix = "arXiv",
    doi = "10.1103/PhysRevD.70.083501",
    journal = "Phys. Rev. D",
    volume = "70",
    pages = "083501",
    year = "2004",
    note = "[Erratum: Phys.Rev.D 73, 089903 (2006)]"
}

@article{Barger:2010gv,
    author = "Barger, Vernon and Keung, Wai-Yee and Marfatia, Danny",
    title = "{Electromagnetic properties of dark matter: Dipole moments and charge form factor}",
    eprint = "1007.4345",
    archivePrefix = "arXiv",
    primaryClass = "hep-ph",
    doi = "10.1016/j.physletb.2010.12.008",
    journal = "Phys. Lett. B",
    volume = "696",
    pages = "74--78",
    year = "2011"
}

@article{Banks:2010eh,
    author = "Banks, Tom and Fortin, Jean-Francois and Thomas, Scott",
    title = "{Direct Detection of Dark Matter Electromagnetic Dipole Moments}",
    eprint = "1007.5515",
    archivePrefix = "arXiv",
    primaryClass = "hep-ph",
    reportNumber = "RUNHETC-2010-19, SCIPP-10-14, UCSD-PTH-10-06",
    month = "7",
    year = "2010"
}

@article{DelNobile:2012tx,
    author = "Del Nobile, Eugenio and Kouvaris, Chris and Panci, Paolo and Sannino, Francesco and Virkajarvi, Jussi",
    title = "{Light Magnetic Dark Matter in Direct Detection Searches}",
    eprint = "1203.6652",
    archivePrefix = "arXiv",
    primaryClass = "hep-ph",
    reportNumber = "CP3-ORIGINS-2012-007, DIAS-2012-8",
    doi = "10.1088/1475-7516/2012/08/010",
    journal = "JCAP",
    volume = "08",
    pages = "010",
    year = "2012"
}

@article{Weiner:2012cb,
    author = "Weiner, Neal and Yavin, Itay",
    title = "{How Dark Are Majorana WIMPs? Signals from MiDM and Rayleigh Dark Matter}",
    eprint = "1206.2910",
    archivePrefix = "arXiv",
    primaryClass = "hep-ph",
    doi = "10.1103/PhysRevD.86.075021",
    journal = "Phys. Rev. D",
    volume = "86",
    pages = "075021",
    year = "2012"
}

@article{Weiner:2012gm,
    author = "Weiner, Neal and Yavin, Itay",
    title = "{UV completions of magnetic inelastic and Rayleigh dark matter for the Fermi Line(s)}",
    eprint = "1209.1093",
    archivePrefix = "arXiv",
    primaryClass = "hep-ph",
    doi = "10.1103/PhysRevD.87.023523",
    journal = "Phys. Rev. D",
    volume = "87",
    number = "2",
    pages = "023523",
    year = "2013"
}

@article{Lisanti:2009am,
    author = "Lisanti, Mariangela and Wacker, Jay G.",
    title = "{Parity Violation in Composite Inelastic Dark Matter Models}",
    eprint = "0911.4483",
    archivePrefix = "arXiv",
    primaryClass = "hep-ph",
    reportNumber = "SLAC-PUB-13841",
    doi = "10.1103/PhysRevD.82.055023",
    journal = "Phys. Rev. D",
    volume = "82",
    pages = "055023",
    year = "2010"
}

@article{Alves:2009nf,
    author = "Alves, Daniele S. M. and Behbahani, Siavosh R. and Schuster, Philip and Wacker, Jay G.",
    title = "{Composite Inelastic Dark Matter}",
    eprint = "0903.3945",
    archivePrefix = "arXiv",
    primaryClass = "hep-ph",
    reportNumber = "SLAC-PUB-14773, SU-ITP-09-13",
    doi = "10.1016/j.physletb.2010.08.006",
    journal = "Phys. Lett. B",
    volume = "692",
    pages = "323--326",
    year = "2010"
}

@article{SpierMoreiraAlves:2010err,
    author = "Spier Moreira Alves, Daniele and Behbahani, Siavosh R. and Schuster, Philip and Wacker, Jay G.",
    title = "{The Cosmology of Composite Inelastic Dark Matter}",
    eprint = "1003.4729",
    archivePrefix = "arXiv",
    primaryClass = "hep-ph",
    reportNumber = "SLAC-PUB-14024, SLAC-PUB14024",
    doi = "10.1007/JHEP06(2010)113",
    journal = "JHEP",
    volume = "06",
    pages = "113",
    year = "2010"
}

@article{Chang:2010en,
    author = "Chang, Spencer and Weiner, Neal and Yavin, Itay",
    title = "{Magnetic Inelastic Dark Matter}",
    eprint = "1007.4200",
    archivePrefix = "arXiv",
    primaryClass = "hep-ph",
    doi = "10.1103/PhysRevD.82.125011",
    journal = "Phys. Rev. D",
    volume = "82",
    pages = "125011",
    year = "2010"
}

@article{Kumar:2011iy,
    author = "Kumar, Kunal and Menon, Arjun and Tait, Tim M. P.",
    title = "{Magnetic Fluffy Dark Matter}",
    eprint = "1111.2336",
    archivePrefix = "arXiv",
    primaryClass = "hep-ph",
    doi = "10.1007/JHEP02(2012)131",
    journal = "JHEP",
    volume = "02",
    pages = "131",
    year = "2012"
}

@article{Kribs:2009fy,
    author = "Kribs, Graham D. and Roy, Tuhin S. and Terning, John and Zurek, Kathryn M.",
    title = "{Quirky Composite Dark Matter}",
    eprint = "0909.2034",
    archivePrefix = "arXiv",
    primaryClass = "hep-ph",
    reportNumber = "FERMILAB-PUB-09-425-T",
    doi = "10.1103/PhysRevD.81.095001",
    journal = "Phys. Rev. D",
    volume = "81",
    pages = "095001",
    year = "2010"
}

@article{Buckley:2012ky,
    author = "Buckley, Matthew R. and Neil, Ethan T.",
    title = "{Thermal dark matter from a confining sector}",
    eprint = "1209.6054",
    archivePrefix = "arXiv",
    primaryClass = "hep-ph",
    reportNumber = "FERMILAB-PUB-12-533-A-T",
    doi = "10.1103/PhysRevD.87.043510",
    journal = "Phys. Rev. D",
    volume = "87",
    number = "4",
    pages = "043510",
    year = "2013"
}

@article{LatticeStrongDynamicsLSD:2013elk,
    author = "Appelquist, T. and others",
    collaboration = "Lattice Strong Dynamics (LSD)",
    title = "{Lattice Calculation of Composite Dark Matter Form Factors}",
    eprint = "1301.1693",
    archivePrefix = "arXiv",
    primaryClass = "hep-ph",
    reportNumber = "LLNL-JRNL-608695, NT-LBL-13-002, UCB-NPAT-13-002, FERMILAB-PUB-13-014-T",
    doi = "10.1103/PhysRevD.88.014502",
    journal = "Phys. Rev. D",
    volume = "88",
    number = "1",
    pages = "014502",
    year = "2013"
}

@article{Appelquist:2015yfa,
    author = "Appelquist, Thomas and others",
    title = "{Stealth Dark Matter: Dark scalar baryons through the Higgs portal}",
    eprint = "1503.04203",
    archivePrefix = "arXiv",
    primaryClass = "hep-ph",
    reportNumber = "INT-PUB-15-004, LLNL-JRNL-667446",
    doi = "10.1103/PhysRevD.92.075030",
    journal = "Phys. Rev. D",
    volume = "92",
    number = "7",
    pages = "075030",
    year = "2015"
}

@article{Appelquist:2015zfa,
    author = "Appelquist, Thomas and others",
    title = "{Detecting Stealth Dark Matter Directly through Electromagnetic Polarizability}",
    eprint = "1503.04205",
    archivePrefix = "arXiv",
    primaryClass = "hep-ph",
    reportNumber = "INT-PUB-15-005, LLNL-JRNL-667121",
    doi = "10.1103/PhysRevLett.115.171803",
    journal = "Phys. Rev. Lett.",
    volume = "115",
    number = "17",
    pages = "171803",
    year = "2015"
}

@article{Pospelov:2000bq,
    author = "Pospelov, Maxim and ter Veldhuis, Tonnis",
    title = "{Direct and indirect limits on the electromagnetic form-factors of WIMPs}",
    eprint = "hep-ph/0003010",
    archivePrefix = "arXiv",
    reportNumber = "TPI-MINN-00-11, UMN-TH-1845-00",
    doi = "10.1016/S0370-2693(00)00358-0",
    journal = "Phys. Lett. B",
    volume = "480",
    pages = "181--186",
    year = "2000"
}

@article{Ovanesyan:2014fha,
    author = "Ovanesyan, Grigory and Vecchi, Luca",
    title = "{Direct detection of dark matter polarizability}",
    eprint = "1410.0601",
    archivePrefix = "arXiv",
    primaryClass = "hep-ph",
    reportNumber = "ACFI-T14-19",
    doi = "10.1007/JHEP07(2015)128",
    journal = "JHEP",
    volume = "07",
    pages = "128",
    year = "2015"
}

@article{Asadi:2022vkc,
    author = "Asadi, Pouya and Kramer, Eric David and Kuflik, Eric and Slatyer, Tracy R. and Smirnov, Juri",
    title = "{Glueballs in a thermal squeezeout model}",
    eprint = "2203.15813",
    archivePrefix = "arXiv",
    primaryClass = "hep-ph",
    reportNumber = "MIT-CTP/5412",
    doi = "10.1007/JHEP07(2022)006",
    journal = "JHEP",
    volume = "07",
    pages = "006",
    year = "2022"
}

@article{Asadi:2021pwo,
    author = "Asadi, Pouya and Kramer, Eric David and Kuflik, Eric and Ridgway, Gregory W. and Slatyer, Tracy R. and Smirnov, Juri",
    title = "{Thermal squeezeout of dark matter}",
    eprint = "2103.09827",
    archivePrefix = "arXiv",
    primaryClass = "hep-ph",
    doi = "10.1103/PhysRevD.104.095013",
    journal = "Phys. Rev. D",
    volume = "104",
    number = "9",
    pages = "095013",
    year = "2021"
}

@article{Asadi:2021yml,
    author = "Asadi, Pouya and Kramer, Eric David and Kuflik, Eric and Ridgway, Gregory W. and Slatyer, Tracy R. and Smirnov, Juri",
    title = "{Accidentally Asymmetric Dark Matter}",
    eprint = "2103.09822",
    archivePrefix = "arXiv",
    primaryClass = "hep-ph",
    reportNumber = "MIT-CTP/5284",
    doi = "10.1103/PhysRevLett.127.211101",
    journal = "Phys. Rev. Lett.",
    volume = "127",
    number = "21",
    pages = "211101",
    year = "2021"
}

@article{Frigerio:2012uc,
    author = "Frigerio, Michele and Pomarol, Alex and Riva, Francesco and Urbano, Alfredo",
    title = "{Composite Scalar Dark Matter}",
    eprint = "1204.2808",
    archivePrefix = "arXiv",
    primaryClass = "hep-ph",
    doi = "10.1007/JHEP07(2012)015",
    journal = "JHEP",
    volume = "07",
    pages = "015",
    year = "2012"
}

@article{Bhattacharya:2013kma,
    author = "Bhattacharya, Subhaditya and Meli\'c, Bla\v{z}enka and Wudka, Jos\'e",
    title = "{Pionic Dark Matter}",
    eprint = "1307.2647",
    archivePrefix = "arXiv",
    primaryClass = "hep-ph",
    doi = "10.1007/JHEP02(2014)115",
    journal = "JHEP",
    volume = "02",
    pages = "115",
    year = "2014"
}

@article{Cline:2013zca,
    author = "Cline, James M. and Liu, Zuowei and Moore, Guy D. and Xue, Wei",
    title = "{Composite strongly interacting dark matter}",
    eprint = "1312.3325",
    archivePrefix = "arXiv",
    primaryClass = "hep-ph",
    doi = "10.1103/PhysRevD.90.015023",
    journal = "Phys. Rev. D",
    volume = "90",
    number = "1",
    pages = "015023",
    year = "2014"
}

@article{Boddy:2014yra,
    author = "Boddy, Kimberly K. and Feng, Jonathan L. and Kaplinghat, Manoj and Tait, Tim M. P.",
    title = "{Self-Interacting Dark Matter from a Non-Abelian Hidden Sector}",
    eprint = "1402.3629",
    archivePrefix = "arXiv",
    primaryClass = "hep-ph",
    reportNumber = "CALT-68-2860, UCI-TR-2013-18",
    doi = "10.1103/PhysRevD.89.115017",
    journal = "Phys. Rev. D",
    volume = "89",
    number = "11",
    pages = "115017",
    year = "2014"
}

@article{Gross:2018zha,
    author = "Gross, Christian and Mitridate, Andrea and Redi, Michele and Smirnov, Juri and Strumia, Alessandro",
    title = "{Cosmological Abundance of Colored Relics}",
    eprint = "1811.08418",
    archivePrefix = "arXiv",
    primaryClass = "hep-ph",
    doi = "10.1103/PhysRevD.99.016024",
    journal = "Phys. Rev. D",
    volume = "99",
    number = "1",
    pages = "016024",
    year = "2019"
}

@article{Dondi:2019olm,
    author = "Dondi, Nicola Andrea and Sannino, Francesco and Smirnov, Juri",
    title = "{Thermal history of composite dark matter}",
    eprint = "1905.08810",
    archivePrefix = "arXiv",
    primaryClass = "hep-ph",
    doi = "10.1103/PhysRevD.101.103010",
    journal = "Phys. Rev. D",
    volume = "101",
    number = "10",
    pages = "103010",
    year = "2020"
}

@article{Hambye:2009fg,
    author = "Hambye, Thomas and Tytgat, Michel H. G.",
    title = "{Confined hidden vector dark matter}",
    eprint = "0907.1007",
    archivePrefix = "arXiv",
    primaryClass = "hep-ph",
    reportNumber = "ULB-TH-09-20",
    doi = "10.1016/j.physletb.2009.11.050",
    journal = "Phys. Lett. B",
    volume = "683",
    pages = "39--41",
    year = "2010"
}

@article{Bai:2010qg,
    author = "Bai, Yang and Hill, Richard J.",
    title = "{Weakly Interacting Stable Pions}",
    eprint = "1005.0008",
    archivePrefix = "arXiv",
    primaryClass = "hep-ph",
    reportNumber = "FERMILAB-PUB-10-001-T, EFI-PREPRINT-10-9",
    doi = "10.1103/PhysRevD.82.111701",
    journal = "Phys. Rev. D",
    volume = "82",
    pages = "111701",
    year = "2010"
}

@article{Kilic:2009mi,
    author = "Kilic, Can and Okui, Takemichi and Sundrum, Raman",
    title = "{Vectorlike Confinement at the LHC}",
    eprint = "0906.0577",
    archivePrefix = "arXiv",
    primaryClass = "hep-ph",
    reportNumber = "UMD-PP-09-037",
    doi = "10.1007/JHEP02(2010)018",
    journal = "JHEP",
    volume = "02",
    pages = "018",
    year = "2010"
}

@article{Buen-Abad:2015ova,
    author = "Buen-Abad, Manuel A. and Marques-Tavares, Gustavo and Schmaltz, Martin",
    title = "{Non-Abelian dark matter and dark radiation}",
    eprint = "1505.03542",
    archivePrefix = "arXiv",
    primaryClass = "hep-ph",
    doi = "10.1103/PhysRevD.92.023531",
    journal = "Phys. Rev. D",
    volume = "92",
    number = "2",
    pages = "023531",
    year = "2015"
}

@article{Soni:2016gzf,
    author = "Soni, Amarjit and Zhang, Yue",
    title = "{Hidden SU(N) Glueball Dark Matter}",
    eprint = "1602.00714",
    archivePrefix = "arXiv",
    primaryClass = "hep-ph",
    reportNumber = "CALT-TH-2016-002",
    doi = "10.1103/PhysRevD.93.115025",
    journal = "Phys. Rev. D",
    volume = "93",
    number = "11",
    pages = "115025",
    year = "2016"
}

@article{Carmona:2015haa,
    author = "Carmona, Adrian and Chala, Mikael",
    title = "{Composite Dark Sectors}",
    eprint = "1504.00332",
    archivePrefix = "arXiv",
    primaryClass = "hep-ph",
    reportNumber = "DESY-15-047",
    doi = "10.1007/JHEP06(2015)105",
    journal = "JHEP",
    volume = "06",
    pages = "105",
    year = "2015"
}

@article{Cohen:2015toa,
    author = "Cohen, Timothy and Lisanti, Mariangela and Lou, Hou Keong",
    title = "{Semivisible Jets: Dark Matter Undercover at the LHC}",
    eprint = "1503.00009",
    archivePrefix = "arXiv",
    primaryClass = "hep-ph",
    doi = "10.1103/PhysRevLett.115.171804",
    journal = "Phys. Rev. Lett.",
    volume = "115",
    number = "17",
    pages = "171804",
    year = "2015"
}

@article{Schwaller:2015gea,
    author = "Schwaller, Pedro and Stolarski, Daniel and Weiler, Andreas",
    title = "{Emerging Jets}",
    eprint = "1502.05409",
    archivePrefix = "arXiv",
    primaryClass = "hep-ph",
    reportNumber = "CERN-PH-TH-2015-031, DESY-15-026",
    doi = "10.1007/JHEP05(2015)059",
    journal = "JHEP",
    volume = "05",
    pages = "059",
    year = "2015"
}

@article{LSD:2014obp,
    author = "Appelquist, T. and others",
    collaboration = "LSD",
    title = "{Composite bosonic baryon dark matter on the lattice: SU(4) baryon spectrum and the effective Higgs interaction}",
    eprint = "1402.6656",
    archivePrefix = "arXiv",
    primaryClass = "hep-lat",
    doi = "10.1103/PhysRevD.89.094508",
    journal = "Phys. Rev. D",
    volume = "89",
    number = "9",
    pages = "094508",
    year = "2014"
}

@article{Hardy:2015boa,
    author = "Hardy, Edward and Lasenby, Robert and March-Russell, John and West, Stephen M.",
    title = "{Signatures of Large Composite Dark Matter States}",
    eprint = "1504.05419",
    archivePrefix = "arXiv",
    primaryClass = "hep-ph",
    doi = "10.1007/JHEP07(2015)133",
    journal = "JHEP",
    volume = "07",
    pages = "133",
    year = "2015"
}

@article{Detmold:2014qqa,
    author = "Detmold, William and McCullough, Matthew and Pochinsky, Andrew",
    title = "{Dark Nuclei I: Cosmology and Indirect Detection}",
    eprint = "1406.2276",
    archivePrefix = "arXiv",
    primaryClass = "hep-ph",
    reportNumber = "MIT-CTP-4554",
    doi = "10.1103/PhysRevD.90.115013",
    journal = "Phys. Rev. D",
    volume = "90",
    number = "11",
    pages = "115013",
    year = "2014"
}

@article{Detmold:2014kba,
    author = "Detmold, William and McCullough, Matthew and Pochinsky, Andrew",
    title = "{Dark nuclei. II. Nuclear spectroscopy in two-color QCD}",
    eprint = "1406.4116",
    archivePrefix = "arXiv",
    primaryClass = "hep-lat",
    reportNumber = "MIT-CTP-4555",
    doi = "10.1103/PhysRevD.90.114506",
    journal = "Phys. Rev. D",
    volume = "90",
    number = "11",
    pages = "114506",
    year = "2014"
}

@article{Krnjaic:2014xza,
    author = "Krnjaic, Gordan and Sigurdson, Kris",
    title = "{Big Bang Darkleosynthesis}",
    eprint = "1406.1171",
    archivePrefix = "arXiv",
    primaryClass = "hep-ph",
    doi = "10.1016/j.physletb.2015.11.001",
    journal = "Phys. Lett. B",
    volume = "751",
    pages = "464--468",
    year = "2015"
}

@article{Asadi:2024tpu,
    author = "Asadi, Pouya and Batz, Austin and Kribs, Graham D.",
    title = "{Noble dark matter: Surprising elusiveness of dark baryons}",
    eprint = "2412.14240",
    archivePrefix = "arXiv",
    primaryClass = "hep-ph",
    doi = "10.1103/PhysRevD.111.095025",
    journal = "Phys. Rev. D",
    volume = "111",
    number = "9",
    pages = "095025",
    year = "2025"
}

@article{Juknevich:2009gg,
    author = "Juknevich, Jose E.",
    title = "{Pure-glue hidden valleys through the Higgs portal}",
    eprint = "0911.5616",
    archivePrefix = "arXiv",
    primaryClass = "hep-ph",
    reportNumber = "RUNHETC-2009-25",
    doi = "10.1007/JHEP08(2010)121",
    journal = "JHEP",
    volume = "08",
    pages = "121",
    year = "2010"
}

@article{Juknevich:2009ji,
    author = "Juknevich, Jose E. and Melnikov, Dmitry and Strassler, Matthew J.",
    title = "{A Pure-Glue Hidden Valley I. States and Decays}",
    eprint = "0903.0883",
    archivePrefix = "arXiv",
    primaryClass = "hep-ph",
    reportNumber = "RUNHETC-2008-18, TAUP-2890-08, ITEP-TH-45-08",
    doi = "10.1088/1126-6708/2009/07/055",
    journal = "JHEP",
    volume = "07",
    pages = "055",
    year = "2009"
}

@article{Bagnasco:1993st,
    author = "Bagnasco, John and Dine, Michael and Thomas, Scott D.",
    title = "{Detecting technibaryon dark matter}",
    eprint = "hep-ph/9310290",
    archivePrefix = "arXiv",
    reportNumber = "SCIPP-93-33",
    doi = "10.1016/0370-2693(94)90830-3",
    journal = "Phys. Lett. B",
    volume = "320",
    pages = "99--104",
    year = "1994"
}

@article{Harnik:2011mv,
    author = "Harnik, Roni and Kribs, Graham D. and Martin, Adam",
    title = "{Quirks at the Tevatron and Beyond}",
    eprint = "1106.2569",
    archivePrefix = "arXiv",
    primaryClass = "hep-ph",
    reportNumber = "FERMILAB-PUB-11-271-T",
    doi = "10.1103/PhysRevD.84.035029",
    journal = "Phys. Rev. D",
    volume = "84",
    pages = "035029",
    year = "2011"
}

@article{Sakharov:1967dj,
    author = "Sakharov, A. D.",
    title = "{Violation of CP Invariance, C asymmetry, and baryon asymmetry of the universe}",
    doi = "10.1070/PU1991v034n05ABEH002497",
    journal = "Pisma Zh. Eksp. Teor. Fiz.",
    volume = "5",
    pages = "32--35",
    year = "1967"
}

@article{Fok:2011yc,
    author = "Fok, R. and Kribs, Graham D.",
    title = "{Chiral Quirkonium Decays}",
    eprint = "1106.3101",
    archivePrefix = "arXiv",
    primaryClass = "hep-ph",
    reportNumber = "FERMILAB-PUB-11-275-T",
    doi = "10.1103/PhysRevD.84.035001",
    journal = "Phys. Rev. D",
    volume = "84",
    pages = "035001",
    year = "2011"
}

@article{Kribs:2018ilo,
    author = "Kribs, Graham D. and Martin, Adam and Ostdiek, Bryan and Tong, Tom",
    title = "{Dark Mesons at the LHC}",
    eprint = "1809.10184",
    archivePrefix = "arXiv",
    primaryClass = "hep-ph",
    doi = "10.1007/JHEP07(2019)133",
    journal = "JHEP",
    volume = "07",
    pages = "133",
    year = "2019"
}

@article{DeLuca:2018mzn,
    author = "De Luca, Valerio and Mitridate, Andrea and Redi, Michele and Smirnov, Juri and Strumia, Alessandro",
    title = "{Colored Dark Matter}",
    eprint = "1801.01135",
    archivePrefix = "arXiv",
    primaryClass = "hep-ph",
    reportNumber = "IFUP-TH-2017, CERN-TH-2017-283, IFUP-TH/2017",
    doi = "10.1103/PhysRevD.97.115024",
    journal = "Phys. Rev. D",
    volume = "97",
    number = "11",
    pages = "115024",
    year = "2018"
}

@article{Contino:2018crt,
    author = "Contino, Roberto and Mitridate, Andrea and Podo, Alessandro and Redi, Michele",
    title = "{Gluequark Dark Matter}",
    eprint = "1811.06975",
    archivePrefix = "arXiv",
    primaryClass = "hep-ph",
    doi = "10.1007/JHEP02(2019)187",
    journal = "JHEP",
    volume = "02",
    pages = "187",
    year = "2019"
}

@article{Contino:2020god,
    author = "Contino, Roberto and Podo, Alessandro and Revello, Filippo",
    title = "{Composite Dark Matter from Strongly-Interacting Chiral Dynamics}",
    eprint = "2008.10607",
    archivePrefix = "arXiv",
    primaryClass = "hep-ph",
    doi = "10.1007/JHEP02(2021)091",
    journal = "JHEP",
    volume = "02",
    pages = "091",
    year = "2021"
}

@article{Garani:2021zrr,
    author = "Garani, Raghuveer and Redi, Michele and Tesi, Andrea",
    title = "{Dark QCD matters}",
    eprint = "2105.03429",
    archivePrefix = "arXiv",
    primaryClass = "hep-ph",
    doi = "10.1007/JHEP12(2021)139",
    journal = "JHEP",
    volume = "12",
    pages = "139",
    year = "2021"
}

@article{Carenza:2022pjd,
    author = "Carenza, Pierluca and Pasechnik, Roman and Salinas, Gustavo and Wang, Zhi-Wei",
    title = "{Glueball Dark Matter Revisited}",
    eprint = "2207.13716",
    archivePrefix = "arXiv",
    primaryClass = "hep-ph",
    doi = "10.1103/PhysRevLett.129.261302",
    journal = "Phys. Rev. Lett.",
    volume = "129",
    number = "26",
    pages = "261302",
    year = "2022"
}

@article{Baldes:2021aph,
    author = "Baldes, Iason and Gouttenoire, Yann and Sala, Filippo and Servant, G\'eraldine",
    title = "{Supercool composite Dark Matter beyond 100 TeV}",
    eprint = "2110.13926",
    archivePrefix = "arXiv",
    primaryClass = "hep-ph",
    reportNumber = "ULB-TH/21-17; DESY 21-172, ULB-TH/21-17, DESY 21-172",
    doi = "10.1007/JHEP07(2022)084",
    journal = "JHEP",
    volume = "07",
    pages = "084",
    year = "2022"
}

@article{Bottaro:2021aal,
    author = "Bottaro, Salvatore and Costa, Marco and Popov, Oleg",
    title = "{Asymmetric accidental composite dark matter}",
    eprint = "2104.14244",
    archivePrefix = "arXiv",
    primaryClass = "hep-ph",
    doi = "10.1007/JHEP11(2021)055",
    journal = "JHEP",
    volume = "11",
    pages = "055",
    year = "2021"
}

@article{Knapen:2021eip,
    author = "Knapen, Simon and Shelton, Jessie and Xu, Dong",
    title = "{Perturbative benchmark models for a dark shower search program}",
    eprint = "2103.01238",
    archivePrefix = "arXiv",
    primaryClass = "hep-ph",
    doi = "10.1103/PhysRevD.103.115013",
    journal = "Phys. Rev. D",
    volume = "103",
    number = "11",
    pages = "115013",
    year = "2021"
}

@article{Batz:2023zef,
    author = "Batz, Austin and Cohen, Timothy and Curtin, David and Gemmell, Caleb and Kribs, Graham D.",
    title = "{Dark sector glueballs at the LHC}",
    eprint = "2310.13731",
    archivePrefix = "arXiv",
    primaryClass = "hep-ph",
    reportNumber = "CERN-TH-2023-194",
    doi = "10.1007/JHEP04(2024)070",
    journal = "JHEP",
    volume = "04",
    pages = "070",
    year = "2024"
}

@article{Morrison:2020yeg,
    author = "Morrison, Logan and Profumo, Stefano and Robinson, Dean J.",
    title = "{Large $N$-ightmare Dark Matter}",
    eprint = "2010.03586",
    archivePrefix = "arXiv",
    primaryClass = "hep-ph",
    doi = "10.1088/1475-7516/2021/05/058",
    journal = "JCAP",
    volume = "05",
    pages = "058",
    year = "2021"
}

@article{Mahbubani:2019pij,
    author = "Mahbubani, Rakhi and Redi, Michele and Tesi, Andrea",
    title = "{Indirect detection of composite asymmetric dark matter}",
    eprint = "1908.00538",
    archivePrefix = "arXiv",
    primaryClass = "hep-ph",
    reportNumber = "CERN-TH-2019-125",
    doi = "10.1103/PhysRevD.101.103037",
    journal = "Phys. Rev. D",
    volume = "101",
    number = "10",
    pages = "103037",
    year = "2020"
}

@article{Eby:2019mgs,
    author = "Eby, Joshua and Fox, Patrick J. and Harnik, Roni and Kribs, Graham D.",
    title = "{Luminous Signals of Inelastic Dark Matter in Large Detectors}",
    eprint = "1904.09994",
    archivePrefix = "arXiv",
    primaryClass = "hep-ph",
    reportNumber = "FERMILAB-PUB-19-147-T",
    doi = "10.1007/JHEP09(2019)115",
    journal = "JHEP",
    volume = "09",
    pages = "115",
    year = "2019"
}

@article{Francis:2018xjd,
    author = "Francis, Anthony and Hudspith, Renwick J. and Lewis, Randy and Tulin, Sean",
    title = "{Dark Matter from Strong Dynamics: The Minimal Theory of Dark Baryons}",
    eprint = "1809.09117",
    archivePrefix = "arXiv",
    primaryClass = "hep-ph",
    reportNumber = "CERN-TH-2018-207",
    doi = "10.1007/JHEP12(2018)118",
    journal = "JHEP",
    volume = "12",
    pages = "118",
    year = "2018"
}

@article{Acharya:2017szw,
    author = "Acharya, Bobby Samir and Fairbairn, Malcolm and Hardy, Edward",
    title = "{Glueball dark matter in non-standard cosmologies}",
    eprint = "1704.01804",
    archivePrefix = "arXiv",
    primaryClass = "hep-ph",
    doi = "10.1007/JHEP07(2017)100",
    journal = "JHEP",
    volume = "07",
    pages = "100",
    year = "2017"
}

@article{Knapen:2016hky,
    author = "Knapen, Simon and Pagan Griso, Simone and Papucci, Michele and Robinson, Dean J.",
    title = "{Triggering Soft Bombs at the LHC}",
    eprint = "1612.00850",
    archivePrefix = "arXiv",
    primaryClass = "hep-ph",
    doi = "10.1007/JHEP08(2017)076",
    journal = "JHEP",
    volume = "08",
    pages = "076",
    year = "2017"
}

@article{Assi:2023cfo,
    author = "Assi, Benoit and Wagman, Michael L.",
    title = "{Baryons, multihadron systems, and composite dark matter in nonrelativistic QCD}",
    eprint = "2305.01685",
    archivePrefix = "arXiv",
    primaryClass = "hep-ph",
    reportNumber = "FERMILAB-PUB-23-127-T",
    doi = "10.1103/PhysRevD.108.096004",
    journal = "Phys. Rev. D",
    volume = "108",
    number = "9",
    pages = "096004",
    year = "2023"
}

@article{Han:2007ae,
    author = "Han, Tao and Si, Zongguo and Zurek, Kathryn M. and Strassler, Matthew J.",
    title = "{Phenomenology of hidden valleys at hadron colliders}",
    eprint = "0712.2041",
    archivePrefix = "arXiv",
    primaryClass = "hep-ph",
    reportNumber = "MADPH-07-1502, SDU-HEP-071201, RUNHETC-2007-31",
    doi = "10.1088/1126-6708/2008/07/008",
    journal = "JHEP",
    volume = "07",
    pages = "008",
    year = "2008"
}

@article{Strassler:2006im,
    author = "Strassler, Matthew J. and Zurek, Kathryn M.",
    title = "{Echoes of a hidden valley at hadron colliders}",
    eprint = "hep-ph/0604261",
    archivePrefix = "arXiv",
    doi = "10.1016/j.physletb.2007.06.055",
    journal = "Phys. Lett. B",
    volume = "651",
    pages = "374--379",
    year = "2007"
}

@article{Kang:2008ea,
    author = "Kang, Junhai and Luty, Markus A.",
    title = "{Macroscopic Strings and 'Quirks' at Colliders}",
    eprint = "0805.4642",
    archivePrefix = "arXiv",
    primaryClass = "hep-ph",
    doi = "10.1088/1126-6708/2009/11/065",
    journal = "JHEP",
    volume = "11",
    pages = "065",
    year = "2009"
}

@article{Knapen:2017kly,
    author = "Knapen, Simon and Lou, Hou Keong and Papucci, Michele and Setford, Jack",
    title = "{Tracking down Quirks at the Large Hadron Collider}",
    eprint = "1708.02243",
    archivePrefix = "arXiv",
    primaryClass = "hep-ph",
    doi = "10.1103/PhysRevD.96.115015",
    journal = "Phys. Rev. D",
    volume = "96",
    number = "11",
    pages = "115015",
    year = "2017"
}

@article{Evans:2018jmd,
    author = "Evans, Jared A. and Luty, Markus A.",
    title = "{Stopping Quirks at the LHC}",
    eprint = "1811.08903",
    archivePrefix = "arXiv",
    primaryClass = "hep-ph",
    doi = "10.1007/JHEP06(2019)090",
    journal = "JHEP",
    volume = "06",
    pages = "090",
    year = "2019"
}

@article{Gouttenoire:2023roe,
    author = "Gouttenoire, Yann and Kuflik, Eric and Liu, Di",
    title = "{Heavy baryon dark matter from SU(N) confinement: Bubble wall velocity and boundary effects}",
    eprint = "2311.00029",
    archivePrefix = "arXiv",
    primaryClass = "hep-ph",
    reportNumber = "LAPTH-055/23",
    doi = "10.1103/PhysRevD.109.035002",
    journal = "Phys. Rev. D",
    volume = "109",
    number = "3",
    pages = "035002",
    year = "2024"
}

@article{Witten:1984rs,
    author = "Witten, Edward",
    title = "{Cosmic Separation of Phases}",
    reportNumber = "PRINT-84-0400 (IAS,PRINCETON)",
    doi = "10.1103/PhysRevD.30.272",
    journal = "Phys. Rev. D",
    volume = "30",
    pages = "272--285",
    year = "1984"
}

@article{Hill:2014yka,
    author = "Hill, Richard J. and Solon, Mikhail P.",
    title = "{Standard Model anatomy of WIMP dark matter direct detection I: weak-scale matching}",
    eprint = "1401.3339",
    archivePrefix = "arXiv",
    primaryClass = "hep-ph",
    reportNumber = "EFI-13-34",
    doi = "10.1103/PhysRevD.91.043504",
    journal = "Phys. Rev. D",
    volume = "91",
    pages = "043504",
    year = "2015"
}

@article{Abe:2024mwa,
    author = "Abe, Tomohiro and Sato, Ryosuke and Yamanaka, Takumu",
    title = "{Composite Dark Matter with Forbidden Annihilation}",
    eprint = "2404.03963",
    archivePrefix = "arXiv",
    primaryClass = "hep-ph",
    reportNumber = "OU-HET-1219",
    month = "4",
    year = "2024"
}

@article{Cirelli:2005uq,
    author = "Cirelli, Marco and Fornengo, Nicolao and Strumia, Alessandro",
    title = "{Minimal dark matter}",
    eprint = "hep-ph/0512090",
    archivePrefix = "arXiv",
    reportNumber = "DFTT40-2005, IFUP-TH-2005-34",
    doi = "10.1016/j.nuclphysb.2006.07.012",
    journal = "Nucl. Phys. B",
    volume = "753",
    pages = "178--194",
    year = "2006"
}

@article{Cirelli:2024ssz,
    author = "Cirelli, Marco and Strumia, Alessandro and Zupan, Jure",
    title = "{Dark Matter}",
    eprint = "2406.01705",
    archivePrefix = "arXiv",
    primaryClass = "hep-ph",
    month = "6",
    year = "2024"
}

@article{Bramante:2016rdh,
        archiveprefix = {arXiv},
        author = {Bramante, Joseph and Fox, Patrick J. and Kribs, Graham D. and Martin, Adam},
        doi = {10.1103/PhysRevD.94.115026},
        eprint = {1608.02662},
        journal = {Phys. Rev.},
        number = {11},
        pages = {115026},
        primaryclass = {hep-ph},
        reportnumber = {FERMILAB-PUB-16-301-T},
        slaccitation = {%%CITATION = ARXIV:1608.02662;%%},
        title = {{Inelastic frontier: Discovering dark matter at high recoil energy}},
        volume = {D94},
        year = {2016}
}

@article{Eby:2023wem,
    author = "Eby, Joshua and Fox, Patrick J. and Kribs, Graham D.",
    title = "{Earth-catalyzed detection of magnetic inelastic dark matter with photons in large underground detectors}",
    eprint = "2312.08478",
    archivePrefix = "arXiv",
    primaryClass = "hep-ph",
    reportNumber = "FERMILAB-PUB-23-781-T, IPMU23-0046",
    doi = "10.1007/JHEP06(2024)165",
    journal = "JHEP",
    volume = "06",
    pages = "165",
    year = "2024"
}

@article{Feldstein:2010su,
    author = "Feldstein, Brian and Graham, Peter W. and Rajendran, Surjeet",
    title = "{Luminous Dark Matter}",
    eprint = "1008.1988",
    archivePrefix = "arXiv",
    primaryClass = "hep-ph",
    reportNumber = "MIT-CTP-4172",
    doi = "10.1103/PhysRevD.82.075019",
    journal = "Phys. Rev. D",
    volume = "82",
    pages = "075019",
    year = "2010"
}

@article{Cheng:2021kjg,
    author = "Cheng, Hsin-Chia and Li, Lingfeng and Salvioni, Ennio",
    title = "{A theory of dark pions}",
    eprint = "2110.10691",
    archivePrefix = "arXiv",
    primaryClass = "hep-ph",
    reportNumber = "CERN-TH-2021-150",
    doi = "10.1007/JHEP01(2022)122",
    journal = "JHEP",
    volume = "01",
    pages = "122",
    year = "2022"
}

@article{Kribs:2018oad,
    author = "Kribs, Graham D. and Martin, Adam and Tong, Tom",
    title = "{Effective Theories of Dark Mesons with Custodial Symmetry}",
    eprint = "1809.10183",
    archivePrefix = "arXiv",
    primaryClass = "hep-ph",
    doi = "10.1007/JHEP08(2019)020",
    journal = "JHEP",
    volume = "08",
    pages = "020",
    year = "2019"
}

@article{LatticeStrongDynamics:2020jwi,
      author         = "Brower, R. C. and Cushman, K. and Fleming, G. T. and Gasbarro, A. and Hasenfratz, A. and Jin, X. Y. and Kribs, G. D. and Neil, E. T. and Osborn, J. C. and Rebbi, C. and Rinaldi, E. and Schaich, D. and Vranas, P. and Witzel, O.", 
    collaboration = "Lattice Strong Dynamics",
    title = "{Stealth dark matter confinement transition and gravitational waves}",
    eprint = "2006.16429",
    archivePrefix = "arXiv",
    primaryClass = "hep-lat",
    reportNumber = "LLNL-JRNL-811356; RIKEN-iTHEMS-Report-20",
    doi = "10.1103/PhysRevD.103.014505",
    journal = "Phys. Rev. D",
    volume = "103",
    number = "1",
    pages = "014505",
    year = "2021"
}

@article{Morningstar:1999rf,
    author = "Morningstar, Colin J. and Peardon, Mike J.",
    title = "{The Glueball spectrum from an anisotropic lattice study}",
    eprint = "hep-lat/9901004",
    archivePrefix = "arXiv",
    reportNumber = "UCSD-PTH-98-36, HLRZ-1998-62",
    doi = "10.1103/PhysRevD.60.034509",
    journal = "Phys. Rev. D",
    volume = "60",
    pages = "034509",
    year = "1999"
}

@article{Mathieu:2008me,
    author = "Mathieu, Vincent and Kochelev, Nikolai and Vento, Vicente",
    title = "{The Physics of Glueballs}",
    eprint = "0810.4453",
    archivePrefix = "arXiv",
    primaryClass = "hep-ph",
    doi = "10.1142/S0218301309012124",
    journal = "Int. J. Mod. Phys. E",
    volume = "18",
    pages = "1--49",
    year = "2009"
}

@article{Strassler:2006qa,
    author = "Strassler, Matthew J.",
    title = "{Possible effects of a hidden valley on supersymmetric phenomenology}",
    eprint = "hep-ph/0607160",
    archivePrefix = "arXiv",
    month = "7",
    year = "2006"
}

@article{Maravin:2007vdd,
    author = "Maravin, Yurii and Gershtein, Yuri",
    collaboration = "D0",
    title = "{Search for long-lived parents of Z boson}",
    reportNumber = "D0-5454",
    month = "7",
    year = "2007"
}

@article{Strassler:2008bv,
    author = "Strassler, Matthew J.",
    title = "{Why Unparticle Models with Mass Gaps are Examples of Hidden Valleys}",
    eprint = "0801.0629",
    archivePrefix = "arXiv",
    primaryClass = "hep-ph",
    month = "1",
    year = "2008"
}

@article{D0:2009mtx,
    author = "Abazov, V. M. and others",
    collaboration = "D0",
    title = "{Search for Resonant Pair Production of long-lived particles decaying to b anti-b in p anti-p collisions at s**(1/2) = 1.96-TeV}",
    eprint = "0906.1787",
    archivePrefix = "arXiv",
    primaryClass = "hep-ex",
    reportNumber = "FERMILAB-PUB-09-299-E",
    doi = "10.1103/PhysRevLett.103.071801",
    journal = "Phys. Rev. Lett.",
    volume = "103",
    pages = "071801",
    year = "2009"
}

@article{Foster:2022ajl,
    author = "Foster, Joshua W. and Kumar, Soubhik and Safdi, Benjamin R. and Soreq, Yotam",
    title = "{Dark Grand Unification in the axiverse: decaying axion dark matter and spontaneous baryogenesis}",
    eprint = "2208.10504",
    archivePrefix = "arXiv",
    primaryClass = "hep-ph",
    reportNumber = "MIT-CTP/5458",
    doi = "10.1007/JHEP12(2022)119",
    journal = "JHEP",
    volume = "12",
    pages = "119",
    year = "2022"
}

@article{Carloni:2010tw,
    author = "Carloni, Lisa and Sjostrand, Torbjorn",
    title = "{Visible Effects of Invisible Hidden Valley Radiation}",
    eprint = "1006.2911",
    archivePrefix = "arXiv",
    primaryClass = "hep-ph",
    reportNumber = "LU-TP-10-17, MCNET-10-11",
    doi = "10.1007/JHEP09(2010)105",
    journal = "JHEP",
    volume = "09",
    pages = "105",
    year = "2010"
}

@article{CMS:2011xlr,
    author = "Chatrchyan, Serguei and others",
    collaboration = "CMS",
    title = "{Search for Light Resonances Decaying into Pairs of Muons as a Signal of New Physics}",
    eprint = "1106.2375",
    archivePrefix = "arXiv",
    primaryClass = "hep-ex",
    reportNumber = "CERN-PH-EP-2011-064, CMS-EXO-11-013",
    doi = "10.1007/JHEP07(2011)098",
    journal = "JHEP",
    volume = "07",
    pages = "098",
    year = "2011"
}

@article{CDF:2011dnt,
    author = "Aaltonen, T. and others",
    collaboration = "CDF",
    title = "{Search for heavy metastable particles decaying to jet pairs in $p\bar{p}$ collisions at $\sqrt{s} = 1.96$ TeV}",
    eprint = "1109.3136",
    archivePrefix = "arXiv",
    primaryClass = "hep-ex",
    reportNumber = "FERMILAB-PUB-11-451-E-PPD, PUB-11-451-E-PPD",
    doi = "10.1103/PhysRevD.85.012007",
    journal = "Phys. Rev. D",
    volume = "85",
    pages = "012007",
    year = "2012"
}

@article{CMS:2012axw,
    author = "Chatrchyan, Serguei and others",
    collaboration = "CMS",
    title = "{Search in Leptonic Channels for Heavy Resonances Decaying to Long-Lived Neutral Particles}",
    eprint = "1211.2472",
    archivePrefix = "arXiv",
    primaryClass = "hep-ex",
    reportNumber = "CMS-EXO-11-101, CERN-PH-EP-2012-304",
    doi = "10.1007/JHEP02(2013)085",
    journal = "JHEP",
    volume = "02",
    pages = "085",
    year = "2013"
}

@article{ATLAS:2012wib,
    author = "Aad, Georges and others",
    collaboration = "ATLAS",
    title = "{A search for prompt lepton-jets in $pp$ collisions at $\sqrt{s}=7$ TeV with the ATLAS detector}",
    eprint = "1212.5409",
    archivePrefix = "arXiv",
    primaryClass = "hep-ex",
    reportNumber = "CERN-PH-EP-2012-319",
    doi = "10.1016/j.physletb.2013.01.034",
    journal = "Phys. Lett. B",
    volume = "719",
    pages = "299--317",
    year = "2013"
}

@article{LHCb:2014jgs,
    author = "Aaij, Roel and others",
    collaboration = "LHCb",
    title = "{Search for long-lived particles decaying to jet pairs}",
    eprint = "1412.3021",
    archivePrefix = "arXiv",
    primaryClass = "hep-ex",
    reportNumber = "LHCB-PAPER-2014-062, CERN-PH-EP-2014-291",
    doi = "10.1140/epjc/s10052-015-3344-6",
    journal = "Eur. Phys. J. C",
    volume = "75",
    number = "4",
    pages = "152",
    year = "2015"
}

@article{Curtin:2014pda,
    author = "Curtin, David and Essig, Rouven and Zhong, Yi-Ming",
    title = "{Uncovering light scalars with exotic Higgs decays to $ b\overline{b}{\mu}^{+}{\mu}^{-} $}",
    eprint = "1412.4779",
    archivePrefix = "arXiv",
    primaryClass = "hep-ph",
    reportNumber = "YITP-SB-14-53",
    doi = "10.1007/JHEP06(2015)025",
    journal = "JHEP",
    volume = "06",
    pages = "025",
    year = "2015"
}

@article{ATLAS:2015mlf,
    author = "Aad, Georges and others",
    collaboration = "ATLAS",
    title = "{Search for pair-produced long-lived neutral particles decaying in the ATLAS hadronic calorimeter in $pp$ collisions at $\sqrt{s}$ = 8 TeV}",
    eprint = "1501.04020",
    archivePrefix = "arXiv",
    primaryClass = "hep-ex",
    reportNumber = "CERN-PH-EP-2014-228",
    doi = "10.1016/j.physletb.2015.02.015",
    journal = "Phys. Lett. B",
    volume = "743",
    pages = "15--34",
    year = "2015"
}

@article{Craig:2015pha,
    author = "Craig, Nathaniel and Katz, Andrey and Strassler, Matt and Sundrum, Raman",
    title = "{Naturalness in the Dark at the LHC}",
    eprint = "1501.05310",
    archivePrefix = "arXiv",
    primaryClass = "hep-ph",
    reportNumber = "UMD-PP-014-028, CERN-PH-TH-2014-263",
    doi = "10.1007/JHEP07(2015)105",
    journal = "JHEP",
    volume = "07",
    pages = "105",
    year = "2015"
}

@article{Cohen:2017pzm,
    author = "Cohen, Timothy and Lisanti, Mariangela and Lou, Hou Keong and Mishra-Sharma, Siddharth",
    title = "{LHC Searches for Dark Sector Showers}",
    eprint = "1707.05326",
    archivePrefix = "arXiv",
    primaryClass = "hep-ph",
    reportNumber = "PUPT-2529",
    doi = "10.1007/JHEP11(2017)196",
    journal = "JHEP",
    volume = "11",
    pages = "196",
    year = "2017"
}

@article{Pierce:2017taw,
    author = "Pierce, Aaron and Shakya, Bibhushan and Tsai, Yuhsin and Zhao, Yue",
    title = "{Searching for confining hidden valleys at LHCb, ATLAS, and CMS}",
    eprint = "1708.05389",
    archivePrefix = "arXiv",
    primaryClass = "hep-ph",
    reportNumber = "MCTP-17-13, PP-017-28",
    doi = "10.1103/PhysRevD.97.095033",
    journal = "Phys. Rev. D",
    volume = "97",
    number = "9",
    pages = "095033",
    year = "2018"
}

@article{Beauchesne:2018myj,
    author = "Beauchesne, Hugues and Bertuzzo, Enrico and Grilli Di Cortona, Giovanni",
    title = "{Dark matter in Hidden Valley models with stable and unstable light dark mesons}",
    eprint = "1809.10152",
    archivePrefix = "arXiv",
    primaryClass = "hep-ph",
    doi = "10.1007/JHEP04(2019)118",
    journal = "JHEP",
    volume = "04",
    pages = "118",
    year = "2019"
}

@article{CMS:2018bvr,
    author = "Sirunyan, Albert M and others",
    collaboration = "CMS",
    title = "{Search for new particles decaying to a jet and an emerging jet}",
    eprint = "1810.10069",
    archivePrefix = "arXiv",
    primaryClass = "hep-ex",
    reportNumber = "CMS-EXO-18-001, CERN-EP-2018-255",
    doi = "10.1007/JHEP02(2019)179",
    journal = "JHEP",
    volume = "02",
    pages = "179",
    year = "2019"
}

@article{Alipour-Fard:2018lsf,
    author = "Alipour-Fard, Samuel and Craig, Nathaniel and Jiang, Minyuan and Koren, Seth",
    title = "{Long Live the Higgs Factory: Higgs Decays to Long-Lived Particles at Future Lepton Colliders}",
    eprint = "1812.05588",
    archivePrefix = "arXiv",
    primaryClass = "hep-ph",
    doi = "10.1088/1674-1137/43/5/053101",
    journal = "Chin. Phys. C",
    volume = "43",
    number = "5",
    pages = "053101",
    year = "2019"
}

@article{Kar:2020bws,
    author = "Kar, Deepak and Sinha, Sukanya",
    title = "{Exploring jet substructure in semi-visible jets}",
    eprint = "2007.11597",
    archivePrefix = "arXiv",
    primaryClass = "hep-ph",
    doi = "10.21468/SciPostPhys.10.4.084",
    journal = "SciPost Phys.",
    volume = "10",
    number = "4",
    pages = "084",
    year = "2021"
}

@article{Lim:2020igi,
    author = "Lim, Sung Hak and Nojiri, Mihoko M.",
    title = "{Morphology for jet classification}",
    eprint = "2010.13469",
    archivePrefix = "arXiv",
    primaryClass = "hep-ph",
    reportNumber = "KEK-TH-2266",
    doi = "10.1103/PhysRevD.105.014004",
    journal = "Phys. Rev. D",
    volume = "105",
    number = "1",
    pages = "014004",
    year = "2022"
}

@article{Murgui:2021eqf,
    author = "Murgui, Clara and Zurek, Kathryn M.",
    title = "{Dark unification: A UV-complete theory of asymmetric dark matter}",
    eprint = "2112.08374",
    archivePrefix = "arXiv",
    primaryClass = "hep-ph",
    doi = "10.1103/PhysRevD.105.095002",
    journal = "Phys. Rev. D",
    volume = "105",
    number = "9",
    pages = "095002",
    year = "2022"
}

@article{Curtin:2022tou,
    author = "Curtin, David and Gemmell, Caleb and Verhaaren, Christopher B.",
    title = "{Simulating glueball production in Nf=0 QCD}",
    eprint = "2202.12899",
    archivePrefix = "arXiv",
    primaryClass = "hep-ph",
    doi = "10.1103/PhysRevD.106.075015",
    journal = "Phys. Rev. D",
    volume = "106",
    number = "7",
    pages = "075015",
    year = "2022"
}

@article{Kar:2022hxn,
    author = "Kar, Deepak and Nzuza, Wandile and Sinha, Sukanya",
    title = "{2B or not $^{2}$B, a study of bottom-quark-philic semi-visible jets}",
    eprint = "2207.01885",
    archivePrefix = "arXiv",
    primaryClass = "hep-ph",
    doi = "10.21468/SciPostPhysCore.7.4.071",
    journal = "SciPost Phys. Core",
    volume = "7",
    pages = "071",
    year = "2024"
}

@article{Curtin:2022oec,
    author = "Curtin, David and Gemmell, Caleb",
    title = "{Indirect detection of Dark Matter annihilating into Dark Glueballs}",
    eprint = "2211.05794",
    archivePrefix = "arXiv",
    primaryClass = "hep-ph",
    doi = "10.1007/JHEP09(2023)010",
    journal = "JHEP",
    volume = "09",
    pages = "010",
    year = "2023"
}

@article{Born:2023vll,
    author = "Born, Susan and Karur, Rohith and Knapen, Simon and Shelton, Jessie",
    title = "{Scouting for dark showers at CMS and LHCb}",
    eprint = "2303.04167",
    archivePrefix = "arXiv",
    primaryClass = "hep-ph",
    doi = "10.1103/PhysRevD.108.035034",
    journal = "Phys. Rev. D",
    volume = "108",
    number = "3",
    pages = "035034",
    year = "2023"
}

@article{Bai:2023yyy,
    author = "Bai, Kehang and Mastandrea, Radha and Nachman, Benjamin",
    title = "{Non-resonant anomaly detection with background extrapolation}",
    eprint = "2311.12924",
    archivePrefix = "arXiv",
    primaryClass = "hep-ph",
    doi = "10.1007/JHEP04(2024)059",
    journal = "JHEP",
    volume = "04",
    pages = "059",
    year = "2024"
}

@article{CMS:2024nca,
    author = "Hayrapetyan, Aram and others",
    collaboration = "CMS",
    title = "{Search for Soft Unclustered Energy Patterns in Proton-Proton Collisions at 13~TeV}",
    eprint = "2403.05311",
    archivePrefix = "arXiv",
    primaryClass = "hep-ex",
    reportNumber = "CMS-EXO-23-002, CERN-EP-2024-054",
    doi = "10.1103/PhysRevLett.133.191902",
    journal = "Phys. Rev. Lett.",
    volume = "133",
    number = "19",
    pages = "191902",
    year = "2024"
}

@article{Niedziela:2024khw,
    author = "Niedziela, Jeremi",
    title = "{SHIFT@LHC: Searches for new physics with shifted interaction on a fixed target at the Large Hadron Collider}",
    eprint = "2406.08557",
    archivePrefix = "arXiv",
    primaryClass = "hep-ph",
    reportNumber = "DESY-24-090",
    doi = "10.1007/JHEP10(2024)204",
    journal = "JHEP",
    volume = "10",
    pages = "204",
    year = "2024"
}

@article{Liebersbach:2024kzc,
    author = "Liebersbach, Samuel and Sandick, Pearl and Shiferaw, Abel and Zhao, Yue",
    title = "{Exploring the hidden valley at MATHUSLA}",
    eprint = "2408.07756",
    archivePrefix = "arXiv",
    primaryClass = "hep-ph",
    doi = "10.1016/j.nuclphysb.2025.117050",
    journal = "Nucl. Phys. B",
    volume = "1018",
    pages = "117050",
    year = "2025"
}

@article{Kulkarni:2024okx,
    author = {Kulkarni, Suchita and Masouminia, M. R. and Pl{\"a}tzer, Simon and Stafford, Dominic},
    title = "{Dark sector showers and hadronisation in Herwig 7}",
    eprint = "2408.10044",
    archivePrefix = "arXiv",
    primaryClass = "hep-ph",
    reportNumber = "IPPP/24/54, IPPP/24/54; PUBDB-2024-06237",
    doi = "10.1140/epjc/s10052-024-13587-8",
    journal = "Eur. Phys. J. C",
    volume = "84",
    number = "11",
    pages = "1210",
    year = "2024"
}

@article{CMS:2025fnr,
    author = "Hayrapetyan, Aram and others",
    collaboration = "CMS",
    title = "{Search for low-mass hidden-valley dark showers with non-prompt muon pairs in proton-proton collisions at $\sqrt{s}$ = 13 TeV}",
    eprint = "2511.11888",
    archivePrefix = "arXiv",
    primaryClass = "hep-ex",
    reportNumber = "CMS-EXO-24-008, CERN-EP-2025-241",
    month = "11",
    year = "2025"
}

@article{Kulkarni:2025rsl,
    author = "Kulkarni, Suchita and Lockyer, Joshua and Strassler, Matthew J.",
    title = "{On the simulation of hidden parton showers in the conformal window}",
    eprint = "2502.18566",
    archivePrefix = "arXiv",
    primaryClass = "hep-ph",
    doi = "10.1007/JHEP09(2025)150",
    journal = "JHEP",
    volume = "09",
    pages = "150",
    year = "2025"
}

@article{Liu:2025bbc,
    author = "Liu, Wei and Lockyer, Joshua and Kulkarni, Suchita",
    title = "{Hidden valley scenario sensitivity in the CMS muon end cap detector}",
    eprint = "2505.03058",
    archivePrefix = "arXiv",
    primaryClass = "hep-ph",
    doi = "10.1103/j4dk-qfy4",
    journal = "Phys. Rev. D",
    volume = "112",
    number = "7",
    pages = "075029",
    year = "2025"
}

@article{Carloni:2011kk,
    author = "Carloni, Lisa and Rathsman, Johan and Sjostrand, Torbjorn",
    title = "{Discerning Secluded Sector gauge structures}",
    eprint = "1102.3795",
    archivePrefix = "arXiv",
    primaryClass = "hep-ph",
    reportNumber = "LU-TP-11-09, MCNET-11-06",
    doi = "10.1007/JHEP04(2011)091",
    journal = "JHEP",
    volume = "04",
    pages = "091",
    year = "2011"
}

@article{Bernreuther:2020vhm,
    author = {Bernreuther, Elias and Finke, Thorben and Kahlhoefer, Felix and Kr{\"a}mer, Michael and M{\"u}ck, Alexander},
    title = "{Casting a graph net to catch dark showers}",
    eprint = "2006.08639",
    archivePrefix = "arXiv",
    primaryClass = "hep-ph",
    reportNumber = "TTK-20-17, P3H-20-025",
    doi = "10.21468/SciPostPhys.10.2.046",
    journal = "SciPost Phys.",
    volume = "10",
    number = "2",
    pages = "046",
    year = "2021"
}

@article{CMS:2021dzg,
    author = "Tumasyan, Armen and others",
    collaboration = "CMS",
    title = "{Search for resonant production of strongly coupled dark matter in proton-proton collisions at 13 TeV}",
    eprint = "2112.11125",
    archivePrefix = "arXiv",
    primaryClass = "hep-ex",
    reportNumber = "CMS-EXO-19-020, CERN-EP-2021-252",
    doi = "10.1007/JHEP06(2022)156",
    journal = "JHEP",
    volume = "06",
    pages = "156",
    year = "2022"
}

@article{Canelli:2021aps,
    author = "Canelli, Florencia and de Cosa, Annapaola and Pottier, Luc Le and Niedziela, Jeremi and Pedro, Kevin and Pierini, Maurizio",
    title = "{Autoencoders for semivisible jet detection}",
    eprint = "2112.02864",
    archivePrefix = "arXiv",
    primaryClass = "hep-ph",
    doi = "10.1007/JHEP02(2022)074",
    journal = "JHEP",
    volume = "02",
    pages = "074",
    year = "2022"
}

@article{Finke:2022lsu,
    author = {Finke, Thorben and Kr{\"a}mer, Michael and Lipp, Maximilian and M{\"u}ck, Alexander},
    title = "{Boosting mono-jet searches with model-agnostic machine learning}",
    eprint = "2204.11889",
    archivePrefix = "arXiv",
    primaryClass = "hep-ph",
    reportNumber = "P3H-22-043, TTK-22-16",
    doi = "10.1007/JHEP08(2022)015",
    journal = "JHEP",
    volume = "08",
    pages = "015",
    year = "2022"
}

@article{Cazzaniga:2022hxl,
    author = "Cazzaniga, Cesare and de Cosa, Annapaola",
    title = "{Leptons lurking in semi-visible jets at the LHC}",
    eprint = "2206.03909",
    archivePrefix = "arXiv",
    primaryClass = "hep-ph",
    doi = "10.1140/epjc/s10052-022-10775-2",
    journal = "Eur. Phys. J. C",
    volume = "82",
    number = "9",
    pages = "793",
    year = "2022"
}

@article{Faucett:2022zie,
    author = "Faucett, Taylor and Hsu, Shih-Chieh and Whiteson, Daniel",
    title = "{Learning to identify semi-visible jets}",
    eprint = "2208.10062",
    archivePrefix = "arXiv",
    primaryClass = "hep-ph",
    doi = "10.1007/JHEP12(2022)132",
    journal = "JHEP",
    volume = "12",
    pages = "132",
    year = "2022"
}

@article{Cohen:2020afv,
    author = "Cohen, Timothy and Doss, Joel and Freytsis, Marat",
    title = "{Jet Substructure from Dark Sector Showers}",
    eprint = "2004.00631",
    archivePrefix = "arXiv",
    primaryClass = "hep-ph",
    doi = "10.1007/JHEP09(2020)118",
    journal = "JHEP",
    volume = "09",
    pages = "118",
    year = "2020"
}

@inproceedings{Beauchesne:2021qrw,
    author = "Beauchesne, Hugues and Grilli di Cortona, Giovanni",
    title = "{Event-level variables for semivisible jets using anomalous jet tagging}",
    booktitle = "{Snowmass 2021}",
    eprint = "2111.12156",
    archivePrefix = "arXiv",
    primaryClass = "hep-ph",
    month = "11",
    year = "2021"
}

@article{Buckley:2022zry,
    author = "Buckley, Andy and Kar, Deepak and Sinha, Sukanya",
    title = "{Towards better discrimination and improved modelling of dark-sector showers}",
    eprint = "2209.14964",
    archivePrefix = "arXiv",
    primaryClass = "hep-ph",
    doi = "10.21468/SciPostPhysProc.15.006",
    journal = "SciPost Phys. Proc.",
    volume = "15",
    pages = "006",
    year = "2024"
}

@article{Beauchesne:2022phk,
    author = "Beauchesne, Hugues and Cazzaniga, Cesare and de Cosa, Annapaola and Doglioni, Caterina and Fitschen, Tobias and di Cortona, Giovanni Grilli and Zhou, Ziyuan",
    title = "{Uncovering tau leptons-enriched semi-visible jets at the LHC}",
    eprint = "2212.11523",
    archivePrefix = "arXiv",
    primaryClass = "hep-ph",
    doi = "10.1140/epjc/s10052-023-11775-6",
    journal = "Eur. Phys. J. C",
    volume = "83",
    number = "7",
    pages = "599",
    year = "2023"
}

@article{Pedro:2023sdp,
    author = "Pedro, Kevin and Shyamsundar, Prasanth",
    title = "{Optimal mass variables for semivisible jets}",
    eprint = "2303.16253",
    archivePrefix = "arXiv",
    primaryClass = "hep-ph",
    reportNumber = "FERMILAB-PUB-23-112-CSAID-PPD-QIS",
    doi = "10.21468/SciPostPhysCore.6.4.067",
    journal = "SciPost Phys. Core",
    volume = "6",
    pages = "067",
    year = "2023"
}

@article{ATLAS:2023swa,
    author = "Aad, Georges and others",
    collaboration = "ATLAS",
    title = "{Search for non-resonant production of semi-visible jets using Run 2 data in ATLAS}",
    eprint = "2305.18037",
    archivePrefix = "arXiv",
    primaryClass = "hep-ex",
    reportNumber = "CERN-EP-2023-084",
    doi = "10.1016/j.physletb.2023.138324",
    journal = "Phys. Lett. B",
    volume = "848",
    pages = "138324",
    year = "2024"
}

@article{Favaro:2023xdl,
    author = {Favaro, Luigi and Kr{\"a}mer, Michael and Modak, Tanmoy and Plehn, Tilman and R{\"u}schkamp, Jan},
    title = "{Semi-visible jets, energy-based models, and self-supervision}",
    eprint = "2312.03067",
    archivePrefix = "arXiv",
    primaryClass = "hep-ph",
    doi = "10.21468/SciPostPhys.18.2.042",
    journal = "SciPost Phys.",
    volume = "18",
    number = "2",
    pages = "042",
    year = "2025"
}

@article{CMS:2025lmn,
    author = "Hayrapetyan, Aram and others",
    collaboration = "CMS",
    title = "{Wasserstein normalized autoencoder for anomaly detection}",
    eprint = "2510.02168",
    archivePrefix = "arXiv",
    primaryClass = "hep-ex",
    reportNumber = "CMS-MLG-24-002, CERN-EP-2025-209",
    month = "10",
    year = "2025"
}

@article{Bhardwaj:2024djv,
    author = "Bhardwaj, Akanksha and Englert, Christoph and Naskar, Wrishik and Ngairangbam, Vishal S. and Spannowsky, Michael",
    title = "{Equivariant, safe and sensitive {\textemdash} graph networks for new physics}",
    eprint = "2402.12449",
    archivePrefix = "arXiv",
    primaryClass = "hep-ph",
    reportNumber = "IPPP/24/07",
    doi = "10.1007/JHEP07(2024)245",
    journal = "JHEP",
    volume = "07",
    pages = "245",
    year = "2024"
}

@article{Cazzaniga:2024mmv,
    author = "Cazzaniga, Cesare and Russo, Alessandro and Sitti, Emre and de Cosa, Annapaola",
    title = "{Phenomenology of photons-enriched semi-visible jets}",
    eprint = "2407.08276",
    archivePrefix = "arXiv",
    primaryClass = "hep-ph",
    doi = "10.1140/epjc/s10052-024-13613-9",
    journal = "Eur. Phys. J. C",
    volume = "84",
    number = "11",
    pages = "1223",
    year = "2024"
}

@article{Liu:2024rbe,
    author = "Liu, Bingxuan and Pedro, Kevin",
    title = "{Semi-visible jets + X: illuminating dark showers with radiation}",
    eprint = "2409.04741",
    archivePrefix = "arXiv",
    primaryClass = "hep-ph",
    reportNumber = "FERMILAB-PUB-24-0563-CSAID-PPD",
    doi = "10.1007/JHEP12(2024)105",
    journal = "JHEP",
    volume = "12",
    pages = "105",
    year = "2024"
}

@article{Carmona:2024tkg,
    author = "Carmona, Adrian and Elahi, Fatemeh and Scherb, Christiane and Schwaller, Pedro",
    title = "{Dark showers from sneaky dark matter}",
    eprint = "2411.15073",
    archivePrefix = "arXiv",
    primaryClass = "hep-ph",
    reportNumber = "MITP/24-083",
    doi = "10.1007/JHEP06(2025)198",
    journal = "JHEP",
    volume = "06",
    pages = "198",
    year = "2025"
}

@article{ATLAS:2025kuz,
    author = "Aad, Georges and others",
    collaboration = "ATLAS",
    title = "{Search for new physics in final states with semivisible jets or anomalous signatures using the ATLAS detector}",
    eprint = "2505.01634",
    archivePrefix = "arXiv",
    primaryClass = "hep-ex",
    reportNumber = "CERN-EP-2025-101",
    doi = "10.1103/44zp-mh1q",
    journal = "Phys. Rev. D",
    volume = "112",
    number = "1",
    pages = "012021",
    year = "2025"
}

@article{Cazzaniga:2025piw,
    author = "Cazzaniga, Cesare and de Cosa, Annapaola and Kahlhoefer, Felix and Maria, Andrea S. and Seidita, Roberto and Sitti, Emre",
    title = "{Probing the Higgs Portal to a Strongly-Interacting Dark Sector at the FCC-ee}",
    eprint = "2510.17675",
    archivePrefix = "arXiv",
    primaryClass = "hep-ph",
    reportNumber = "P3H-25-077, TTP25-036",
    month = "10",
    year = "2025"
}

@article{ATLAS:2022gbw,
    author = "Aad, Georges and others",
    collaboration = "ATLAS",
    title = "{Search for events with a pair of displaced vertices from long-lived neutral particles decaying into hadronic jets in the ATLAS muon spectrometer in pp collisions at $\sqrt s$=13{\,}{\,}TeV}",
    eprint = "2203.00587",
    archivePrefix = "arXiv",
    primaryClass = "hep-ex",
    reportNumber = "CERN-EP-2021-195",
    doi = "10.1103/PhysRevD.106.032005",
    journal = "Phys. Rev. D",
    volume = "106",
    number = "3",
    pages = "032005",
    year = "2022"
}

@article{Renner:2018fhh,
    author = "Renner, Sophie and Schwaller, Pedro",
    title = "{A flavoured dark sector}",
    eprint = "1803.08080",
    archivePrefix = "arXiv",
    primaryClass = "hep-ph",
    doi = "10.1007/JHEP08(2018)052",
    journal = "JHEP",
    volume = "08",
    pages = "052",
    year = "2018"
}

@article{Cheng:2019yai,
    author = "Cheng, Hsin-Chia and Li, Lingfeng and Salvioni, Ennio and Verhaaren, Christopher B.",
    title = "{Light Hidden Mesons through the Z Portal}",
    eprint = "1906.02198",
    archivePrefix = "arXiv",
    primaryClass = "hep-ph",
    reportNumber = "TUM-HEP-1203-19",
    doi = "10.1007/JHEP11(2019)031",
    journal = "JHEP",
    volume = "11",
    pages = "031",
    year = "2019"
}

@article{Mies:2020mzw,
    author = "Mies, Hanna and Scherb, Christiane and Schwaller, Pedro",
    title = "{Collider constraints on dark mediators}",
    eprint = "2011.13990",
    archivePrefix = "arXiv",
    primaryClass = "hep-ph",
    reportNumber = "MITP/20-068,P3H-20-069,TTK-20-41",
    doi = "10.1007/JHEP04(2021)049",
    journal = "JHEP",
    volume = "04",
    pages = "049",
    year = "2021"
}

@article{Archer-Smith:2021ntx,
    author = "Archer-Smith, Paul and Linthorne, Dylan and Stolarski, Daniel",
    title = "{Emerging jets displaced into the future}",
    eprint = "2112.05690",
    archivePrefix = "arXiv",
    primaryClass = "hep-ph",
    doi = "10.1007/JHEP02(2022)027",
    journal = "JHEP",
    volume = "02",
    pages = "027",
    year = "2022"
}

@article{Carrasco:2023loy,
    author = "Carrasco, Juliana and Zurita, Jos{\'e}",
    title = "{Emerging jet probes of strongly interacting dark sectors}",
    eprint = "2307.04847",
    archivePrefix = "arXiv",
    primaryClass = "hep-ph",
    doi = "10.1007/JHEP01(2024)034",
    journal = "JHEP",
    volume = "01",
    pages = "034",
    year = "2024"
}

@article{ATLAS:2024xna,
    author = "Aad, Georges and others",
    collaboration = "ATLAS",
    title = "{The ATLAS trigger system for LHC Run 3 and trigger performance in 2022}",
    eprint = "2401.06630",
    archivePrefix = "arXiv",
    primaryClass = "hep-ex",
    reportNumber = "CERN-EP-2023-299",
    doi = "10.1088/1748-0221/19/06/P06029",
    journal = "JINST",
    volume = "19",
    number = "06",
    pages = "P06029",
    year = "2024"
}

@article{CMS:2024gxp,
    author = "Hayrapetyan, Aram and others",
    collaboration = "CMS",
    title = "{Search for dark QCD with emerging jets in proton-proton collisions at $ \sqrt{s} $ = 13 TeV}",
    eprint = "2403.01556",
    archivePrefix = "arXiv",
    primaryClass = "hep-ex",
    reportNumber = "CMS-EXO-22-015, CERN-EP-2024-049",
    doi = "10.1007/JHEP07(2024)142",
    journal = "JHEP",
    volume = "07",
    pages = "142",
    year = "2024"
}

@article{ATLAS:2025bsz,
    author = "Aad, Georges and others",
    collaboration = "ATLAS",
    title = "{Search for emerging jets in pp collisions at $\sqrt{s} = 13.6$ TeV with the ATLAS experiment}",
    eprint = "2505.02429",
    archivePrefix = "arXiv",
    primaryClass = "hep-ex",
    reportNumber = "CERN-EP-2025-099",
    doi = "10.1088/1361-6633/adfe17",
    journal = "Rept. Prog. Phys.",
    volume = "88",
    number = "9",
    pages = "097801",
    year = "2025"
}

@article{ATLAS:2025lfx,
    author = "Aad, Georges and others",
    collaboration = "ATLAS",
    title = "{Search for emerging jets in $pp$ collisions at $\sqrt{s} = 13$ TeV with the ATLAS experiment}",
    eprint = "2510.12347",
    archivePrefix = "arXiv",
    primaryClass = "hep-ex",
    reportNumber = "CERN-EP-2025-225",
    month = "10",
    year = "2025"
}

@article{Carrasco:2025bct,
    author = "Carrasco, Juliana and Kulkarni, Suchita and Liu, Wei and Lockyer, Joshua and Zurita, Jose",
    title = "{Semi-visible emerging jets}",
    eprint = "2511.02918",
    archivePrefix = "arXiv",
    primaryClass = "hep-ph",
    month = "11",
    year = "2025"
}

@article{Heimel:2018mkt,
    author = "Heimel, Theo and Kasieczka, Gregor and Plehn, Tilman and Thompson, Jennifer M.",
    title = "{QCD or What?}",
    eprint = "1808.08979",
    archivePrefix = "arXiv",
    primaryClass = "hep-ph",
    doi = "10.21468/SciPostPhys.6.3.030",
    journal = "SciPost Phys.",
    volume = "6",
    number = "3",
    pages = "030",
    year = "2019"
}

@article{Bernreuther:2022jlj,
    author = {Bernreuther, Elias and B{\"o}se, Kai and Ferber, Torben and Hearty, Christopher and Kahlhoefer, Felix and Morandini, Alessandro and Schmidt-Hoberg, Kai},
    title = "{Forecasting dark showers at Belle II}",
    eprint = "2203.08824",
    archivePrefix = "arXiv",
    primaryClass = "hep-ph",
    reportNumber = "P3H-22-033, TTK-22-12, FERMILAB-PUB-22-140-T, TTP22-019, DESY-22-047",
    doi = "10.1007/JHEP12(2022)005",
    journal = "JHEP",
    volume = "12",
    pages = "005",
    year = "2022"
}

@article{Cohen:2023mya,
    author = "Cohen, Timothy and Roloff, Jennifer and Scherb, Christiane",
    title = "{Dark sector showers in the Lund jet plane}",
    eprint = "2301.07732",
    archivePrefix = "arXiv",
    primaryClass = "hep-ph",
    reportNumber = "CERN-TH-2023-007",
    doi = "10.1103/PhysRevD.108.L031501",
    journal = "Phys. Rev. D",
    volume = "108",
    number = "3",
    pages = "L031501",
    year = "2023"
}

@article{Lu:2023gjk,
    author = "Lu, Chih-Ting and Lv, Huifang and Shen, Wei and Wu, Lei and Zhang, Jia",
    title = "{Probing dark QCD sector through the Higgs portal with machine learning at the LHC}",
    eprint = "2304.03237",
    archivePrefix = "arXiv",
    primaryClass = "hep-ph",
    doi = "10.1007/JHEP08(2023)187",
    journal = "JHEP",
    volume = "08",
    pages = "187",
    year = "2023"
}

@article{Anzalone:2023ugq,
    author = "Anzalone, Luca and Chhibra, Simranjit Singh and Maier, Benedikt and Chernyavskaya, Nadezda and Pierini, Maurizio",
    title = "{Triggering dark showers with conditional dual auto-encoders}",
    eprint = "2306.12955",
    archivePrefix = "arXiv",
    primaryClass = "hep-ex",
    doi = "10.1088/2632-2153/ad652b",
    journal = "Mach. Learn. Sci. Tech.",
    volume = "5",
    number = "3",
    pages = "035064",
    year = "2024"
}

@article{Chhibra:2023tyf,
    author = "Chhibra, Simranjit Singh and Chernyavskaya, Nadezda and Maier, Benedikt and Pierini, Maurzio and Hasan, Syed",
    title = "{Autoencoders for real-time SUEP detection}",
    eprint = "2306.13595",
    archivePrefix = "arXiv",
    primaryClass = "hep-ex",
    doi = "10.1140/epjp/s13360-024-05028-y",
    journal = "Eur. Phys. J. Plus",
    volume = "139",
    number = "3",
    pages = "281",
    year = "2024"
}

@article{Cheng:2024hvq,
    author = "Cheng, Hsin-Chia and Jiang, Xu-Hui and Li, Lingfeng and Salvioni, Ennio",
    title = "{Dark showers from Z-dark Z' mixing}",
    eprint = "2401.08785",
    archivePrefix = "arXiv",
    primaryClass = "hep-ph",
    doi = "10.1007/JHEP04(2024)081",
    journal = "JHEP",
    volume = "04",
    pages = "081",
    year = "2024"
}

@article{CMS:2024bvl,
    author = "Hayrapetyan, Aram and others",
    collaboration = "CMS",
    title = "{Search for long-lived particles decaying in the CMS muon detectors in proton-proton collisions at s=13{\,}{\,}TeV}",
    eprint = "2402.01898",
    archivePrefix = "arXiv",
    primaryClass = "hep-ex",
    reportNumber = "CMS-EXO-21-008, CERN-EP-2024-008",
    doi = "10.1103/PhysRevD.110.032007",
    journal = "Phys. Rev. D",
    volume = "110",
    number = "3",
    pages = "032007",
    year = "2024"
}

@article{Cheng:2024aco,
    author = "Cheng, Hsin-Chia and Jiang, Xu-Hui and Li, Lingfeng",
    title = "{Phenomenology of electroweak portal dark showers: high energy direct probes and low energy complementarity}",
    eprint = "2408.13304",
    archivePrefix = "arXiv",
    primaryClass = "hep-ph",
    doi = "10.1007/JHEP01(2025)149",
    journal = "JHEP",
    volume = "01",
    pages = "149",
    year = "2025"
}

@article{Buckley:2025hty,
    author = "Buckley, Andy and Butterworth, Jon and Corpe, Louie and Doglioni, Caterina and Kar, Deepak and Prat, Clarisse and Sinha, Sukanya and Wilson-Edwards, Danielle",
    title = "{Probing the sensitivity of semi-visible jets to current LHC measurements using the CONTUR toolkit}",
    eprint = "2502.11237",
    archivePrefix = "arXiv",
    primaryClass = "hep-ph",
    month = "2",
    year = "2025"
}

@article{Lu:2025cty,
    author = "Lu, Chih-Ting and Xi, Changbin",
    title = "{Searching for leptophilic composite asymmetric dark sector at e+e- colliders}",
    eprint = "2509.12504",
    archivePrefix = "arXiv",
    primaryClass = "hep-ph",
    doi = "10.1103/xsmv-f97b",
    journal = "Phys. Rev. D",
    volume = "112",
    number = "9",
    pages = "095027",
    year = "2025"
}

@article{Bernreuther:2025xqk,
    author = "Bernreuther, Elias and Hemme, Nicoline and Kahlhoefer, Felix and Kulkarni, Suchita and Ovchynnikov, Maksym",
    title = "{Sub-GeV dark matter and multi-decay signatures from dark showers at beam-dump experiments}",
    eprint = "2510.23696",
    archivePrefix = "arXiv",
    primaryClass = "hep-ph",
    reportNumber = "CERN-TH-2025-215, P3H-25-083, TTP25-039",
    month = "10",
    year = "2025"
}

@article{ATLAS:2024xbu,
    author = "Aad, Georges and others",
    collaboration = "ATLAS",
    title = "{Search for dark mesons decaying to top and bottom quarks in proton-proton collisions at $ \sqrt{\textrm{s}} $ = 13 TeV with the ATLAS detector}",
    eprint = "2405.20061",
    archivePrefix = "arXiv",
    primaryClass = "hep-ex",
    reportNumber = "CERN-EP-2024-150",
    doi = "10.1007/JHEP09(2024)005",
    journal = "JHEP",
    volume = "09",
    pages = "005",
    year = "2024"
}

@techreport{CMS:2025caz,
      collaboration = "CMS",
      title         = "{Search for s-channel production of lepton-enriched
                       semivisible jets in proton-proton collisions at 13 TeV}",
      institution   = "CERN",
      reportNumber  = "CMS-PAS-EXO-24-029",
      address       = "Geneva",
      year          = "2025",
      url           = "https://cds.cern.ch/record/2940795",
}

@article{Dreyer:2020brq,
    author = "Dreyer, Fr{\'e}d{\'e}ric A. and Qu, Huilin",
    title = "{Jet tagging in the Lund plane with graph networks}",
    eprint = "2012.08526",
    archivePrefix = "arXiv",
    primaryClass = "hep-ph",
    reportNumber = "OUTP-20-15P",
    doi = "10.1007/JHEP03(2021)052",
    journal = "JHEP",
    volume = "03",
    pages = "052",
    year = "2021"
}

@misc{suep_generator,
  title = {SUEP Generatior},
  howpublished = {\url{https://gitlab.com/simonknapen/suep_generator}},
  year = {2019},
  month = {Dec}
}

@article{Burdman:2008ek,
    author = "Burdman, Gustavo and Chacko, Z. and Goh, Hock-Seng and Harnik, Roni and Krenke, Christopher A.",
    title = "{The Quirky Collider Signals of Folded Supersymmetry}",
    eprint = "0805.4667",
    archivePrefix = "arXiv",
    primaryClass = "hep-ph",
    reportNumber = "UCB-PTH-08-09, SLAC-PUB-13251",
    doi = "10.1103/PhysRevD.78.075028",
    journal = "Phys. Rev. D",
    volume = "78",
    pages = "075028",
    year = "2008"
}

@article{Harnik:2008ax,
    author = "Harnik, Roni and Wizansky, Tommer",
    title = "{Signals of New Physics in the Underlying Event}",
    eprint = "0810.3948",
    archivePrefix = "arXiv",
    primaryClass = "hep-ph",
    reportNumber = "SLAC-PUB-13483, SU-ITP-26-08",
    doi = "10.1103/PhysRevD.80.075015",
    journal = "Phys. Rev. D",
    volume = "80",
    pages = "075015",
    year = "2009"
}

@article{Cai:2008au,
    author = "Cai, Haiying and Cheng, Hsin-Chia and Terning, John",
    title = "{A Quirky Little Higgs Model}",
    eprint = "0812.0843",
    archivePrefix = "arXiv",
    primaryClass = "hep-ph",
    doi = "10.1088/1126-6708/2009/05/045",
    journal = "JHEP",
    volume = "05",
    pages = "045",
    year = "2009"
}

@article{Craig:2010au,
    author = "Craig, Nathaniel and March-Russell, John",
    title = "{Axion-Assisted Electroweak Baryogenesis}",
    eprint = "1007.0019",
    archivePrefix = "arXiv",
    primaryClass = "hep-ph",
    reportNumber = "SU-ITP-10-18, OUTP-10-10P",
    month = "7",
    year = "2010"
}

@article{D0:2010kkd,
    author = "Abazov, V. M. and others",
    collaboration = "D0",
    title = "{Search for New Fermions ('Quirks') at the Fermilab Tevatron Collider}",
    eprint = "1008.3547",
    archivePrefix = "arXiv",
    primaryClass = "hep-ex",
    reportNumber = "FERMILAB-PUB-10-324-E",
    doi = "10.1103/PhysRevLett.105.211803",
    journal = "Phys. Rev. Lett.",
    volume = "105",
    pages = "211803",
    year = "2010"
}

@article{Martin:2010kk,
    author = "Martin, Stephen P.",
    title = "{Quirks in Supersymmetry with Gauge Coupling Unification}",
    eprint = "1012.2072",
    archivePrefix = "arXiv",
    primaryClass = "hep-ph",
    reportNumber = "FERMILAB-PUB-10-733-T",
    doi = "10.1103/PhysRevD.83.035019",
    journal = "Phys. Rev. D",
    volume = "83",
    pages = "035019",
    year = "2011"
}

@article{Farina:2017cts,
    author = "Farina, Marco and Low, Matthew",
    title = "{Constraining Quirky Tracks with Conventional Searches}",
    eprint = "1703.00912",
    archivePrefix = "arXiv",
    primaryClass = "hep-ph",
    doi = "10.1103/PhysRevLett.119.111801",
    journal = "Phys. Rev. Lett.",
    volume = "119",
    number = "11",
    pages = "111801",
    year = "2017"
}

@article{Cheng:2018gvu,
    author = "Cheng, Hsin-Chia and Li, Lingfeng and Salvioni, Ennio and Verhaaren, Christopher B.",
    title = "{Singlet Scalar Top Partners from Accidental Supersymmetry}",
    eprint = "1803.03651",
    archivePrefix = "arXiv",
    primaryClass = "hep-ph",
    reportNumber = "TUM-HEP-1134-18",
    doi = "10.1007/JHEP05(2018)057",
    journal = "JHEP",
    volume = "05",
    pages = "057",
    year = "2018"
}

@article{Li:2019wce,
    author = "Li, Jinmian and Li, Tianjun and Pei, Junle and Zhang, Wenxing",
    title = "{Uncovering quirk signal via energy loss inside tracker}",
    eprint = "1911.02223",
    archivePrefix = "arXiv",
    primaryClass = "hep-ph",
    doi = "10.1103/PhysRevD.102.056006",
    journal = "Phys. Rev. D",
    volume = "102",
    number = "5",
    pages = "056006",
    year = "2020"
}

@article{Li:2020aoq,
    author = "Li, Jinmian and Li, Tianjun and Pei, Junle and Zhang, Wenxing",
    title = "{The quirk trajectory}",
    eprint = "2002.07503",
    archivePrefix = "arXiv",
    primaryClass = "hep-ph",
    doi = "10.1140/epjc/s10052-020-8209-y",
    journal = "Eur. Phys. J. C",
    volume = "80",
    number = "7",
    pages = "651",
    year = "2020"
}

@article{Barela:2023exp,
    author = "Barela, Mario W. and Capdevilla, Rodolfo",
    title = "{Di-Higgs Signatures in Neutral Naturalness}",
    eprint = "2310.02317",
    archivePrefix = "arXiv",
    primaryClass = "hep-ph",
    reportNumber = "FERMILAB-PUB-23-472",
    doi = "10.1007/JHEP02(2024)050",
    journal = "JHEP",
    volume = "2024",
    number = "02",
    pages = "050",
    year = "2024"
}

@article{Li:2023jrt,
    author = "Li, Jinmian and Liao, Xufei and Ni, Jian and Pei, Junle",
    title = "{Detection prospects of long-lived quirk pairs at the LHC far detectors}",
    eprint = "2311.15486",
    archivePrefix = "arXiv",
    primaryClass = "hep-ph",
    doi = "10.1103/PhysRevD.109.095005",
    journal = "Phys. Rev. D",
    volume = "109",
    number = "9",
    pages = "095005",
    year = "2024"
}

@article{Sha:2024hzq,
    author = "Sha, Qiyu and Murnane, Daniel and Fieg, Max and Tong, Shelley and Zakharyan, Mark and Fang, Yaquan and Whiteson, Daniel",
    title = "{Learning to Reconstruct Quirky Tracks}",
    eprint = "2410.00269",
    archivePrefix = "arXiv",
    primaryClass = "hep-ex",
    month = "9",
    year = "2024"
}

@article{Forsyth:2025wks,
    author = "Forsyth, Joshua and Low, Matthew and Tenney, Carson and Verhaaren, Christopher B.",
    title = "{Visible collider signals of natural quirks}",
    eprint = "2504.02940",
    archivePrefix = "arXiv",
    primaryClass = "hep-ph",
    doi = "10.1103/34xj-vz82",
    journal = "Phys. Rev. D",
    volume = "111",
    number = "11",
    pages = "115021",
    year = "2025"
}

@article{Curtin:2025ngf,
    author = "Curtin, David and Dreyer, Sascha and Fust{\'e} Costa, Max and Heim, Sarah and Kasieczka, Gregor and Moureaux, Louis and Rousso, David and Shih, David and Sommerhalder, Manuel",
    title = "{Soft-unclustered-energy patterns from quirks}",
    eprint = "2506.11192",
    archivePrefix = "arXiv",
    primaryClass = "hep-ph",
    reportNumber = "DESY-25-081-0, DESY-25-081, CERN-TH-2025-119",
    doi = "10.1103/g7c7-6qh2",
    journal = "Phys. Rev. D",
    volume = "113",
    number = "1",
    pages = "015010",
    year = "2026"
}

@article{Asadi:2025btr,
    author = "Asadi, Pouya and Kribs, Graham D. and Luty, Markus A.",
    title = "{Quirks Live in Cool Universes}",
    eprint = "2512.20696",
    archivePrefix = "arXiv",
    primaryClass = "hep-ph",
    reportNumber = "CERN-TH-2025-243",
    month = "12",
    year = "2025"
}

@article{CidVidal:2026ist,
    author = "Cid Vidal, Xabier and G{\'o}mez, Miguel Fern{\'a}ndez and Low, Matthew and Cal, Alejandro Novo and Tsai, Yuhsin and V{\'a}zquez Sierra, Carlos",
    title = "{Searching for Quirks at LHCb}",
    eprint = "2601.09023",
    archivePrefix = "arXiv",
    primaryClass = "hep-ph",
    month = "1",
    year = "2026"
}

@article{Hur:2007uz,
    author = "Hur, Taeil and Jung, Dong-Won and Ko, P. and Lee, Jae Yong",
    title = "{Electroweak symmetry breaking and cold dark matter from strongly interacting hidden sector}",
    eprint = "0709.1218",
    archivePrefix = "arXiv",
    primaryClass = "hep-ph",
    doi = "10.1016/j.physletb.2010.12.047",
    journal = "Phys. Lett. B",
    volume = "696",
    pages = "262--265",
    year = "2011"
}

@article{Ryttov:2008xe,
    author = "Ryttov, Thomas A. and Sannino, Francesco",
    title = "{Ultra Minimal Technicolor and its Dark Matter TIMP}",
    eprint = "0809.0713",
    archivePrefix = "arXiv",
    primaryClass = "hep-ph",
    reportNumber = "CERN-PH-TH-2008-188",
    doi = "10.1103/PhysRevD.78.115010",
    journal = "Phys. Rev. D",
    volume = "78",
    pages = "115010",
    year = "2008"
}

@article{Nussinov:1985xr,
    author = "Nussinov, S.",
    title = "{TECHNOCOSMOLOGY: COULD A TECHNIBARYON EXCESS PROVIDE A 'NATURAL' MISSING MASS CANDIDATE?}",
    reportNumber = "CLNS-85/703",
    doi = "10.1016/0370-2693(85)90689-6",
    journal = "Phys. Lett. B",
    volume = "165",
    pages = "55--58",
    year = "1985"
}

@article{Agashe:2004ci,
    author = "Agashe, Kaustubh and Servant, Geraldine",
    title = "{Warped unification, proton stability and dark matter}",
    eprint = "hep-ph/0403143",
    archivePrefix = "arXiv",
    reportNumber = "ANL-HEP-PR-04-19, EFI-04-06",
    doi = "10.1103/PhysRevLett.93.231805",
    journal = "Phys. Rev. Lett.",
    volume = "93",
    pages = "231805",
    year = "2004"
}

@article{Agashe:2007jb,
    author = "Agashe, Kaustubh and Falkowski, Adam and Low, Ian and Servant, Geraldine",
    title = "{KK Parity in Warped Extra Dimension}",
    eprint = "0712.2455",
    archivePrefix = "arXiv",
    primaryClass = "hep-ph",
    reportNumber = "ANL-HEP-PR-07-104, CERN-PH-TH-2007-247, SPHT-T07-153, SU-4252-863, UMD-PP-07-008",
    doi = "10.1088/1126-6708/2008/04/027",
    journal = "JHEP",
    volume = "04",
    pages = "027",
    year = "2008"
}

@article{Cheng:2003ju,
    author = "Cheng, Hsin-Chia and Low, Ian",
    title = "{TeV symmetry and the little hierarchy problem}",
    eprint = "hep-ph/0308199",
    archivePrefix = "arXiv",
    reportNumber = "HUTP-03-A051",
    doi = "10.1088/1126-6708/2003/09/051",
    journal = "JHEP",
    volume = "09",
    pages = "051",
    year = "2003"
}

@article{Hur:2011sv,
    author = "Hur, Taeil and Ko, P.",
    title = "{Scale invariant extension of the standard model with strongly interacting hidden sector}",
    eprint = "1103.2571",
    archivePrefix = "arXiv",
    primaryClass = "hep-ph",
    reportNumber = "KIAS-PREPRINT:-P11010",
    doi = "10.1103/PhysRevLett.106.141802",
    journal = "Phys. Rev. Lett.",
    volume = "106",
    pages = "141802",
    year = "2011"
}

@article{Lewis:2011zb,
    author = "Lewis, Randy and Pica, Claudio and Sannino, Francesco",
    title = "{Light Asymmetric Dark Matter on the Lattice: SU(2) Technicolor with Two Fundamental Flavors}",
    eprint = "1109.3513",
    archivePrefix = "arXiv",
    primaryClass = "hep-ph",
    reportNumber = "CP3-ORIGINS-2011-28, DIAS-2011-15",
    doi = "10.1103/PhysRevD.85.014504",
    journal = "Phys. Rev. D",
    volume = "85",
    pages = "014504",
    year = "2012"
}

@article{Huo:2015nwa,
    author = "Huo, Ran and Matsumoto, Shigeki and Sming Tsai, Yue-Lin and Yanagida, Tsutomu T.",
    title = "{A scenario of heavy but visible baryonic dark matter}",
    eprint = "1506.06929",
    archivePrefix = "arXiv",
    primaryClass = "hep-ph",
    reportNumber = "IPMU15-0093",
    doi = "10.1007/JHEP09(2016)162",
    journal = "JHEP",
    volume = "09",
    pages = "162",
    year = "2016"
}

@article{DiMauro:2026owr,
    author = "Di Mauro, Mattia and Gemmell, Caleb and Batz, Austin and Curtin, David and Donato, Fiorenza and Fornengo, Nicolao and Kribs, Graham D.",
    title = "{Enhanced Cosmic-Ray Antinuclei Fluxes with Dark Matter Annihilation into SUEPs}",
    eprint = "2602.15132",
    archivePrefix = "arXiv",
    primaryClass = "hep-ph",
    reportNumber = "CERN-TH-2026-012",
    month = "2",
    year = "2026"
}

@article{Brax:2019koq,
    author = "Brax, Philippe and Fichet, Sylvain and Tanedo, Philip",
    title = "{The Warped Dark Sector}",
    eprint = "1906.02199",
    archivePrefix = "arXiv",
    primaryClass = "hep-ph",
    reportNumber = "UCR-TR-2019-FLIP-NCC-1804",
    doi = "10.1016/j.physletb.2019.135012",
    journal = "Phys. Lett. B",
    volume = "798",
    pages = "135012",
    year = "2019"
}

@article{Costantino:2020msc,
    author = "Costantino, Alexandria and Fichet, Sylvain and Tanedo, Philip",
    title = "{Effective Field Theory in AdS: Continuum Regime, Soft Bombs, and IR Emergence}",
    eprint = "2002.12335",
    archivePrefix = "arXiv",
    primaryClass = "hep-th",
    reportNumber = "UCR-TR-2020-FLIP-IG-11",
    doi = "10.1103/PhysRevD.102.115038",
    journal = "Phys. Rev. D",
    volume = "102",
    number = "11",
    pages = "115038",
    year = "2020"
}

@article{Chaffey:2021tmj,
    author = "Chaffey, Ian and Fichet, Sylvain and Tanedo, Philip",
    title = "{Continuum-Mediated Self-Interacting Dark Matter}",
    eprint = "2102.05674",
    archivePrefix = "arXiv",
    primaryClass = "hep-ph",
    reportNumber = "UCR-TR-2021-FLIP-MSE-6",
    doi = "10.1007/JHEP06(2021)008",
    journal = "JHEP",
    volume = "06",
    pages = "008",
    year = "2021"
}

@article{Barron:2021btf,
    author = "Barron, Jared and Curtin, David and Kasieczka, Gregor and Plehn, Tilman and Spourdalakis, Aris",
    title = "{Unsupervised hadronic SUEP at the LHC}",
    eprint = "2107.12379",
    archivePrefix = "arXiv",
    primaryClass = "hep-ph",
    doi = "10.1007/JHEP12(2021)129",
    journal = "JHEP",
    volume = "12",
    pages = "129",
    year = "2021"
}

@article{DiPetrillo:2022qsb,
    author = "Di Petrillo, K. F. and Farr, J. N. and Guo, C. and Holmes, T. R. and Nelson, J. and Pachal, K.",
    title = "{Optimizing trigger-level track reconstruction for sensitivity to exotic signatures}",
    eprint = "2211.05720",
    archivePrefix = "arXiv",
    primaryClass = "hep-ex",
    reportNumber = "FERMILAB-PUB-22-843-PPD",
    doi = "10.1007/JHEP02(2023)034",
    journal = "JHEP",
    volume = "02",
    pages = "034",
    year = "2023"
}

@article{Tsai:2020vpi,
    author = "Tsai, Yu-Dai and McGehee, Robert and Murayama, Hitoshi",
    title = "{Resonant Self-Interacting Dark Matter from Dark QCD}",
    eprint = "2008.08608",
    archivePrefix = "arXiv",
    primaryClass = "hep-ph",
    reportNumber = "FERMILAB-PUB-20-365-AE-T",
    doi = "10.1103/PhysRevLett.128.172001",
    journal = "Phys. Rev. Lett.",
    volume = "128",
    number = "17",
    pages = "172001",
    year = "2022"
}

@article{Hochberg:2014kqa,
    author = "Hochberg, Yonit and Kuflik, Eric and Murayama, Hitoshi and Volansky, Tomer and Wacker, Jay G.",
    title = "{Model for Thermal Relic Dark Matter of Strongly Interacting Massive Particles}",
    eprint = "1411.3727",
    archivePrefix = "arXiv",
    primaryClass = "hep-ph",
    doi = "10.1103/PhysRevLett.115.021301",
    journal = "Phys. Rev. Lett.",
    volume = "115",
    number = "2",
    pages = "021301",
    year = "2015"
}

@article{Bernreuther:2023kcg,
    author = "Bernreuther, Elias and Hemme, Nicoline and Kahlhoefer, Felix and Kulkarni, Suchita",
    title = "{Dark matter relic density in strongly interacting dark sectors with light vector mesons}",
    eprint = "2311.17157",
    archivePrefix = "arXiv",
    primaryClass = "hep-ph",
    reportNumber = "FERMILAB-PUB-23-744-T, TTP23-057, P3H-23-096",
    doi = "10.1103/PhysRevD.110.035009",
    journal = "Phys. Rev. D",
    volume = "110",
    number = "3",
    pages = "035009",
    year = "2024"
}

@article{Faraggi:2000pv,
    author = "Faraggi, Alon E. and Pospelov, Maxim",
    title = "{Selfinteracting dark matter from the hidden heterotic string sector}",
    eprint = "hep-ph/0008223",
    archivePrefix = "arXiv",
    reportNumber = "TPI-MINN-00-42, UMN-TH-1918-00, OUTP-00-37P",
    doi = "10.1016/S0927-6505(01)00121-9",
    journal = "Astropart. Phys.",
    volume = "16",
    pages = "451--461",
    year = "2002"
}

@article{Baldes:2020kam,
    author = "Baldes, Iason and Gouttenoire, Yann and Sala, Filippo",
    title = "{String Fragmentation in Supercooled Confinement and Implications for Dark Matter}",
    eprint = "2007.08440",
    archivePrefix = "arXiv",
    primaryClass = "hep-ph",
    reportNumber = "DESY-20-122",
    doi = "10.1007/JHEP04(2021)278",
    journal = "JHEP",
    volume = "04",
    pages = "278",
    year = "2021"
}

@article{Forestell:2016qhc,
    author = "Forestell, Lindsay and Morrissey, David E. and Sigurdson, Kris",
    title = "{Non-Abelian Dark Forces and the Relic Densities of Dark Glueballs}",
    eprint = "1605.08048",
    archivePrefix = "arXiv",
    primaryClass = "hep-ph",
    doi = "10.1103/PhysRevD.95.015032",
    journal = "Phys. Rev. D",
    volume = "95",
    number = "1",
    pages = "015032",
    year = "2017"
}

@article{Forestell:2017wov,
    author = "Forestell, Lindsay and Morrissey, David E. and Sigurdson, Kris",
    title = "{Cosmological Bounds on Non-Abelian Dark Forces}",
    eprint = "1710.06447",
    archivePrefix = "arXiv",
    primaryClass = "hep-ph",
    doi = "10.1103/PhysRevD.97.075029",
    journal = "Phys. Rev. D",
    volume = "97",
    number = "7",
    pages = "075029",
    year = "2018"
}

@article{Carenza:2023eua,
    author = "Carenza, Pierluca and Ferreira, Tassia and Pasechnik, Roman and Wang, Zhi-Wei",
    title = "{Glueball dark matter, precisely}",
    eprint = "2306.09510",
    archivePrefix = "arXiv",
    primaryClass = "hep-ph",
    doi = "10.1103/PhysRevD.108.123027",
    journal = "Phys. Rev. D",
    volume = "108",
    number = "12",
    pages = "123027",
    year = "2023"
}

@article{Bai:2013xga,
    author = "Bai, Yang and Schwaller, Pedro",
    title = "{Scale of dark QCD}",
    eprint = "1306.4676",
    archivePrefix = "arXiv",
    primaryClass = "hep-ph",
    doi = "10.1103/PhysRevD.89.063522",
    journal = "Phys. Rev. D",
    volume = "89",
    number = "6",
    pages = "063522",
    year = "2014"
}

@article{GarciaGarcia:2015pnn,
    author = "Garcia Garcia, Isabel and Lasenby, Robert and March-Russell, John",
    title = "{Twin Higgs Asymmetric Dark Matter}",
    eprint = "1505.07410",
    archivePrefix = "arXiv",
    primaryClass = "hep-ph",
    doi = "10.1103/PhysRevLett.115.121801",
    journal = "Phys. Rev. Lett.",
    volume = "115",
    number = "12",
    pages = "121801",
    year = "2015"
}

@article{Farina:2015uea,
    author = "Farina, Marco",
    title = "{Asymmetric Twin Dark Matter}",
    eprint = "1506.03520",
    archivePrefix = "arXiv",
    primaryClass = "hep-ph",
    doi = "10.1088/1475-7516/2015/11/017",
    journal = "JCAP",
    volume = "11",
    pages = "017",
    year = "2015"
}

@article{Farina:2016ndq,
    author = "Farina, Marco and Monteux, Angelo and Shin, Chang Sub",
    title = "{Twin mechanism for baryon and dark matter asymmetries}",
    eprint = "1604.08211",
    archivePrefix = "arXiv",
    primaryClass = "hep-ph",
    doi = "10.1103/PhysRevD.94.035017",
    journal = "Phys. Rev. D",
    volume = "94",
    number = "3",
    pages = "035017",
    year = "2016"
}

@article{Lonsdale:2017mzg,
    author = "Lonsdale, Stephen J. and Schroor, Martine and Volkas, Raymond R.",
    title = "{Asymmetric Dark Matter and the hadronic spectra of hidden QCD}",
    eprint = "1704.05213",
    archivePrefix = "arXiv",
    primaryClass = "hep-ph",
    doi = "10.1103/PhysRevD.96.055027",
    journal = "Phys. Rev. D",
    volume = "96",
    number = "5",
    pages = "055027",
    year = "2017"
}

@article{Lonsdale:2018xwd,
    author = "Lonsdale, Stephen J. and Volkas, Raymond R.",
    title = "{Comprehensive asymmetric dark matter model}",
    eprint = "1801.05561",
    archivePrefix = "arXiv",
    primaryClass = "hep-ph",
    doi = "10.1103/PhysRevD.97.103510",
    journal = "Phys. Rev. D",
    volume = "97",
    number = "10",
    pages = "103510",
    year = "2018"
}

@article{Ibe:2018juk,
    author = "Ibe, Masahiro and Kamada, Ayuki and Kobayashi, Shin and Nakano, Wakutaka",
    title = "{Composite Asymmetric Dark Matter with a Dark Photon Portal}",
    eprint = "1805.06876",
    archivePrefix = "arXiv",
    primaryClass = "hep-ph",
    doi = "10.1007/JHEP11(2018)203",
    journal = "JHEP",
    volume = "11",
    pages = "203",
    year = "2018"
}

@article{Ritter:2022opo,
    author = "Ritter, Alexander C. and Volkas, Raymond R.",
    title = "{Exploring the cosmological dark matter coincidence using infrared fixed points}",
    eprint = "2210.11011",
    archivePrefix = "arXiv",
    primaryClass = "hep-ph",
    doi = "10.1103/PhysRevD.107.015029",
    journal = "Phys. Rev. D",
    volume = "107",
    number = "1",
    pages = "015029",
    year = "2023"
}

@article{Alonso-Alvarez:2023bat,
    author = "Alonso-{\'A}lvarez, Gonzalo and Curtin, David and Rasovic, Andrija and Yuan, Zhihan",
    title = "{Baryogenesis through asymmetric reheating in the mirror twin Higgs}",
    eprint = "2311.06341",
    archivePrefix = "arXiv",
    primaryClass = "hep-ph",
    doi = "10.1007/JHEP05(2024)069",
    journal = "JHEP",
    volume = "05",
    pages = "069",
    year = "2024"
}

@article{Ritter:2024sqv,
    author = "Ritter, Alexander C. and Volkas, Raymond R.",
    title = "{Explaining the cosmological dark matter coincidence in asymmetric dark QCD}",
    eprint = "2404.05999",
    archivePrefix = "arXiv",
    primaryClass = "hep-ph",
    doi = "10.1103/PhysRevD.110.015032",
    journal = "Phys. Rev. D",
    volume = "110",
    number = "1",
    pages = "015032",
    year = "2024"
}

@article{Cox:2025wxk,
    author = "Cox, Peter and P{\'e}rez, Rafael E. and Volkas, Raymond R.",
    title = "{A new idea for relating the asymmetric dark matter mass scale to the proton mass}",
    eprint = "2512.14119",
    archivePrefix = "arXiv",
    primaryClass = "hep-ph",
    month = "12",
    year = "2025"
}

@article{Halverson:2016nfq,
    author = "Halverson, James and Nelson, Brent D. and Ruehle, Fabian",
    title = "{String Theory and the Dark Glueball Problem}",
    eprint = "1609.02151",
    archivePrefix = "arXiv",
    primaryClass = "hep-ph",
    reportNumber = "DESY-16-170",
    doi = "10.1103/PhysRevD.95.043527",
    journal = "Phys. Rev. D",
    volume = "95",
    number = "4",
    pages = "043527",
    year = "2017"
}

@article{Halverson:2018olu,
    author = "Halverson, James and Nelson, Brent D. and Ruehle, Fabian and Salinas, Gustavo",
    title = "{Dark Glueballs and their Ultralight Axions}",
    eprint = "1805.06011",
    archivePrefix = "arXiv",
    primaryClass = "hep-ph",
    doi = "10.1103/PhysRevD.98.043502",
    journal = "Phys. Rev. D",
    volume = "98",
    number = "4",
    pages = "043502",
    year = "2018"
}

@article{Li:2026nse,
    author = "Li, Ji-Wei and Pasechnik, Roman and Wang, Wei and Wang, Zhi-Wei",
    title = "{Dark Glueball Direct Detection}",
    eprint = "2602.18753",
    archivePrefix = "arXiv",
    primaryClass = "hep-ph",
    month = "2",
    year = "2026"
}

@article{CMS:2014tzs,
    author = "Chatrchyan, Serguei and others",
    collaboration = "CMS",
    title = "{Search for New Physics in the Multijet and Missing Transverse Momentum Final State in Proton-Proton Collisions at $\sqrt{s}$= 8 TeV}",
    eprint = "1402.4770",
    archivePrefix = "arXiv",
    primaryClass = "hep-ex",
    reportNumber = "CMS-SUS-13-012, CERN-PH-EP-2014-015",
    doi = "10.1007/JHEP06(2014)055",
    journal = "JHEP",
    volume = "06",
    pages = "055",
    year = "2014"
}

@article{ATLAS:2014jxt,
    author = "Aad, Georges and others",
    collaboration = "ATLAS",
    title = "{Search for squarks and gluinos with the ATLAS detector in final states with jets and missing transverse momentum using $\sqrt{s}=8$ TeV proton--proton collision data}",
    eprint = "1405.7875",
    archivePrefix = "arXiv",
    primaryClass = "hep-ex",
    reportNumber = "CERN-PH-EP-2014-093",
    doi = "10.1007/JHEP09(2014)176",
    journal = "JHEP",
    volume = "09",
    pages = "176",
    year = "2014"
}

@article{Fox:2011qd,
    author = "Fox, Patrick J. and Liu, Jia and Tucker-Smith, David and Weiner, Neal",
    title = "{An Effective Z'}",
    eprint = "1104.4127",
    archivePrefix = "arXiv",
    primaryClass = "hep-ph",
    reportNumber = "FERMILAB-PUB-11-179-T",
    doi = "10.1103/PhysRevD.84.115006",
    journal = "Phys. Rev. D",
    volume = "84",
    pages = "115006",
    year = "2011"
}

@article{ATLAS:2019fgd,
    author = "Aad, Georges and others",
    collaboration = "ATLAS",
    title = "{Search for new resonances in mass distributions of jet pairs using 139 fb$^{-1}$ of $pp$ collisions at $\sqrt{s}=13$ TeV with the ATLAS detector}",
    eprint = "1910.08447",
    archivePrefix = "arXiv",
    primaryClass = "hep-ex",
    reportNumber = "CERN-EP-2019-162",
    doi = "10.1007/JHEP03(2020)145",
    journal = "JHEP",
    volume = "03",
    pages = "145",
    year = "2020"
}

@article{CMS:2019gwf,
    author = "Sirunyan, Albert M and others",
    collaboration = "CMS",
    title = "{Search for high mass dijet resonances with a new background prediction method in proton-proton collisions at $\sqrt{s} =$ 13 TeV}",
    eprint = "1911.03947",
    archivePrefix = "arXiv",
    primaryClass = "hep-ex",
    reportNumber = "CMS-EXO-19-012, CERN-EP-2019-222",
    doi = "10.1007/JHEP05(2020)033",
    journal = "JHEP",
    volume = "05",
    pages = "033",
    year = "2020"
}

@article{ATLAS:2023tkt,
    author = "Aad, Georges and others",
    collaboration = "ATLAS",
    title = "{Combination of searches for invisible decays of the Higgs boson using 139 fb{\ensuremath{-}}1 of proton-proton collision data at s=13 TeV collected with the ATLAS experiment}",
    eprint = "2301.10731",
    archivePrefix = "arXiv",
    primaryClass = "hep-ex",
    reportNumber = "CERN-EP-2022-289",
    doi = "10.1016/j.physletb.2023.137963",
    journal = "Phys. Lett. B",
    volume = "842",
    pages = "137963",
    year = "2023"
}

@article{CMS:2014wda,
    author = "Khachatryan, Vardan and others",
    collaboration = "CMS",
    title = "{Search for Long-Lived Neutral Particles Decaying to Quark-Antiquark Pairs in Proton-Proton Collisions at $\sqrt{s} =$ 8 TeV}",
    eprint = "1411.6530",
    archivePrefix = "arXiv",
    primaryClass = "hep-ex",
    reportNumber = "CMS-EXO-12-038, CERN-PH-EP-2014-256",
    doi = "10.1103/PhysRevD.91.012007",
    journal = "Phys. Rev. D",
    volume = "91",
    number = "1",
    pages = "012007",
    year = "2015"
}

@article{ATLAS:2014fzk,
    author = "Aad, Georges and others",
    collaboration = "ATLAS",
    title = "{Search for long-lived neutral particles decaying into lepton jets in proton-proton collisions at $ \sqrt{s}=8 $ TeV with the ATLAS detector}",
    eprint = "1409.0746",
    archivePrefix = "arXiv",
    primaryClass = "hep-ex",
    reportNumber = "CERN-PH-EP-2014-209",
    doi = "10.1007/JHEP11(2014)088",
    journal = "JHEP",
    volume = "11",
    pages = "088",
    year = "2014"
}

@article{LHCNewPhysicsWorkingGroup:2011mji,
    author = "Alves, Daniele",
    editor = "Arkani-Hamed, Nima and others",
    collaboration = "LHC New Physics Working Group",
    title = "{Simplified Models for LHC New Physics Searches}",
    eprint = "1105.2838",
    archivePrefix = "arXiv",
    primaryClass = "hep-ph",
    reportNumber = "SLAC-PUB-15045, FERMILAB-PUB-11-842-A-PPD",
    doi = "10.1088/0954-3899/39/10/105005",
    journal = "J. Phys. G",
    volume = "39",
    pages = "105005",
    year = "2012"
}

@article{Abdallah:2014hon,
    author = "Abdallah, Jalal and others",
    title = "{Simplified Models for Dark Matter and Missing Energy Searches at the LHC}",
    eprint = "1409.2893",
    archivePrefix = "arXiv",
    primaryClass = "hep-ph",
    reportNumber = "FERMILAB-FN-0990-PPD",
    month = "9",
    year = "2014"
}

@article{Abdallah:2015ter,
    author = "Abdallah, Jalal and others",
    title = "{Simplified Models for Dark Matter Searches at the LHC}",
    eprint = "1506.03116",
    archivePrefix = "arXiv",
    primaryClass = "hep-ph",
    reportNumber = "FERMILAB-PUB-15-283-CD, CERN-PH-TH-2015-139",
    doi = "10.1016/j.dark.2015.08.001",
    journal = "Phys. Dark Univ.",
    volume = "9-10",
    pages = "8--23",
    year = "2015"
}

@article{Grzadkowski:2010es,
    author = "Grzadkowski, B. and Iskrzynski, M. and Misiak, M. and Rosiek, J.",
    title = "{Dimension-Six Terms in the Standard Model Lagrangian}",
    eprint = "1008.4884",
    archivePrefix = "arXiv",
    primaryClass = "hep-ph",
    reportNumber = "IFT-9-2010, TTP10-35",
    doi = "10.1007/JHEP10(2010)085",
    journal = "JHEP",
    volume = "10",
    pages = "085",
    year = "2010"
}

@article{Pomarol:2013zra,
    author = "Pomarol, Alex and Riva, Francesco",
    title = "{Towards the Ultimate SM Fit to Close in on Higgs Physics}",
    eprint = "1308.2803",
    archivePrefix = "arXiv",
    primaryClass = "hep-ph",
    doi = "10.1007/JHEP01(2014)151",
    journal = "JHEP",
    volume = "01",
    pages = "151",
    year = "2014"
}

@article{Falkowski:2015fla,
    author = "Falkowski, Adam",
    title = "{Effective field theory approach to LHC Higgs data}",
    eprint = "1505.00046",
    archivePrefix = "arXiv",
    primaryClass = "hep-ph",
    reportNumber = "LPT-ORSAY-15-33",
    doi = "10.1007/s12043-016-1251-5",
    journal = "Pramana",
    volume = "87",
    number = "3",
    pages = "39",
    year = "2016"
}

@article{Mueller:1982cq,
    author = "Mueller, Alfred H.",
    title = "{Multiplicity and Hadron Distributions in QCD Jets: Nonleading Terms}",
    reportNumber = "CU-TP-247",
    doi = "10.1016/0550-3213(83)90176-1",
    journal = "Nucl. Phys. B",
    volume = "213",
    pages = "85--108",
    year = "1983"
}

@article{Bierlich:2022pfr,
    author = "Bierlich, Christian and others",
    title = "{A comprehensive guide to the physics and usage of PYTHIA 8.3}",
    eprint = "2203.11601",
    archivePrefix = "arXiv",
    primaryClass = "hep-ph",
    reportNumber = "LU-TP 22-16, MCNET-22-04, FERMILAB-PUB-22-227-SCD",
    doi = "10.21468/SciPostPhysCodeb.8",
    journal = "SciPost Phys. Codeb.",
    volume = "2022",
    pages = "8",
    year = "2022"
}

@article{Redi:2018muu,
    author = "Redi, Michele and Tesi, Andrea",
    title = "{Cosmological Production of Dark Nuclei}",
    eprint = "1812.08784",
    archivePrefix = "arXiv",
    primaryClass = "hep-ph",
    doi = "10.1007/JHEP04(2019)108",
    journal = "JHEP",
    volume = "04",
    pages = "108",
    year = "2019"
}

@article{Buttazzo:2019mvl,
    author = "Buttazzo, Dario and Di Luzio, Luca and Ghorbani, Parsa and Gross, Christian and Landini, Giacomo and Strumia, Alessandro and Teresi, Daniele and Wang, Jin-Wei",
    title = "{Scalar gauge dynamics and Dark Matter}",
    eprint = "1911.04502",
    archivePrefix = "arXiv",
    primaryClass = "hep-ph",
    reportNumber = "DESY 19-197, DESY-19-197",
    doi = "10.1007/JHEP01(2020)130",
    journal = "JHEP",
    volume = "01",
    pages = "130",
    year = "2020"
}

@article{Carvunis:2020exc,
    author = "Carvunis, Alexandre and Guadagnoli, Diego and Reboud, M{\'e}ril and Stangl, Peter",
    title = "{Composite Dark Matter and a horizontal symmetry}",
    eprint = "2007.11931",
    archivePrefix = "arXiv",
    primaryClass = "hep-ph",
    reportNumber = "LAPTH-036/20",
    doi = "10.1007/JHEP02(2021)056",
    journal = "JHEP",
    volume = "02",
    pages = "056",
    year = "2021"
}

@article{McKeen:2024trt,
    author = "McKeen, David and Mizuta, Riku and Morrissey, David E. and Shamma, Michael",
    title = "{Dark matter from dark glueball dominance}",
    eprint = "2406.18635",
    archivePrefix = "arXiv",
    primaryClass = "hep-ph",
    doi = "10.1103/PhysRevD.111.015044",
    journal = "Phys. Rev. D",
    volume = "111",
    number = "1",
    pages = "015044",
    year = "2025"
}

@article{Fleming:2024flc,
    author = "Fleming, George T. and Kribs, Graham D. and Neil, Ethan T. and Schaich, David and Vranas, Pavlos M.",
    title = "{Hyperstealth dark matter and long-lived particles}",
    eprint = "2412.14540",
    archivePrefix = "arXiv",
    primaryClass = "hep-ph",
    reportNumber = "LLNL-JRNL-2001590, FERMILAB-PUB-24-0954-T",
    doi = "10.1103/9nv9-rqq7",
    journal = "Phys. Rev. D",
    volume = "112",
    number = "7",
    pages = "075004",
    year = "2025"
}

@article{Kuwahara:2025eeo,
    author = "Kuwahara, Takumi and Uchida, Yoshiki",
    title = "{Composite Asymmetric Dark Matter from Primordial Black Holes}",
    eprint = "2511.16354",
    archivePrefix = "arXiv",
    primaryClass = "hep-ph",
    month = "11",
    year = "2025"
}

@article{Chacko:2018vss,
    author = "Chacko, Zackaria and Curtin, David and Geller, Michael and Tsai, Yuhsin",
    title = "{Cosmological Signatures of a Mirror Twin Higgs}",
    eprint = "1803.03263",
    archivePrefix = "arXiv",
    primaryClass = "hep-ph",
    reportNumber = "FERMILAB-PUB-18-075-T",
    doi = "10.1007/JHEP09(2018)163",
    journal = "JHEP",
    volume = "09",
    pages = "163",
    year = "2018"
}

@article{Bittar:2023kdl,
    author = "Bittar, Pedro and Burdman, Gustavo and Kiriliuk, Larissa",
    title = "{Baryogenesis and dark matter in the mirror twin Higgs}",
    eprint = "2307.04662",
    archivePrefix = "arXiv",
    primaryClass = "hep-ph",
    doi = "10.1007/JHEP11(2023)043",
    journal = "JHEP",
    volume = "11",
    pages = "043",
    year = "2023"
}

@article{Bahr:2008pv,
    author = "Bahr, M. and others",
    title = "{Herwig++ Physics and Manual}",
    eprint = "0803.0883",
    archivePrefix = "arXiv",
    primaryClass = "hep-ph",
    reportNumber = "CERN-PH-TH-2008-038, CAVENDISH-HEP-08-03, KA-TP-05-2008, DCPT-08-22, IPPP-08-11, CP3-08-05",
    doi = "10.1140/epjc/s10052-008-0798-9",
    journal = "Eur. Phys. J. C",
    volume = "58",
    pages = "639--707",
    year = "2008"
}

@article{Bellm:2015jjp,
    author = "Bellm, Johannes and others",
    title = "{Herwig 7.0/Herwig++ 3.0 release note}",
    eprint = "1512.01178",
    archivePrefix = "arXiv",
    primaryClass = "hep-ph",
    reportNumber = "CERN-PH-TH-2015-289, MAN-HEP-2015-15, IFJPAN-IV-2015-13, KA-TP-18-2015, DCPT-15-142, MCNET-15-28, IPPP-15-71, HERWIG-2015-01",
    doi = "10.1140/epjc/s10052-016-4018-8",
    journal = "Eur. Phys. J. C",
    volume = "76",
    number = "4",
    pages = "196",
    year = "2016"
}

@article{Bright-Thonney:2023gdl,
    author = "Bright-Thonney, Samuel and Nachman, Benjamin and Thaler, Jesse",
    title = "{Infrared-safe energy weighting does not guarantee small nonperturbative effects}",
    eprint = "2311.07652",
    archivePrefix = "arXiv",
    primaryClass = "hep-ph",
    reportNumber = "MIT-CTP 5641",
    doi = "10.1103/PhysRevD.110.014029",
    journal = "Phys. Rev. D",
    volume = "110",
    number = "1",
    pages = "014029",
    year = "2024"
}

@article{Manohar:1994kq,
    author = "Manohar, Aneesh V. and Wise, Mark B.",
    title = "{Power suppressed corrections to hadronic event shapes}",
    eprint = "hep-ph/9406392",
    archivePrefix = "arXiv",
    reportNumber = "UCSD-PTH-94-11, CALT-68-1937",
    doi = "10.1016/0370-2693(94)01504-6",
    journal = "Phys. Lett. B",
    volume = "344",
    pages = "407--412",
    year = "1995"
}

@article{Dreyer:2018nbf,
    author = "Dreyer, Fr{\'e}d{\'e}ric A. and Salam, Gavin P. and Soyez, Gr{\'e}gory",
    title = "{The Lund Jet Plane}",
    eprint = "1807.04758",
    archivePrefix = "arXiv",
    primaryClass = "hep-ph",
    reportNumber = "CERN-TH-2018-151",
    doi = "10.1007/JHEP12(2018)064",
    journal = "JHEP",
    volume = "12",
    pages = "064",
    year = "2018"
}

@article{Geller:2018biy,
    author = "Geller, Michael and Iwamoto, Sho and Lee, Gabriel and Shadmi, Yael and Telem, Ofri",
    title = "{Dark quarkonium formation in the early universe}",
    eprint = "1802.07720",
    archivePrefix = "arXiv",
    primaryClass = "hep-ph",
    doi = "10.1007/JHEP06(2018)135",
    journal = "JHEP",
    volume = "06",
    pages = "135",
    year = "2018"
}

@article{Robson:1977pm,
    author = "Robson, D.",
    title = "{A Basic Guide for the Glueball Spotter}",
    reportNumber = "LTH 33",
    doi = "10.1016/0550-3213(77)90110-9",
    journal = "Nucl. Phys. B",
    volume = "130",
    pages = "328--348",
    year = "1977"
}

@article{Donoghue:1980hw,
    author = "Donoghue, John F. and Johnson, K. and Li, Bing An",
    title = "{Low Mass Glueballs in the Meson Spectrum}",
    reportNumber = "UMHEP-139, MIT-CTP-891",
    doi = "10.1016/0370-2693(81)90560-8",
    journal = "Phys. Lett. B",
    volume = "99",
    pages = "416--420",
    year = "1981"
}

@article{Cornwall:1982zn,
    author = "Cornwall, J. M. and Soni, A.",
    title = "{Glueballs as Bound States of Massive Gluons}",
    reportNumber = "UCLA-82-TEP-3, NSF-ITP-82-61",
    doi = "10.1016/0370-2693(83)90481-1",
    journal = "Phys. Lett. B",
    volume = "120",
    pages = "431",
    year = "1983"
}

@article{Jaffe:1985qp,
    author = "Jaffe, R. L. and Johnson, K. and Ryzak, Z.",
    title = "{Qualitative Features of the Glueball Spectrum}",
    reportNumber = "MIT-CTP-1271",
    doi = "10.1016/0003-4916(86)90035-7",
    journal = "Annals Phys.",
    volume = "168",
    pages = "344",
    year = "1986"
}

@article{Ochs:2013gi,
    author = "Ochs, Wolfgang",
    title = "{The Status of Glueballs}",
    eprint = "1301.5183",
    archivePrefix = "arXiv",
    primaryClass = "hep-ph",
    reportNumber = "MPP-2013-44",
    doi = "10.1088/0954-3899/40/4/043001",
    journal = "J. Phys. G",
    volume = "40",
    pages = "043001",
    year = "2013"
}

@article{Berg:1982kp,
    author = "Berg, B. and Billoire, A.",
    title = "{Glueball Spectroscopy in Four-Dimensional SU(3) Lattice Gauge Theory. 1.}",
    reportNumber = "DESY-82-079",
    doi = "10.1016/0550-3213(83)90620-X",
    journal = "Nucl. Phys. B",
    volume = "221",
    pages = "109--140",
    year = "1983"
}

@article{Berg:1983qd,
    author = "Berg, B. and Billoire, A.",
    title = "{Glueball Spectroscopy in Four-dimensional SU(3) Lattice Gauge Theory. 2.}",
    reportNumber = "SACLAY-SPH-T-83-042",
    doi = "10.1016/0550-3213(83)90199-2",
    journal = "Nucl. Phys. B",
    volume = "226",
    pages = "405--416",
    year = "1983"
}

@article{Michael:1988jr,
    author = "Michael, Christopher and Teper, M.",
    title = "{The Glueball Spectrum in SU(3)}",
    reportNumber = "LTH-218",
    doi = "10.1016/0550-3213(89)90156-9",
    journal = "Nucl. Phys. B",
    volume = "314",
    pages = "347--362",
    year = "1989"
}

@article{Bali:1993fb,
    author = "Bali, G. S. and Schilling, K. and Hulsebos, A. and Irving, A. C. and Michael, Christopher and Stephenson, P. W.",
    collaboration = "UKQCD",
    title = "{A Comprehensive lattice study of SU(3) glueballs}",
    eprint = "hep-lat/9304012",
    archivePrefix = "arXiv",
    reportNumber = "LTH-303, WUB-93-10",
    doi = "10.1016/0370-2693(93)90948-H",
    journal = "Phys. Lett. B",
    volume = "309",
    pages = "378--384",
    year = "1993"
}

@article{Morningstar:1997ff,
    author = "Morningstar, Colin J. and Peardon, Mike J.",
    title = "{Efficient glueball simulations on anisotropic lattices}",
    eprint = "hep-lat/9704011",
    archivePrefix = "arXiv",
    reportNumber = "UCSD-PTH-97-05, UK-97-02",
    doi = "10.1103/PhysRevD.56.4043",
    journal = "Phys. Rev. D",
    volume = "56",
    pages = "4043--4061",
    year = "1997"
}

@article{Gregory:2012hu,
    author = "Gregory, E. and Irving, A. and Lucini, B. and McNeile, C. and Rago, A. and Richards, C. and Rinaldi, E.",
    title = "{Towards the glueball spectrum from unquenched lattice QCD}",
    eprint = "1208.1858",
    archivePrefix = "arXiv",
    primaryClass = "hep-lat",
    doi = "10.1007/JHEP10(2012)170",
    journal = "JHEP",
    volume = "10",
    pages = "170",
    year = "2012"
}

@article{Hochberg:2014dra,
    author = "Hochberg, Yonit and Kuflik, Eric and Volansky, Tomer and Wacker, Jay G.",
    title = "{Mechanism for Thermal Relic Dark Matter of Strongly Interacting Massive Particles}",
    eprint = "1402.5143",
    archivePrefix = "arXiv",
    primaryClass = "hep-ph",
    doi = "10.1103/PhysRevLett.113.171301",
    journal = "Phys. Rev. Lett.",
    volume = "113",
    pages = "171301",
    year = "2014"
}

@article{Pappadopulo:2016pkp,
    author = "Pappadopulo, Duccio and Ruderman, Joshua T. and Trevisan, Gabriele",
    title = "{Dark matter freeze-out in a nonrelativistic sector}",
    eprint = "1602.04219",
    archivePrefix = "arXiv",
    primaryClass = "hep-ph",
    doi = "10.1103/PhysRevD.94.035005",
    journal = "Phys. Rev. D",
    volume = "94",
    number = "3",
    pages = "035005",
    year = "2016"
}

@article{Farina:2016llk,
    author = "Farina, Marco and Pappadopulo, Duccio and Ruderman, Joshua T. and Trevisan, Gabriele",
    title = "{Phases of Cannibal Dark Matter}",
    eprint = "1607.03108",
    archivePrefix = "arXiv",
    primaryClass = "hep-ph",
    doi = "10.1007/JHEP12(2016)039",
    journal = "JHEP",
    volume = "12",
    pages = "039",
    year = "2016"
}

@article{Buen-Abad:2018mas,
    author = "Buen-Abad, Manuel A. and Emami, Razieh and Schmaltz, Martin",
    title = "{Cannibal Dark Matter and Large Scale Structure}",
    eprint = "1803.08062",
    archivePrefix = "arXiv",
    primaryClass = "hep-ph",
    doi = "10.1103/PhysRevD.98.083517",
    journal = "Phys. Rev. D",
    volume = "98",
    number = "8",
    pages = "083517",
    year = "2018"
}

@article{DAgnolo:2015ujb,
    author = "D'Agnolo, Raffaele Tito and Ruderman, Joshua T.",
    title = "{Light Dark Matter from Forbidden Channels}",
    eprint = "1505.07107",
    archivePrefix = "arXiv",
    primaryClass = "hep-ph",
    doi = "10.1103/PhysRevLett.115.061301",
    journal = "Phys. Rev. Lett.",
    volume = "115",
    number = "6",
    pages = "061301",
    year = "2015"
}

@article{Csaki:2021gfm,
    author = "Cs{\'a}ki, Csaba and Hong, Sungwoo and Kurup, Gowri and Lee, Seung J. and Perelstein, Maxim and Xue, Wei",
    title = "{Continuum dark matter}",
    eprint = "2105.07035",
    archivePrefix = "arXiv",
    primaryClass = "hep-ph",
    doi = "10.1103/PhysRevD.105.035025",
    journal = "Phys. Rev. D",
    volume = "105",
    number = "3",
    pages = "035025",
    year = "2022"
}

@article{Csaki:2021xpy,
    author = "Cs{\'a}ki, Csaba and Hong, Sungwoo and Kurup, Gowri and Lee, Seung J. and Perelstein, Maxim and Xue, Wei",
    title = "{Z-Portal Continuum Dark Matter}",
    eprint = "2105.14023",
    archivePrefix = "arXiv",
    primaryClass = "hep-ph",
    doi = "10.1103/PhysRevLett.128.081807",
    journal = "Phys. Rev. Lett.",
    volume = "128",
    number = "8",
    pages = "081807",
    year = "2022"
}

@article{Hong:2022gzo,
    author = "Hong, Sungwoo and Kurup, Gowri and Perelstein, Maxim",
    title = "{Dark matter from a conformal Dark Sector}",
    eprint = "2207.10093",
    archivePrefix = "arXiv",
    primaryClass = "hep-ph",
    doi = "10.1007/JHEP02(2023)221",
    journal = "JHEP",
    volume = "02",
    pages = "221",
    year = "2023"
}

@article{Hong:2024zsn,
    author = "Hong, Sungwoo and Perelstein, Maxim and Youn, Taewook",
    title = "{Conformal freeze-in from neutrino portal}",
    eprint = "2412.00181",
    archivePrefix = "arXiv",
    primaryClass = "hep-ph",
    doi = "10.1007/JHEP04(2025)089",
    journal = "JHEP",
    volume = "04",
    pages = "089",
    year = "2025"
}

@article{Ferrante:2023bcz,
    author = "Ferrante, Steven and Ismail, Ameen and Lee, Seung J. and Lee, Yunha",
    title = "{Forbidden conformal dark matter at a GeV}",
    eprint = "2308.16219",
    archivePrefix = "arXiv",
    primaryClass = "hep-ph",
    doi = "10.1007/JHEP11(2023)186",
    journal = "JHEP",
    volume = "11",
    pages = "186",
    year = "2023"
}

@article{Ferrante:2023fpx,
    author = "Ferrante, Steven and Lee, Seung J. and Perelstein, Maxim",
    title = "{Collider signatures of near-continuum dark matter}",
    eprint = "2306.13009",
    archivePrefix = "arXiv",
    primaryClass = "hep-ph",
    doi = "10.1007/JHEP05(2024)215",
    journal = "JHEP",
    volume = "05",
    pages = "215",
    year = "2024"
}

@article{Ferrante:2025ofe,
    author = "Ferrante, Steven and Luo, Lillian and Perelstein, Maxim and Youn, Taewook",
    title = "{Collider Searches for Near-Continuum Dark Matter}",
    eprint = "2510.17989",
    archivePrefix = "arXiv",
    primaryClass = "hep-ph",
    month = "10",
    year = "2025"
}

@article{Kaplan:2009ag,
    author = "Kaplan, David E. and Luty, Markus A. and Zurek, Kathryn M.",
    title = "{Asymmetric Dark Matter}",
    eprint = "0901.4117",
    archivePrefix = "arXiv",
    primaryClass = "hep-ph",
    reportNumber = "FERMILAB-PUB-09-345-A-T",
    doi = "10.1103/PhysRevD.79.115016",
    journal = "Phys. Rev. D",
    volume = "79",
    pages = "115016",
    year = "2009"
}

@article{Brzeminski:2023wza,
    author = "Brzeminski, Dawid and Hook, Anson",
    title = "{A Dynamical Explanation of the Dark Matter-Baryon Coincidence}",
    eprint = "2310.07777",
    archivePrefix = "arXiv",
    primaryClass = "hep-ph",
    doi = "10.1103/PhysRevLett.132.201001",
    journal = "Phys. Rev. Lett.",
    volume = "132",
    number = "20",
    pages = "201001",
    year = "2024"
}

@article{Banerjee:2024xhn,
    author = "Banerjee, Abhishek and Brzeminski, Dawid and Hook, Anson",
    title = "{Predicting the Dark Matter -- Baryon Abundance Ratio}",
    eprint = "2410.22412",
    archivePrefix = "arXiv",
    primaryClass = "hep-ph",
    month = "10",
    year = "2024"
}

@article{Chacko:2005pe,
    author = "Chacko, Z. and Goh, Hock-Seng and Harnik, Roni",
    title = "{The Twin Higgs: Natural electroweak breaking from mirror symmetry}",
    eprint = "hep-ph/0506256",
    archivePrefix = "arXiv",
    doi = "10.1103/PhysRevLett.96.231802",
    journal = "Phys. Rev. Lett.",
    volume = "96",
    pages = "231802",
    year = "2006"
}

@article{Gresham:2017cvl,
    author = "Gresham, Moira I. and Lou, Hou Keong and Zurek, Kathryn M.",
    title = "{Early Universe synthesis of asymmetric dark matter nuggets}",
    eprint = "1707.02316",
    archivePrefix = "arXiv",
    primaryClass = "hep-ph",
    doi = "10.1103/PhysRevD.97.036003",
    journal = "Phys. Rev. D",
    volume = "97",
    number = "3",
    pages = "036003",
    year = "2018"
}

@article{Terning:2019hgj,
    author = "Terning, John and Verhaaren, Christopher B. and Zora, Kyle",
    title = "{Composite Twin Dark Matter}",
    eprint = "1902.08211",
    archivePrefix = "arXiv",
    primaryClass = "hep-ph",
    doi = "10.1103/PhysRevD.99.095020",
    journal = "Phys. Rev. D",
    volume = "99",
    number = "9",
    pages = "095020",
    year = "2019"
}

@article{Ibe:2019ena,
    author = "Ibe, Masahiro and Kamada, Ayuki and Kobayashi, Shin and Kuwahara, Takumi and Nakano, Wakutaka",
    title = "{Baryon-Dark Matter Coincidence in Mirrored Unification}",
    eprint = "1907.03404",
    archivePrefix = "arXiv",
    primaryClass = "hep-ph",
    reportNumber = "IPMU19-0094, CTPU-PTC-19-19",
    doi = "10.1103/PhysRevD.100.075022",
    journal = "Phys. Rev. D",
    volume = "100",
    number = "7",
    pages = "075022",
    year = "2019"
}

@article{Feng:2020urb,
    author = "Feng, Wan-Zhe and Yu, Jiang-Hao",
    title = "{Twin cogenesis}",
    eprint = "2005.06471",
    archivePrefix = "arXiv",
    primaryClass = "hep-ph",
    doi = "10.1088/1572-9494/acbb5b",
    journal = "Commun. Theor. Phys.",
    volume = "75",
    number = "4",
    pages = "045201",
    year = "2023"
}

@article{Bodas:2024idn,
    author = "Bodas, Arushi and Buen-Abad, Manuel A. and Hook, Anson and Sundrum, Raman",
    title = "{A closer look in the mirror: reflections on the matter/dark matter coincidence}",
    eprint = "2401.12286",
    archivePrefix = "arXiv",
    primaryClass = "hep-ph",
    reportNumber = "FERMILAB-PUB-24-0023-T-V",
    doi = "10.1007/JHEP06(2024)052",
    journal = "JHEP",
    volume = "06",
    pages = "052",
    year = "2024"
}

@article{Hodges:1993yb,
    author = "Hodges, H. M.",
    title = "{Mirror baryons as the dark matter}",
    doi = "10.1103/PhysRevD.47.456",
    journal = "Phys. Rev. D",
    volume = "47",
    pages = "456--459",
    year = "1993"
}

@article{Chacko:2021vin,
    author = "Chacko, Zackaria and Curtin, David and Geller, Michael and Tsai, Yuhsin",
    title = "{Direct detection of mirror matter in Twin Higgs models}",
    eprint = "2104.02074",
    archivePrefix = "arXiv",
    primaryClass = "hep-ph",
    doi = "10.1007/JHEP11(2021)198",
    journal = "JHEP",
    volume = "11",
    pages = "198",
    year = "2021"
}
\bibliographystyle{utphys}

\end{document}